\documentclass[draft]{agujournal2019}
\usepackage{url}
\usepackage{lineno}
\usepackage[inline]{trackchanges}
\usepackage{soul}
\usepackage{amsmath}
\usepackage{amssymb}
\usepackage{mathrsfs}
\usepackage{caption}
\usepackage{color}
\addeditor{Victor Afigbo}

\draftfalse

\journalname{JGR: Planets}

\begin{document}

\title{Unveiling What Makes Saturn Ring \\[0.5ex] \large Quantifying the Amplitudes of Saturn's Planetary Normal-mode Oscillations and trends in C-ring Properties using Kronoseismology (VII)}

\authors{V. M. Afigbo\affil{1}, M. M. Hedman\affil{1}, P. D. Nicholson\affil{2}, R. G. French\affil{3}, C. R. Mankovich\affil{4}, R. G. Jerousek\affil{5}, J. Dewberry\affil{6}}

\affiliation{1}{Department of Physics, University of Idaho, Moscow, ID, USA}
\affiliation{2}{Department of Astronomy, Cornell University, Ithaca, NY, USA}
\affiliation{3}{Department of Astronomy, Wellesley College, Wellesley, MA, USA}
\affiliation{4}{Jet Propulsion Laboratory, California Institute of Technology, Pasadena, CA, USA}
\affiliation{5}{Department of Physics, University of Central Florida, Orlando, FL, USA}
\affiliation{6}{Canadian Institute for Theoretical Astrophysics, Toronto, ON, Canada}

\correspondingauthor{Victor Afigbo}{afig0258@vandals.uidaho.edu}

\begin{keypoints}

\item C-ring density wave profiles are fit to linear density wave models in order to obtain information about Saturn's interior and the C-ring.
\item 
The opacity and viscosity of the C-ring changes around 84,000 km, likely due to a change in the typical particle mass density.
\item Saturn's normal mode spectrum is a complex function of angular degree and frequency, indicating multiple excitation sources inside Saturn. 

\end{keypoints}

\begin{abstract}
Certain spiral density waves in Saturn's rings are generated through resonances with planetary normal modes, making them valuable probes of Saturn's internal structure. Previous research has primarily focused on the rotation rates of these waves. However, other characteristics of these waves also contain valuable information about the planet's interior. In this work, we investigate the amplitudes of the waves across the C-ring by analyzing high signal-to-noise profiles derived from phase-corrected averages of occultation profiles obtained by Cassini's Visual and Infrared Mapping Spectrometer (VIMS). By fitting these wave profiles to linear density wave models, we estimate the ring's surface mass density, mass extinction coefficient, and effective kinematic viscosity at 34 locations in the C-ring, as well as the amplitude of the gravitational potential perturbations associated with 6 satellite resonances and 28 planetary normal mode resonances.
Our estimates of the C-ring's mass extinction coefficient indicate that the typical particle mass density is around 0.3 g/cm$^3$ interior to 84,000 km, but can get as low as 0.03 g/cm$^3$ exterior to 84,000 km. We also find the ring's viscosity is reduced in the outer C-ring, which is consistent with the exceptionally high porosity of the particles in this region. 
Meanwhile, we find the amplitudes of Saturn's normal modes are complex functions of frequency, $\ell$ and $m$, implying that multiple factors influence how efficiently these modes are excited. 
This analysis identified two primary sources of these normal-mode oscillations: a deep source located close to Saturn's core, and a shallow source residing near the surface.
\end{abstract}

\section*{Plain Language Summary}
Saturn's C-ring contains waves, which are patterns generated mostly by the planet's internal vibrations. 
We analyze these patterns using data from the Cassini spacecraft in order to obtain new information about both the planet and its C-ring. We find systematic differences in the ring's properties which suggest that the ring particles are much more porous in the outer part of the C-ring. 
In addition, these patterns allow us to estimate the strength of the planet's internal vibrations. 
These amplitudes exhibit trends that suggest they may be excited by at least two different sources: a deep one near the planet's core and a shallow one closer to the surface. 

\section{Introduction}

Saturn has a variety of internal oscillations that produce detectable structures in its surrounding rings \cite{Marley1993PlanetaryAM}. These structures, known as spiral density and bending waves \cite{1984prin.conf..513S}, encode a wide variety of information about both the properties of the rings themselves and about the internal structure of the planet. Previous analyses of these waves have determined the azimuthal wave numbers and oscillation frequencies of over twenty-five planetary normal modes \cite{2013AJ....146...12H, Hedman_2014, Hedman_2018, 2019Icar..319..599F, French2021Icar}, which have already provided important insights into Saturn's internal structure and rotation state \cite{Fuller_2014, Mankovich2019, Mankovich2021, Dewberry_2022, Mankovich23}. However, more information can be extracted about both Saturn and its rings from other aspects of the observable waves. 

First of all, the amplitudes of the ring waves can be used to estimate the amplitudes of individual planetary normal mode oscillations, which depend upon how those modes are excited and damped within Saturn. In principle, there are a variety of ways that normal modes could be excited inside a giant planet, including sufficiently deep and intense storms \cite{Markham2018}. However, thus far, analyses of normal mode contributions to both Jupiter's and Saturn's gravitational fields have needed to assume that the amplitudes of normal modes are smooth and relatively simple functions of frequency in order to match the currently available spacecraft tracking data \cite{Markham_2020, 2022NatCo..13.4632D}. By contrast, preliminary studies of the ring waves indicated that the normal mode amplitudes could vary in rather complex ways \cite{Hedman2019}. Since different normal modes penetrate to different depths within the planet \cite{Marley1993PlanetaryAM}, more detailed and quantitative investigations of these wave amplitudes should clarify where and how planetary normal modes are being excited inside Saturn.

Furthermore, the waves generated by planetary normal modes are broadly distributed throughout the C-ring, and so can provide valuable information about the surface mass density and viscosity of this ring. The C-ring extends between 74,490 and 91,980 km and lies interior to  the much denser B-ring. It contains a variety of structures, including gaps, ringlets, regions of elevated optical depths called plateaux, and broad undulations in optical depth \cite{2009sfch.book..375C}. Prior studies of C-ring waves identified some interesting trends in the ring's surface mass density \cite{Rosen1991Icarus, BailliColwellLissauerEsposito, Hedman_2014}, but the available data on the inner half of the C-ring has been very limited because of the small number of identified waves in this region. A consistent analysis of all the density waves within the C-ring, including the many newly discovered waves in the inner C-ring \cite{2019Icar..319..599F} is therefore warranted. 

In this paper, we perform a comprehensive analysis of all the known density waves in Saturn's C-ring using high signal-to-noise profiles derived from stellar occultation observations made by the Visual and Infrared Mapping Spectrometer (VIMS) onboard the Cassini spacecraft.
In Section~\ref{Theory}, we provide a theoretical background for both planetary normal modes and for spiral density waves relevant to this project. Next, in Section~\ref{method}, we lay out what methods we used to obtain the required wave profiles and extract relevant parameters that can constrain properties of the planet's interior and its rings.  
The results of this analysis are presented in Section~\ref{results} and their implications for the rings and the planet are discussed in Section~\ref{discussion}.

\section{Theoretical Background}
\label{Theory}

Prior to describing how we analyzed the observations of the C-ring waves, it is useful to first review the basic theory of both planetary normal modes and spiral density waves. First, in Section~\ref{normalmodes} we describe how normal mode oscillations in the planet can give rise to perturbations in the planet's gravitational field. Then in Section~\ref{waves} we describe how these perturbations generate and influence the observable properties of density waves within the rings. 

\subsection{The Saturnian oscillations}
\label{normalmodes}

Oscillations inside fluid planets like Saturn can be decomposed into normal modes \cite{unno1989nonradial} whose properties depend on the object's internal structure. These normal modes involve perturbations in the planet's density, pressure, and gravitational potential. Each of these  oscillation modes corresponds to a density perturbation inside the planet $\rho'$ that can be expressed in terms of spherical harmonics:

\begin{equation}
\rho'(r,\theta,\phi, t)=
 \sum_{n=-\infty}^{\infty} \sum_{\ell=2}^{\infty}\sum_{m=-\ell}^{\ell}   
 \rho_{\ell mn}'(r)Y_{\ell m}(\theta,\phi^c_{\ell mn}(\phi,t)), 
\label{rhoeq}
\end{equation}
where $r$ is the radius, $Y_{\ell m}(\theta,\phi^c_{\ell mn})$ are the standard (real) spherical harmonics, with $\theta$ being the colatitude and $\phi^c_{\ell mn}(\phi,t)$ being the azimuth in the frame co-rotating with the normal mode in the planet. For the prograde-rotating modes ($m > 0$) that can form resonances in the rings, this angle can be written in terms of the fixed azimuth angle $\phi$ as $\phi^c_{\ell mn}=\phi-(\sigma_{\ell mn}t/m+\phi_{\ell mn})$, where $\sigma_{\ell mn}$ is the mode oscillation frequency \cite{Marley1993PlanetaryAM} and $\phi_{\ell mn}$ is the phase of the oscillation at $t=0$. Note that each mode is labelled by a combination of the spherical harmonic indices $\ell$ and $m$, as well as a third index $n$ that accounts for the fact that modes having the same $\ell$ and $m$ values can have different radial orders or propagation types, and hence, different frequencies. For example, at $\ell=m=2$, the spectrum may include an f-mode with $n=0$, g-modes with $n=-1,-2,\ldots$, and p-modes with $n=1,2,\ldots$, all with distinct frequencies \cite{TakataMasao2006}. The modes that are expected to most strongly affect the observable gravity field are those most similar to the so-called f-modes \cite{1981NASSP.450..263L, Marley1993PlanetaryAM, 2019ApJ...871....1M}. Strictly speaking, rotation as rapid as Saturn's couples variability with a fixed frequency and azimuthal order $m$ across multiple harmonic degrees $\ell$ \cite{Reese2006}, so identifying each mode with a single spherical harmonic is an approximation. We discuss the relevance of such coupling further in Section~\ref{sec:modes}.

Since we will be examining trends with $\ell$ and $\ell-m$, it is worth summarizing what those parameters physically represent. The angular degree $\ell$ measures the number of circles bounding the regions of positive and negative perturbations, while the azimuthal order $m$ shows the number of great circles passing through the polar axis, bounding these positive and negative perturbations. Thus when the angular order $m$ is zero, the harmonics are called zonal and oscillate only in the latitudinal direction, and when $\ell=m$ the harmonics are called sectoral and oscillate only in the longitudinal direction.  For other values of $m$, the harmonics oscillate in both the longitudinal and latitudinal directions. In general, the quantity $\ell-m$ accounts for the number of latitudinal circles forming latitudinal boundaries, while $2m$ accounts for the number of zero crossings in the longitudinal direction. 

Each normal mode inside the planet generates a corresponding perturbation in the planet's gravitational field that can affect its rings.

Consistent with previous works \cite{Marley1993PlanetaryAM}, we express the total gravitational potential of a planet as: 

\begin{equation}
    \Phi(t) = \Phi_{0} + \Phi'(t).
\end{equation}
Here, $\Phi_{0}$ is the unperturbed gravitational potential, which we will express as \cite{1978ppi..book.....Z, Marley1993PlanetaryAM}: 

\begin{equation}
\Phi_{0} = -\frac{GM}{r}\left\{1 - \sum_{\ell=1}^{\infty}\left(\frac{R_s}{r}\right)^{\ell}J_{\ell}P_{\ell}(\cos \theta) 
\right\},
\end{equation}

where G is the gravitational constant, M is the planet's mass, $r$ is the distance from the planet's center to a reference point, $R_s$ is the planet's (equatorial) radius. $J_{\ell}$ is the $\ell$-th multipole gravitational moment, which is a measure of how the external gravity field departs from spherical symmetry \cite{1978ppi..book.....Z, 1984prin.conf..513S,  Marley1993PlanetaryAM, Guillot2015, tremaine2023dynamics}. $P_{\ell}(\cos \theta)$ is the standard Legendre polynomial expressed as \cite{Guillot2015, https://doi.org/10.1029/2018GC007529, tremaine2023dynamics}:

\begin{equation}
    P_{\ell}(\cos \theta) = \frac{1}{2^\ell \ell!} \frac{d^\ell}{d(\cos\theta)^\ell} \left[(\cos^2\theta - 1)^\ell\right],
\end{equation}
where $\theta$ is the given colatitude. Note that since Saturn is a fluid planet in hydrostatic equilibrium, we neglect any permanent non-axisymmetric terms in this part of the potential. While Saturn does show evidence for gravitational anomalies that rotate at close to the planet's rotation rate \cite{Hedman_2022}, these perturbations can be neglected here.

By contrast, the time-variable part of the field $\Phi{'}$ can contain components with any combination of $\ell$ and $m$, and can even contain multiple components with the same $\ell$ and $m$ but different frequencies. This component of the field can be written in the following form \cite{Marley1993PlanetaryAM}:

%\begin{equation}
%\Phi{'}(t) = \frac{GM}{r}
%\sum_{n=-\infty}^{\infty} \left\{ - \sum_{\ell=0}^{\infty}\left(\frac{R_s}{r}\right)^{\ell}J_{\ell n}^{'}P_{\ell}(\cos \theta) +\sum_{\ell=0}^{\infty}\sum_{m=0}^{\ell}\left(\frac{R_s}{r}\right)^{\ell}P_{\ell m}(\cos \theta)[{C}_{\ell m n}^{'}\cos (m\phi^c_{\ell mn}) + {S}_{\ell m n}^{'}\sin(m\phi^c_{\ell mn})]\right\},
%\end{equation}

\begin{align}
\Phi{'}(t) &= \frac{GM}{r}
\sum_{n=-\infty}^{\infty} \Bigg\{ 
    - \sum_{\ell=0}^{\infty} \left(\frac{R_s}{r}\right)^{\ell} J_{\ell n}^{'} P_{\ell}(\cos \theta)  \nonumber \\
    &\quad +\sum_{\ell=0}^{\infty} \sum_{m=0}^{\ell} \left(\frac{R_s}{r}\right)^{\ell} P_{\ell m}(\cos \theta) 
    \Big[{C}_{\ell m n}^{'}\cos (m\phi^c_{\ell mn}) + {S}_{\ell m n}^{'}\sin(m\phi^c_{\ell mn}) \Big]
\Bigg\},
\end{align}

\noindent where again $n$ is an index that identifies different normal modes with the same $\ell$ and $m$ but different frequencies. Retaining only the f-modes ($n=0$, $\ell\geq2$) expected to dominate the time-dependent part of the external potential, this becomes,

\begin{equation}
\Phi'(t) = \frac{GM}{r}
\left\{ - \sum_{\ell=2}^{\infty}\left(\frac{R_s}{r}\right)^{\ell}J_{\ell 0}'P_{\ell}(\cos \theta) +\sum_{\ell=2}^{\infty}\sum_{m=2}^{\ell}\left(\frac{R_s}{r}\right)^{\ell}P_{\ell m}(\cos \theta)[{C}_{\ell m 0}'\cos (m\phi^c_{\ell m0}) + {S}_{\ell m 0}'\sin(m\phi^c_{\ell m0})]\right\}.
\end{equation}
Note that, unlike \citeA{Marley1993PlanetaryAM}, we denote the associated Legendre polynomials as $P_{\ell m}$ because we use the real versions of these functions, and this notation is consistent with this approach \cite{https://doi.org/10.1029/2018GC007529}. Note that for f-modes driving spiral waves in the rings, only values of $m \geq 2$ are relevant, so that we can henceforth neglect retrograde oscillations (negative $m$) and $m = 1$. We also neglect $m = 0$ oscillations in this study, which eliminates the $J_{\ell n}'$ terms from the equation. While such oscillations could in principle exist inside the planet and drive axisymmetric waves in rings \cite{Hedman2019}, no such wave has yet been attributed to such planetary oscillations. 

In addition, we can more explicitly document the time-dependence of the prograde non-axisymmetric normal modes by  re-writing the azimuth angle $\phi^c_{\ell m 0}$ explicitly in terms of $\phi$, $t$, $\sigma_{\ell m 0}$, and $\phi_{\ell m 0}$. With these changes the above expression becomes: 

\begin{equation}
\Phi'(t) = \frac{GM}{r}
\sum_{\ell=2}^{\infty}\sum_{m=2}^{\ell}\left(\frac{R_s}{r}\right)^{\ell}P_{\ell m}(\cos \theta)[{C}_{\ell m 0}'\cos (m\phi-\sigma_{\ell m 0}t-m\phi_{\ell m 0}) + {S}_{\ell m 0}'\sin(m\phi-\sigma_{\ell m 0}t-m\phi_{\ell m 0})].
\end{equation}
Furthermore, this expression can be simplified slightly by choosing the phase parameter $\phi_{\ell m 0}$ so that $S'_{\ell m 0}=0$, in which case we can re-write it as \cite{1978ppi..book.....Z}:

\begin{equation}
\Phi'(t) = \frac{GM}{r}
\sum_{\ell=2}^{\infty}\sum_{m=2}^{\ell} A_{\ell m 0}'\left(\frac{R_s}{r}\right)^{\ell}P_{\ell m}(\cos \theta) \cos (m\phi-\sigma_{\ell m 0}t-m\phi_{\ell m 0}).
\label{Aeq1}
\end{equation}

While Equation~\ref{Aeq1} provides a consistent way to express the planet's gravitational field, the $A'_{\ell m0}$ coefficients are not ideal parameters for quantifying the perturbations in the planet's gravity field. In particular, if we use Equation 12 of \citeA{Marley1993PlanetaryAM} to express these coefficients in terms of the density perturbations $\rho'_{\ell m0}$, we obtain the following equation:

\begin{equation}
A'_{\ell m0}=\sqrt{\frac{4\pi}{2\ell+1}\frac{(\ell-m)!}{(\ell+m)!}}\int_0^{R_s} \frac{\rho'_{\ell m0}(r) r^{\ell+2}}{M R_s^\ell} dr. 
\end{equation}
While the integral is a sensible measure of the corresponding perturbation to the planet's internal structure, the prefactor arises solely from how the spherical harmonics and Legendre polynomials are normalized and can vary by several orders of magnitude over the range of $\ell$ and $m$ values observed in Saturn's normal modes. 

In this sort of scenario, it is preferable to use the normalized form of the Legendre polynomials $\bar{P}_{\ell m}$, which are commonly used in the seismology community \cite{2007gdsy.book...11J, 2015JASS...32..247S, https://doi.org/10.1029/2018GC007529} and can be derived from the unnormalized Legendre coefficients using the following equation: 

\begin{equation}
    \bar{P}_{\ell m} (\mu) = \sqrt{\frac{(2-\delta_{0m})(2\ell+1)}{4\pi} \frac{(\ell-m)!}{(\ell+m)!}} P_{\ell m} (\mu),
\end{equation}
where  $\delta_{0m}$ is the Kronecker delta function. Note that this specific normalization ensures the corresponding real spherical harmonics are proper orthonormal functions \cite{https://doi.org/10.1029/2018GC007529}.

Re-writing Equation~\ref{Aeq1} in terms of these normalized Legendre polynomials, we obtain the following expression:

\begin{equation}
\Phi'(t) = \frac{GM}{r} \sum_{\ell=2}^{\infty}\sum_{m=2}^{\ell}\left(\frac{R_s}{r}\right)^{\ell}\sqrt{\frac{4\pi}{(2-\delta_{0m})(2\ell+1)}\frac{(\ell+m)!}{{(\ell-m)!}}}A'_{\ell m0}\bar{P}_{\ell m}(\cos\theta)\cos(m\phi-\sigma_{\ell m0}t-m\phi_{\ell m0}),
\end{equation}
or, more compactly:

\begin{equation}
\Phi'(t) = \frac{GM}{r} \sum_{\ell=2}^{\infty}\sum_{m=2}^{\ell}\left(\frac{R_s}{r}\right)^{\ell}\mathscr{A}'_{\ell m 0}\bar{P}_{\ell m}(\cos\theta)\cos(m\phi-\sigma_{\ell m0}t-m\phi_{\ell m0}),
\end{equation}
and

\begin{equation}
\Phi'(t) = \frac{GM}{r} \sum_{\ell=2}^{\infty}\sum_{m=2}^{\ell}\left(\frac{R_s}{r}\right)^{\ell}\mathscr{A}'_{\ell m 0} Y_{\ell m}(\theta,\phi^c_{\ell m0}) = 
\sum_{\ell=2}^{\infty}\sum_{m=2}^{\ell}{\Phi}'_{\ell m 0}(r) Y_{\ell m}(\theta,\phi^c_{\ell m0}),
%m\phi-\sigma_{\ell mn}t-\phi_{\ell mn}).
\label{Aeq2}
\end{equation}
where $\phi^c_{\ell m0}$ is the same (time-dependent) angle used in Equation~\ref{rhoeq}, and $\Phi'_{\ell m0}(r)$ are simply a more compact expression for the spherical harmonic coefficients. Note that the coefficients for this new expansion are given by the following formula:

\begin{equation}
\mathscr{A}'_{\ell m 0}=\sqrt{\frac{4\pi}{(2-\delta_{0 m})(2\ell+1)}\frac{(\ell+m)!}{{(\ell-m)!}}}A'_{\ell m0},
\label{normamp}
\end{equation}
which means that these coefficients have a much more straightforward relationship with the density perturbations inside the planet, 

\begin{equation}
\mathscr{A}'_{\ell m 0}=\frac{4\pi}{2\ell+1}\int_0^{R_s} \frac{\rho'_{\ell m0}(r) r^{\ell+2}}{M R_s^\ell} dr. 
\end{equation}
It is important to note that this particular relationship assumes that the spherical harmonics used in Equations~\ref{rhoeq} and ~\ref{Aeq2} are both real, which cancels out the factor of $2-\delta_{0 m}$. 

\subsection{The Theory of spiral density waves: Linear density wave model}
\label{waves}

The perturbations to the planet's gravitational field generated by the internal normal modes can generate spiral density waves within the rings. The observable properties of these waves depend on both the $\mathscr{A}'_{\ell m 0}$ coefficients derived above and characteristics of the ring such as its surface mass density and effective viscosity \cite{1984prin.conf..513S, 1990AJ....100.1339N, 2007Icar..189...14T}.   

Fortunately,  a mature theory for such density waves has already been developed for waves generated by the gravitational perturbations from various satellites \cite{1984prin.conf..513S, 1990AJ....100.1339N, 2007Icar..189...14T}, and this theory can be extended to also apply to waves generated by planetary normal modes \cite{Marley1993PlanetaryAM, 2013AJ....146...12H}. Here we summarize key aspects of this model.

Spiral density waves arise at locations that correspond to Lindblad resonances with the perturbation in the gravitational field. For a perturbation with an azimuthal wavenumber $m$ rotating with a pattern speed $\Omega_P$, these resonances occur where the ring particle's orbital mean motion $n$ and radial epicyclic frequency $\kappa=n-\dot{\varpi}$ satisfy the relationship $m(n-\Omega_{P}) = \pm \kappa = \pm(n-\dot{\varpi})$, or $(m \mp 1)n \pm \dot{\varpi} = m\Omega_{P} = \sigma_{\ell m n}$. This expression yields two different resonant locations, but in practice, it is useful to treat these two solutions as corresponding to values of $m$ with positive and negative signs \cite{2013AJ....146...12H}. With this notational change, each resonance corresponds to a location that satisfies the equation $(m - 1)n + \dot{\varpi} = m\Omega_{P}$, which simplifies many of the following equations. Note that this is simply a convenient convention and does not imply that the mode or the wave is retrograde, as $\Omega_{P}$ is still positive.

The spiral density wave generated at these resonances manifests as fluctuations in the ring's surface mass density that consist of $|m|$ spiral arms rotating around the planet at a rate of $\Omega_{P}$. The perturbed surface mass density $\sigma(r,\phi,t)$, normalized to the background surface mass density in areas outside the waves' region, $\sigma_0$, is represented by the following expression (Equation B2 in \citeA{1990AJ....100.1339N}):

\begin{equation}
\frac{\sigma(r,\phi,t)}{\sigma_0} = 1 + \Re\{A_{L}e^{i(\phi_L-\pi/2)}[\pi^{-1/2}+2\xi e^{-i\pi/2}H(\xi)]\}e^{-(\xi/\xi_{D})^{3}},
\label{lineardensitywaveeqgn}
\end{equation}
where $\phi$ and $t$ are the (inertial) longitude and time of observation, respectively. $A_{L}$ is the dimensionless amplitude factor that depends on the strength of the perturbation in the planet's gravitational field that induces the wave (see below), while $\phi_L = |m|(\phi-\Omega_P t)$ is the wave phase. $\xi$ is a dimensionless quantity that specifies the distance from the resonant radius, $r_{L}$, and it is given by the following expression:

\begin{equation}
\xi = \left[\frac{\mathcal{D}_{L}r_{L}}{4\pi G \sigma_{0}}\right]^{\frac{1}{2}}\left(\frac{r-r_{L}}{r_{L}}\right),
\end{equation}

where again $G$ is the gravitational constant, $\mathcal{D}_{L}$ = $3|m-1|n_L^2 + J_{2}(\frac{R_{s}}{r_{L}})^{2}[\frac{21}{2}-\frac{9}{2}|m-1|]n_L^2$ \cite[the second term in $\mathcal{D}_{L}$ being a small correction except when $m=1$]{2007Icar..189...14T}, $r$ is the radial distance from the center of the planet, $r_{L}$ is the radius of the exact resonance with the perturbation, and $n_L$ is the mean motion of ring particles at that location. $n_L^2 \simeq {G M_P}/{r_L^3}$. $H(\xi)=\pi^{-1/2}e^{-i\xi^2}\int^\xi_{-\infty} e^{i\eta^2} d\eta$ is a standard Fresnel Integral, and finally $\xi_{D}$ is the dimensionless damping parameter that is sensitive to the effective ring viscosity, $\nu$ via the expression:

\begin{equation}
    \nu = \frac{9}{7 \Omega_L\xi_{D}^{3}}\left(\frac{r_{L}}{\mathcal{D}_{L}}\right)^{\frac{1}{2}}(2\pi G \sigma_{0})^{\frac{3}{2}},
\label{viscoeq}
\end{equation}
which matches Equation 10 from \citeA{1990AJ....100.1339N}. 

Using De Moivre's theorem and the principles of Cauchy's integral calculus (residue theorem) 
\cite{1968hmfw.book.....A, Arfken2005, NIST2011}, Equation~\ref{lineardensitywaveeqgn} can be approximated as: 

\begin{equation}
\frac{\sigma(r,\phi,t)}{\sigma_{0}} = 1 + {\{A_{L}\xi\left[\,\text{sgn}({m})\text{sgn}(\xi) + 1\right]\cos{(\phi_L - 3\pi/4 - \xi^{2})}}\}e^{-\left(\frac{|\xi|}{\xi_{D}}\right)^3}.
\label{lineardensitywavemodel}
\end{equation}
We can  therefore express the density variations $\Delta \sigma=\sigma-\sigma_0$ as:

\begin{equation}
\frac{\Delta\sigma(r,\phi,t)}{\sigma_{0}} = A_{L}\xi e^{-\left(\frac{|\xi|}{\xi_{D}}\right)^3}\cos(\phi_L-3\pi/4-\xi^2)[1+\text{sgn}(m)\text{sgn}(\xi)],
\end{equation}
where the factor of $[1+\text{sgn}(m)\text{sgn}({\xi})]$ accounts for the fact that the waves only exist on one side of the resonance.

Re-writing the above expression explicitly in terms of $r-r_L$  gives:

\begin{equation}
\frac{\Delta\sigma(r,\phi,t)}{\sigma_{0}} =A_{L}
\frac{(r-r_L)}{r_f}e^{-\left(\frac{|r-r_L|}{\xi_D r_f}\right)^3}\cos{\left(\phi_L - 3\pi/4 - \left(\frac{r-r_L}{r_f}\right)^{2}\right)}[1+\,\text{sgn}({m})\text{sgn}\left(\frac{r-r_L}{r_f}\right)],
\label{sigeq1}
\end{equation}
where $r_f = r_L\sqrt{\frac{4\pi G \sigma_{0}}{\mathcal{D}_{L}r_{L}}}$ and $\xi_D r_f$ is the physical damping length of the wave. So the adjustable parameters of our density wave model are $A_{L}$, $r_{L}$, $r_{f}$, $\xi_{D}$ and the local wave phase $\phi_L$. Note, it is workable to remove a factor of $r_f$ from the signum function, since $r_f$ is always positive. Also note that with this particular parametrization, the mean surface mass density is determined by $r_f$:

\begin{equation}
\sigma_0=\frac{\mathcal{D}_{L} r_f^2}{4\pi G r_L}.
\label{sigeqf}
\end{equation}

Of particular interest for this investigation is the amplitude $A_{L}$, which is directly proportional to the amplitude of the driving term in the perturbation, more like a ``forcing function" for density waves, which is designated $\Psi_{\ell m}^{'}$ in \citeA{Marley1993PlanetaryAM} and  $\Psi_L$ in \citeA{1984prin.conf..513S}, whose notation we will follow here.

In order to determine the constant of proportionality between $A_{L}$ and $\Psi_L$, it is useful to consider a variant of the amplitude of the surface mass density variations $S(r)$ given in \citeA{1984prin.conf..513S}, which can be expressed in our notation as: 

\begin{equation}
\frac{S(r)}{\sigma_{0}} =A_{L}%\sqrt{\frac{3|m-1|M_P}{4\pi \sigma_{0} r_L^2}}
\frac{(r-r_L)}{r_f}e^{-\left(\frac{|r-r_L|}{r_{D}}\right)^3}[1+\,\text{sgn}({m})\text{sgn}({r-r_L})].
\label{sigeq2}
\end{equation}
For comparison, Equation 54 of \citeA{1984prin.conf..513S} says that the amplitude of the surface mass density variation near the resonance is given by the following expression:

\begin{equation}
    \frac{S(r)}{\sigma_{0}} = \sqrt{\frac{2\pi}{|\epsilon|^{3}}} \frac{|\Psi_L|}{r_{L}^{2}|\mathcal{D}_{L}|} \left(\frac{r-r_L}{r_L}\right),
\end{equation}

where the dimensionless parameter $\epsilon$ is given by Equation (45b) of \citeA{1984prin.conf..513S}:

\begin{equation}
    \epsilon = \frac{2\pi G \sigma_{0}}{r_{L}\mathcal{D}_{L}} =  \frac{2\pi G \sigma_{0}}{3|m-1|n^{2}r_{L}} =  \frac{2\pi \sigma_{0} r_L^2}{3|m-1|M_P},
\end{equation}
and the second equality  is applicable for $m \neq 1$. 

Combining the previous two equations allows us to re-express the amplitude of the surface mass density variations as follows:

\begin{equation}
    \frac{S(r)}{\sigma_{0}} =  \frac{|\Psi_L|}{\sqrt{\pi}G\sigma_0r_L} \left(\frac{r-r_L}{r_f}\right).
\label{APsieq1}
\end{equation}

Comparing Equations~\ref{sigeq2} and~\ref{APsieq1}, and recalling that the latter is only applicable where $r-r_L \ll r_D$, means that we can derive the following relationship between $A_L$ and $\Psi_L$:

\begin{equation}
2A_L = \frac{\Psi_L}{\sqrt{\pi}G\sigma_{0} r_{L}},
\label{Psieq}
\end{equation}
where the factor of 2 on the left-hand side of Equation~\ref{Psieq} arises from evaluating the signum function. Note that this expression yields an equation for the wave amplitude that differs by a factor of $\sqrt{\pi/2}$ from Equation 60 in \citeA{Hedman_2022}, which reflects an error in that paper. 

In general, the perturbation $\Psi_L$ is the following function of the corresponding component of the perturbed potential from Equation 45a of \citeA{1984prin.conf..513S} and Equation 19 of  \citeA{Marley1993PlanetaryAM}:

\begin{equation}
    \Psi_{L} = r \frac{{d \Phi}_{m}'}{dr} \pm 2|m|\Phi_{m}',
\end{equation}
where $\Phi_{m}'$ is a component of the perturbed potential in the ring plane derived using an expansion of the following form:

\begin{equation}
\Phi'(r,\theta, \phi, t)=\sum_m \Phi'_m(r, \theta) \cos(m(\phi-\Omega_P t)+\phi_o),
\label{phiexp2}
\end{equation}
where $\Omega_P$ is the pattern speed of the perturbation and $\phi_o$ is a phase offset. See also \citeA{Gold1982}.

For a first-order resonance with a satellite, these terms in the potential can be evaluated in terms of the mass of the satellite $\mathcal{M}_s$, yielding \cite{1990AJ....100.1339N}:

\begin{equation}
    \Psi_{L} = \frac{G\mathcal{M}_s}{a_s}\left[\alpha\frac{db^m_{1/2}}{d\alpha}+2mb^m_{1/2}\right],
\end{equation}
where $a_s$ is the semi-major axis of the satellite, $\alpha=r_L/a_s$, and $b^m_{1/2}(\alpha)$ are Laplace coefficients \cite{BrouwerClemence, murray_dermott_2000}. Hence, for these sorts of resonances, the amplitude of the wave is given by:

\begin{equation}
    A_L = \frac{\mathcal{M}_s}{2\sqrt{\pi}\sigma_0r_La_s}\left[\alpha\frac{db^m_{1/2}}{d\alpha}+2mb^m_{1/2}\right],
\end{equation}
while more complex expressions are needed for higher-order satellite resonances \cite{TiscarenoandHarris2018}.

For resonances with planetary normal modes, $\Psi_{L}$ is proportional to the corresponding spherical harmonic coefficient in the planet's gravitational potential. However, we must take care to note the differences in the expansions used in Equations~\ref{Aeq2} and~\ref{phiexp2}. Using Equation 20 of \citeA{Marley1993PlanetaryAM} and Equation~\ref{Aeq2} above, this gives:

\begin{equation}
   \Psi_L = (2|m| +\ell+1)\Phi{'}_{m}=(2|m| +\ell+1)\frac{GM_P}{r}\left(\frac{R_s}{r}\right)^\ell\mathscr{A}'_{\ell m0}\bar{P}_{\ell m}(0),
\end{equation}
where the factor of $\bar{P}_{\ell m}(0)$ arises because the spherical harmonic component in Equation~\ref{Aeq2} needs to be evaluated in the ring plane to produce each of the terms shown in Equation~\ref{phiexp2}. Hence, for these sorts of resonances, the amplitude of the wave is:

\begin{equation}
   A_L =(2|m| +\ell+1)\frac{M_P}{2\sqrt{\pi}\sigma_0r_L^2}\left(\frac{R_s}{r_L}\right)^\ell\bar{P}_{\ell m}(0) \mathscr{A}'_{\ell m0}.
\end{equation}

The coefficient of the relevant component of the gravitational potential responsible for generating the wave can therefore be derived from the wave amplitude using the following formula: 

\begin{equation}
   \mathscr{A}'_{\ell m0} =\frac{2\sqrt{\pi}\sigma_0r_L^2}{(2|m|+\ell+1)M_P \bar{P}_{\ell m}(0)}\left(\frac{r_L}{R_S}\right)^\ell A_L=\frac{3|m-1|r_f^2}{2\sqrt{\pi}(2|m|+\ell+1) \bar{P}_{\ell m}(0)r_L^2}\left(\frac{r_L}{R_S}\right)^\ell A_L,
   \label{almneq}
\end{equation}
where the second equality uses Equation~\ref{sigeqf} to express the mean surface mass density in terms of the observable parameter $r_f$.  

\section{Methods}
\label{method}

The previous section demonstrates that the observable properties of the density waves ($A_L, r_f, \xi_D$) depend upon the amplitudes of the oscillations inside the planet ($\mathscr{A}'_{\ell m0}$), as well as the local surface mass density ($\sigma_0$) and effective viscosity of the rings ($\nu$). Those observable parameters can therefore be used to constrain those physical properties of the planet and the rings. In this section, we will describe the methods we used to estimate these parameters for most of the density waves in Saturn's C-ring. We begin by describing the VIMS occultation data used for this study in Section~\ref{data}. We then describe the custom wavelet algorithms we used to transform the data from the individual occultations into high signal-to-noise profiles of each wave in Section~\ref{wavelet}. Finally, we describe how we fitted these profiles and extracted estimates of relevant parameters and their uncertainties in Section~\ref{fit}. 

\subsection{Data acquisition}
\label{data}

The raw data for this study are derived from seventy-seven (77) stellar occultations observed by the Visual and Infrared Mapping Spectrometer (VIMS) onboard the Cassini spacecraft. Details of the instrument can be found in \citeA{Brown2004SSR} and an overview of the occultation observations of the rings is provided by \citeA{Nichlson2020Icarus}. 

During each occultation, VIMS monitored the brightness of the star as it passed behind the rings as viewed from the spacecraft, providing a high-resolution profile of the rings' optical depths. As with previous analyses of these data \cite{2013AJ....146...12H, Hedman_2014, Hedman2019, 2019Icar..319..599F, French2021Icar, Hedman_2022}, we use the appropriate SPICE kernels \cite{Acton1996PSS} to convert the timing information encoded with each occultation into estimates of the radius and (inertial) longitude where the starlight passed through the rings. These calculations include timing offsets that ensure circular ring features are aligned to within 150 m \cite{French2017Icarus}. In addition, we corrected the radius scale by a factor of $\delta r_T$ to account for the widespread effects of the Titan 1:0 apsidal resonance \cite{Nichoslon2014Icarus}. Consistent with \citeA{French2021Icar}, our correction factor was $\delta r_T = -382 {\rm  km}^2 \cos(\phi-\phi_T)/(r-77861.5 {\rm km})$, where $r$ and $\phi$ are the observed radius and longitude in the rings and $\phi_T$ is the corresponding longitude of Titan \cite{Nichoslon2014Icarus}.

After determining the geometry of each occultation, we converted the measured star brightness as a function of time between 2.87 $\mu$m and 3.00 $\mu$m into profiles of the transmission through the rings $T$ versus ring radius. Since VIMS' response function is highly linear \cite{Brown2004SSR}, we calculated $T$ as the ratio of the observed ring signal at any radius to the average signal in a region where the star is not blocked by the rings. Also, to facilitate the following wavelet analysis, all the occultation data are binned and/or interpolated onto a uniform grid of ring radii sampled every 100 meters. These data, along with the corresponding observed times and longitudes, were saved to a series of standardized files listed in Table~\ref{partab:C1} that provided the input data into the Python code used for this analysis.

In order to make the various occultation profiles directly comparable to one another, we first convert the observed transmission $T$ to normal optical depth $\tau_n=-\sin|B|\ln(T)$, where $B$ is the ring opening angle to the star. For low optical depth rings like the C-ring, $\tau_n$ should be independent of the viewing geometry. 

For each wave, we only considered a subset of the available occultations. First of all, we automatically excluded any occultation where the average normal optical depth in a region near the wave deviates from the median value of this parameter by more than 0.1 (parameter {\sl mthres} in Table~\ref{partab:C1}, this removes occultations with problematic background levels), and any occultation where the maximum radial resolution was above 1 km (parameter {\sl rthres} in Table~\ref{partab:C1}). We also examined the remaining optical depth profiles and deliberately excluded occultations with obvious issues like data gaps, cosmic ray spikes, and occultations where the star was poorly centered in the pixel which showed an oscillatory signal outside the rings. The specific list of occultations used for each wave is provided in Tables~\ref{obstab:C2} and~\ref{obstab2:C3}.

\subsection{Wavelet analysis}
\label{wavelet}

Wavelet-based analysis has already proven to be a useful way to isolate signals from density waves that are not apparent in individual occultation profiles \cite{Bijaoui1999, Hedman2019, 2019Icar..319..599F, French2021Icar, Guo2022_9785993, Hedman_2022, Wang_2023}. The novel aspect of this particular analysis is that we are using wavelet-based tools to obtain high signal-to-noise profiles of the waves that can provide useful estimates of normal mode amplitudes and ring properties. This sort of analysis can be broken down into three steps. First, we apply a continuous wavelet transform to each occultation profile. Second, we apply appropriate phase corrections to each wavelet and average the phase-corrected wavelets together to obtain a single, high signal-to-noise wavelet transform. Third, we transform the average wavelet back into a single high signal-to-noise optical depth profile of the appropriate wave.
In order to ensure that the wavelets and the final profile are properly normalized, we use a dedicated Python script for performing each of these steps, whose algorithms are described in detail below. 

\subsubsection{The Continuous Wavelet Transformation}

Consider a signal $y$ that is a function of ring radius $r$, say $y(r)$. A continuous wavelet transformation involves convolving this signal with a normalized wavelet function $\psi(r) \in L^2(\mathbb{R})$ that has zero average  \cite{1989wtfm.conf....2G, 1990ITIT...36..961D, Vaidyanathan1994, Burrus1998, 1998BAMS...79...61T, mertins2001signal, SteinShakarchi2011, pereyra2012harmonic}, and whose dependence on radius involves rescaling parameters $s$ and $R$ such that:

\begin{equation} 
\psi_{s,R}(r) = f(s) \psi\left(\frac{r-R}{s}\right).
\label{scale}
\end{equation}
A common choice for the prefactor $f(s)$ is $\frac{1}{\sqrt{|s|}}$, which ensures all functions with $s\in\mathbb{R}$ have the same energy \cite{1989wtfm.conf....2G, 1990ITIT...36..961D, Burrus1998, 1998BAMS...79...61T, mertins2001signal, 2022PhRvR...4c3078R}. However, in this case, we are primarily interested in being able to recover a normalized profile from the wavelet, and so for the sake of accuracy in results and simplicity in that step of the process (see below), we instead choose $f(s)=1/s$. The Continuous Wavelet Transformation of the signal $y(r)$ for a given set of $s$ and $R$ can therefore be expressed as: 

\begin{equation} 
\mathcal{Y}(s,R) 
= \langle y, \psi_{s,R} \rangle 
=\int_{-\infty}^{+\infty} y(r) \psi_{s,R}^{*}(r) \, dr.
\end{equation}
Here, $\psi_{s,R}^{*}(r)$ is the complex conjugate of $\psi_{s,R}(r)$. 

For this particular analysis, we use the standard Morlet wavelet, comprising a plane wave modulated by a Gaussian \cite{1989wtfm.conf....2G, 1992tlw..conf.....D, 1992AnRFM..24..395F, 1998BAMS...79...61T, mertins2001signal, SteinShakarchi2011, pereyra2012harmonic}:

\begin{equation} 
\psi(r)= \pi^{-1/4}e^{i\omega_{0}r}e^{-r^2/2},
\end{equation}
where $\omega_{0}=6$ is a dimensionless frequency that satisfies the admissibility condition \cite{1989wtfm.conf....2G, 1992AnRFM..24..395F, 1998BAMS...79...61T, mertins2001signal, SteinShakarchi2011, pereyra2012harmonic}. The value of 6 for this parameter, $\omega_{0}$, has been a standard choice in previous analyses of waves in Saturn's rings \cite{2007Icar..189...14T, 2013AJ....146...12H}. 

Note that while Morlet wavelets lack compact support, they are infinitely differentiable. This property makes them ideal for multiresolution analysis through dilation and translation, which is precisely why they are often used in this context.

In practice, computing convolutions for all the required values of $s$ and $R$ is slow, so our algorithm instead uses Fast Fourier Transforms (FFTs) to compute the required wavelet transformations. That is, we first evaluate the FFT of $y(r)$, which we can designate $\hat{y}(k)$. Then for each spatial scale $s$ we multiply this transform by the Fourier Transform of the Morlet wavelet:

\begin{equation}
\hat{\psi}(sk)=\pi^{-1/4}e^{-(sk-\omega_0)^2/2}.
\end{equation}
Finally, we perform the inverse FFT on each of these products $\hat{y}(k)\hat\psi(sk)$, which yields the desired wavelet transform $\mathcal{Y}(s,R)$, so long as we choose $f(s)=1/s$ in Equation~\ref{scale}.

\subsubsection{Phase-correcting the Wavelet Transforms}
Transforming the occultation profiles into wavelets enables us to account for the variable phases of the density waves among the different profiles. The wavelet derived from profile $i$ is a complex array that can be expressed as:
\cite{Hedman2019}, 

\begin{equation}
    \mathcal{Y}_i(s,R) = 
    Y_{i}e^{i\varphi_{L,i}},
\end{equation}
where $Y_{i}$ and $\varphi_{L,i}$ are real quantities. For the signals from the density waves, the wavelet phase is equivalent to the wave phase in Equation~\ref{sigeq1}, so $\varphi_{L,i} = \phi_{L,i}-3\pi/4 -(R-r_L)^2/r_f^2$ with $\phi_{L,i}=|m|(\phi_i-\Omega_Pt_i)$, where $\phi_i$ and $t_i$ are the observed longitude and time for the specific occultation and $\Omega_P$ is the pattern speed of the wave. Note that for a stable density wave, $\phi_{L,i}$ is the only part of $\mathcal{Y}_i$ that should vary from occultation to occultation, with both $Y_i$ and the radius-dependent part of $\varphi_{L,i}$ being constant among all the occultations.

Since the longitudes $\phi_{i}$ and times $t_{i}$ are known quantities for each occultation, and the values of $m$ and $\Omega_P$ have been determined for each wave, we can estimate $\phi_{L,i}$ for each wavelet transform of each wave, and compute the corresponding phase-corrected wavelet as:

\begin{equation}
    \mathcal{Y}^\phi_{i}(s,R) = \mathcal{Y}_{i}(s,R)e^{-i\phi_{L,i}} = Y_ie^{i(\varphi_{L,i} - \phi_{L,i})}.
\end{equation}

Since $\varphi_{L,i}-\phi_{L,i}$ should be the same for all the occultation profiles, the wave signal should be preserved in the average phase-corrected wavelet:

\begin{equation}
    \overline{\mathcal{Y}}^{\phi}(s,R) = \frac{1}{N} \sum_{i=1}^{N}\mathcal{Y}^{\phi}_{i}(s,R).
\end{equation}
By contrast, any signal without similar phase attributes will average to zero, so this quantity should isolate the wave signal from both background trends and fluctuations due to noise in each profile \cite{Hedman2019}.

\subsubsection{Reconstructing a single profile from the average phase-corrected wavelet}

A profile can be reconstructed from the average phase-corrected wavelet using a variety of techniques
\cite{Caldern1964IntermediateSA, 1998BAMS...79...61T, SteinShakarchi2011, pereyra2012harmonic, 2022PhRvR...4c3078R}.

In this particular situation, this reconstruction is best done using a version of Morlet's technique \cite{1992AnRFM..24..395F,2022PhRvR...4c3078R}. This yields the following expression for the reconstructed profile:

\begin{equation}
\bar{y}_\phi(R)= \mathscr{C}\int_0^\infty \Re\left(\bar{\mathcal{Y}}^\phi(s,R)\right)\frac{ds}{s},
\label{reconstruct}
\end{equation}
where $\mathscr{C} = \frac{\sqrt{2}}{\pi^{1/4}}\omega_0$. See~\ref{reconstructionconstant} for a detailed calculation showing that this formula yields the required normalization for the wave profile. 

In practice, we evaluate this integral by computing the wavelet for a series of uniformly-spaced scales $s_i$ between 0.001 km and 15 km with spacing $\Delta s$ and calculating the discrete sum:

\begin{equation}
\bar{y}_\phi(R)=\frac{\sqrt{2}}{\pi^{1/4}}\omega_0\sum_i \Re\left(\bar{\mathcal{Y}}^\phi(s_i,R)\right)\frac{\Delta s}{s_i}.
\label{reconstruct2}
\end{equation}
The finite range of scales used in this sum acts to filter out residual signals at very long and short wavelengths, but does not significantly affect the wave signal itself.

\subsection{Wave-fitting routine}
\label{fit}

The final step of our analytical procedures is to fit the reconstructed wave profile and extract estimates of the relevant ring parameters and gravitational harmonic amplitudes. 

\subsubsection{Fitting wave parameters}

Equation~\ref{sigeq1} provides an explicit prediction for the fractional variations in the ring's surface mass density $\Delta \sigma/\sigma_0$, though the occultation data do not directly measure this quantity. However, if we make the common assumption that the normal optical depth of the ring is directly proportional to the surface mass density $\tau_n=\kappa\sigma$, then $\Delta \sigma/\sigma_0=\Delta \tau_n/\tau_0$, where $\Delta \tau_n/\tau_0$ are the fractional variations in the optical depth due to the wave and $\kappa$ is known as the ring's mass extinction coefficient. In practice, this quantity corresponds to the ratio of the profile $\bar{y}_\phi(r)$ derived in the previous subsection to the average normal optical depth profile $\bar{\tau}_n(r)$. Hence, for most of the reconstructed waves, we estimate $\Delta \tau_n/\tau_0$ as $\bar{y}_\phi(r)/\bar{\tau}_n(r)$.  However, two of the waves (W74.51 and W76.46) fall very close to the edges of narrow gaps \cite{2019Icar..319..599F, French2021Icar} where the mean ring optical depth is zero and so the above ratio becomes ill-defined. For these two waves, we instead estimate $\Delta \tau_n/\tau_0$ as $\bar{y}_\phi(r)/\tau_{av}$, where $\tau_{av}$ is the radially-averaged value of the $\bar{\tau}_n(r)$ profile within the wave.

If we denote the fractional optical depth variations as $y$ and the radial displacements from the nominal resonance location as $x-x_{r}$, then we can fit these data to the following function using the {\tt scipy.optimize.curve\_fit} program in the Scipy Python package \cite{2020SciPy-NMeth} using the Trust Region Reflective Method (TRF) \cite{More1983, PressWilliam1986, DennisSchnabelRobert, nocedal2006numerical, Gill2019}:

\begin{equation}
    y(x) = A_{L}\frac{(x-x_{r})}{r_{f}}e^{-\left(\frac{|x-x_{r}|}{\xi_{D}r_{f}}\right)^{3}}\cos \left(\phi_L - 3\pi/4 - \left(\frac{x-x_{r}}{r_{f}}\right)^{2} \right)
    \left[1 + \mathrm{sgn}(m) \mathrm{sgn}\left(\frac{x-x_{r}}{r_{f}}\right)\right],
\label{wavefiteg}
\end{equation}

where $m$ is assumed to be the appropriate signed value for each wave, and  $A_{L}$, $x_{r}$, $r_{f}$, $\xi_{D}$, and $\phi_L$ are all fit parameters. $x_{r}$ is the correction to the resonance radius $r_{L}$. In order to ensure fit convergence, we impose bounds on each of these parameters for each wave (see Table~\ref{boundtabMAIN:C4} for the values used for each wave).

We then use the above fit parameters to compute the following physical parameters:

{\bf Ring surface mass density, $\sigma_{0}$.}

This parameter is given explicitly by Equation~\ref{sigeqf}, which can be re-written approximately (for a case when $m \ne 1$ ) as:

\begin{equation}
\label{sigmasigma0}
 \sigma_{0} \simeq \frac{3 |m-1| M_{P}r_{f}^{2}}{4\pi r_{L}^{4}}, 
\end{equation}
where $m$ is the azimuthal order of the wave, and $r_{L}$ is the resonant radius of the wave.  $M_{P}$ is the mass of Saturn (here taken to be $5.6846 \times 10^{26}$ kg). 

{\bf Ring mass extinction coefficient, $\kappa$.} 

This parameter is simply the ratio of the normal optical depth to the surface mass density: 

\begin{equation}
    \kappa = \frac{\tau_{av}}{\sigma_{0}},
\end{equation}
where $\sigma_0$ is given by the above formula and $\tau_{av}$  is the average normal optical depth in the region containing the wave. This parameter provides useful information about particle properties beyond those captured by $\sigma_0$ (see below).

{\bf Ring kinematic viscosity, $\nu$.}

This parameter is given by Equation~\ref{viscoeq}, which can be re-expressed in terms of $\xi_D$ and $\sigma_0$ (for a case when $m \ne 1$ ) by making substitutions for $\mathcal{D}_L$ and $n_{L}$, 

\begin{equation}
    \nu = \frac{9}{7 \xi_{D}^{3}}\left(\frac{Gr_{L}^{7}}{M_{P}^{2}
    \left[3|m-1|+J_2(R_{s}/r_L)^2
\left(\frac{21}{2}-\frac{9}{2}|m-1|\right)\right]}\right)^{\frac{1}{2}}(2\pi\sigma_{0})^{\frac{3}{2}},
\label{viscoeqnew}
\end{equation}
 
 where we assume $G = 6.674 \times 10^{-11} \mathrm{m^{3} kg^{-1} s^{-2}}$, $R_{s} = 60330$ km (nominal radius for Saturn) and  $J_{2} = 0.01629$ \cite{Jacobson2022}.

{\bf Potential Disturbance $\Psi_{L}$ and Saturn's normal mode amplitude, $\mathscr{A}'_{\ell m0}$.}

Finally, the potential disturbance and the normal mode amplitude coefficient can be derived from the parameter $A_{L}$ and the background surface mass density $\sigma_0$ using Equations~\ref{Psieq} and~\ref{almneq}. More specifically, the potential disturbance is given by the following formula:

\begin{equation}
\Psi_{L}  =2\sqrt{\pi} A_{L} G \sigma_0 r_L.
\end{equation}
For those waves generated by normal modes inside the planet, Equation~\ref{almneq} is used. 

\subsubsection{Error Estimation}

Estimating uncertainties on these various parameters is nontrivial because the phase-corrections and averaging of occultation profiles with different resolutions introduces correlations among the data points that are difficult to model. We therefore instead estimate the uncertainties in these parameters using a variant of bootstrap resampling.
Bootstrap resampling is a powerful statistical technique that involves drawing multiple samples from the observed data to simulate a distribution of parameter values or estimate the uncertainty or variability in a parameter without relying on assumptions about the underlying population distribution \cite{Efron1994AnIT, davison_hinkley_1997, bevington2003data, IskanderZA, Chernick2007BootstrapMA, BenkeNREP}.

We initially considered independent resampled datasets consisting of disjoint sets of occultations, but we found that the signal-to-noise for certain waves was too low for us to obtain sensible estimates on the fit parameters from such samples. 

We therefore instead chose to consider datasets that excluded a relatively small fraction of the occultations.
Let $\mathbf{Y} = \{y_1, y_2, \ldots, y_n\}$ represent the original occultation dataset. We then generated data sets that excluded the occultation profiles $y_i$ where  $i \mod 4$ or  $i \mod 5$ had each possible value $j$, which we designate as $\mathbf{Y}_j$. 

For each of these datasets $\mathbf{Y}_{j}$, we generate a phase-corrected average profile and fit it to obtain estimates of all the fit parameters $A_{L}$, $x_{r}$, $r_{f}$, $\xi_{D}$, and $\phi_L$. Let us denote a generic parameter derived from a fit to data set $\mathbf{Y}_j$ as $\vartheta^*_j$. 

After obtaining the parameter estimates $\vartheta^{*}_{j}$ for all resampled datasets, we can calculate the mean ($\bar{\vartheta}^{*}$) and standard deviation  ($\varsigma^{*}$) of these estimates using the standard formulae: 

\begin{equation}
\bar{\vartheta}^{*} = \frac{1}{N_{j}}\sum_{j=1}^{N_{j}}\vartheta^{*}_{j},
\end{equation}
and

\begin{equation}
\varsigma^{*} = \sqrt{\frac{1}{N_{j}-1}\sum_{j=1}^{N_{j}}(\vartheta^{*}_{j} - \bar{\vartheta}^{*})^2}.
\end{equation}
For this analysis, we use $\varsigma^*$ to estimate the uncertainty in our various parameters. Note that we use the standard deviation and not the standard error of the mean because the sample datasets $\mathbf{Y}_j$ are not independent and so their dispersion represents how much the parameters can change due to excluding small fractions of the data. Also note that our final estimate of the parameters is the value derived from the full data set $\mathbf{Y}$ and not $\bar{\vartheta}^*$.

Finally, we can use the standard error propagation formulas \cite{taylor1982introduction, weisstein2000error, bevington2003data, LUO201723, BenkeNREP} to translate the estimates and uncertainties in the various fit parameters into uncertainties for the physical parameters $\sigma_0, \kappa, \nu, \Psi_{L}$ and $\mathscr{A}'_{\ell m0}$. Designating the set of fit parameter estimates as $\vartheta_k$ and their uncertainties as $\sigma_{\vartheta_k}$, and assuming the covariance among the fit parameters is negligible, the uncertainty for a generic physical parameter $w$ is given by:  

\begin{equation}
\frac{\sigma_w}{w} \approx \sqrt{\sum_{k=1}^{n} \left(\frac{\partial w}{\partial \vartheta_k}\right)^2 \left(\frac{\sigma_{\vartheta_k}}{\vartheta_k}\right)^2},
\end{equation}
where the relevant derivatives are evaluated using the formulas above.

\section{Results}
\label{results}

This section describes the results of our wave analysis. Section~\ref{satres} examines the waves generated by resonances with Saturn's various moons, which serve to validate our procedures. Section~\ref{planetres} then provides a brief overview of the waves generated by planetary normal modes. 

\begin{table}[ht]
\caption{Table of values comparing estimates of ring properties derived using details from spiral density waves excited by Saturn's satellite resonances.}
\label{tab:satparcomp}
\centering
\resizebox{\textwidth}{!}{\begin{tabular}{|lcc|cc|cccr}
\hline
Resonance Name & $r_{L}$ (km) & $m$ & \multicolumn{2}{|c|}{$\xi_D$} & \multicolumn{3}{|c}{$\sigma_{0}$ (g/cm$^{2}$)}  \\
               &              &     & This work & \cite{BailliColwellLissauerEsposito} & This work & \cite{BailliColwellLissauerEsposito} & \cite{Hedman_2022}  \\
\hline
Mimas 4:1 & 74890.070 & 2 & 2.796 $\pm$ 0.065 & 4.23   & 1.060 $\pm$ 0.431 & 0.58 $\pm$ 0.09 & - \\
Pan 2:1 & 85105.020 & 2 & 4.370 $\pm$ 0.209  & -  & 1.780 $\pm$ 0.032 & - & - \\

Atlas 2:1 & 87645.680 & 2 & 2.253 $\pm$ 0.072 & 5.42  & 0.250 $\pm$ 0.013 & 0.22 $\pm$ 0.03 & - \\

Prometheus 4:2 & 88434.120 & 3 & 5.803 $\pm$ 0.078  & - & 1.274 $\pm$ 0.015 & - & 1.42 $\pm$ 0.41 \\

Mimas 6:2 & 89884.000 & 3 & 4.861 $\pm$ 0.148 & 6.61  & 1.151 $\pm$ 0.057 & 1.31 $\pm$ 0.20 & 1.28 $\pm$ 0.01$^a$ \\      

Pandora 4:2 & 89893.680 & 3 & 6.915 $\pm$ 0.115 & 6.69 & 1.460 $\pm$ 0.003 & 1.42 $\pm$ 0.21 & 1.28 $\pm$ 0.01$^a$ \\   
\hline
\end{tabular}}

$^a$ This is the average surface mass density of the region containing both waves.

\end{table}

\begin{table}
\caption{Estimates of gravitational potential from theoretical model for spiral density waves excited by Saturn's satellite resonances.}
\label{tab:satampcomp}
\centering
%\begin{tabular}{lllll}
\resizebox{\textwidth}{!}{\begin{tabular}{lcccccccccccr}
\hline
Resonance Name & Radial Range (km) & $r_{L}$ (km) & $m$ & $\Psi_{m,d}' (m^{2}s^{-2})$ & $\Psi_{m,p}' (m^{2}s^{-2})$\\
\hline
Mimas 4:1 & 74880.00 - 74895.00 & 74890.07 & 2 & 0.043118 $\pm$ 0.028284 & 0.035125 \\
Pan 2:1 & 85090.00 - 85130.00 & 85105.02 & 2 & 0.005607 $\pm$ 0.000662 & 0.005273 \\
Atlas 2:1 & 87640.00 - 87650.00 & 87645.68 & 2 & 0.009018 $\pm$ 0.000873 & 0.006767 \\
Prometheus 4:2 & 88420.00 - 88440.00 & 88434.12 & 3 & 0.001895 $\pm$ 0.000064 & 0.001717 \\
Mimas 6:2 & 89870.00 - 89890.00 & 89884.00 & 3 & 0.003265 $\pm$ 0.000312 & 0.003209 \\
Pandora 4:2 & 89887.00 - 89897.00 & 89893.68 & 3 & 0.002374 $\pm$ 0.000114 & 0.002755 \\
\hline
\end{tabular}}

The subscripts ``d" and ``p" in $\Psi_{m,d}'$ and $\Psi_{m,p}'$ represent ``data" and ``predicted," respectively.

\end{table}

\subsection{Validating procedures with waves generated by satellite resonances}
\label{satres}

There are six mean-motion resonances with satellites that generate reasonably clear density waves within the C-ring \cite{TiscarenoandHarris2018}. In order of distance from Saturn's center, these are the Mimas 4:1, Pan 2:1, Atlas 2:1, Prometheus 4:2, Mimas 6:2, and Pandora 4:2. We can therefore compare our derived  estimates of gravitational perturbations $\Psi_L$ for these waves with theoretical predictions and compare our derived  estimates of the ring's local surface mass density and viscosity with previous estimates \cite{BailliColwellLissauerEsposito, Hedman_2022}. These comparisons allow us to validate our procedures and therefore have more confidence in the parameters derived from the planetary normal mode resonances.

\begin{figure}[tbp]
\hspace{0in}
\centering 
\includegraphics[width=1\textwidth]{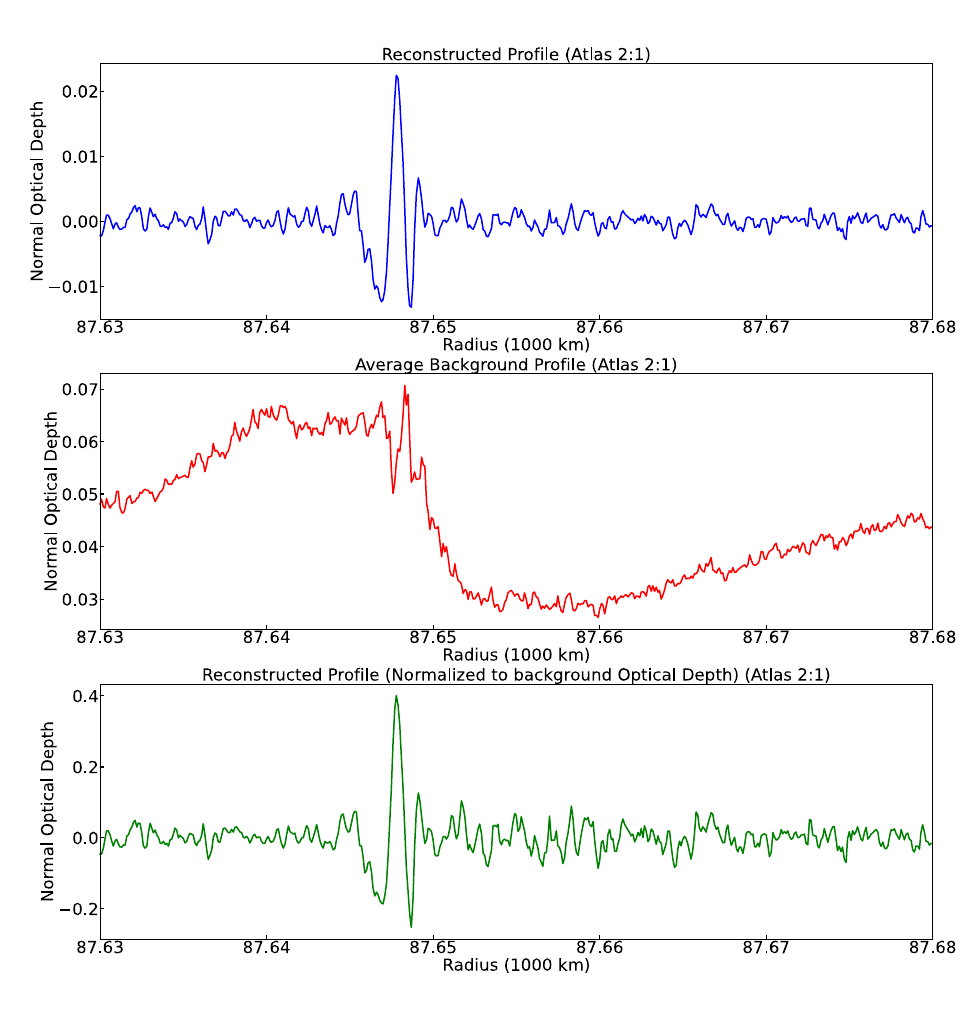}
\caption{(Top) The optical depth profile derived from the average phase-corrected wavelet of the Atlas 2:1 wave $\bar{y}_\phi$. (Middle). The average optical depth profile $\bar{\tau}_n$ of this same region. (Bottom) The fractional optical depth variations derived from the ratio of the above two profiles.} \label{fig:atlas21profs}
\end{figure}

Figure~\ref{fig:atlas21profs} illustrates the average profiles generated by the procedures described in Section~\ref{method} for the Atlas 2:1 wave. The top plot shows the normal optical depth profile derived from the phase-corrected average wavelet $\bar{y}_\phi$. The middle plot represents the background average normal optical depth profile $\bar{\tau}_n$ associated with the signal. Finally, the bottom plot shows the ratio of the above two profiles, which corresponds to the fractional optical depth variations $\Delta \tau_n/\bar{\tau}_n$ associated with the wave. Note that the background optical depth variations are absent from the top profile because the phase-corrected average eliminates those trends, while the signal from the wave itself is suppressed in the $\bar{\tau}_n$ profile due to how those profiles were averaged. This is already evidence that our techniques can properly isolate the wave signals.

\begin{figure}[tbp]
\hspace{0in}
\centering 
\includegraphics[width=1\textwidth]{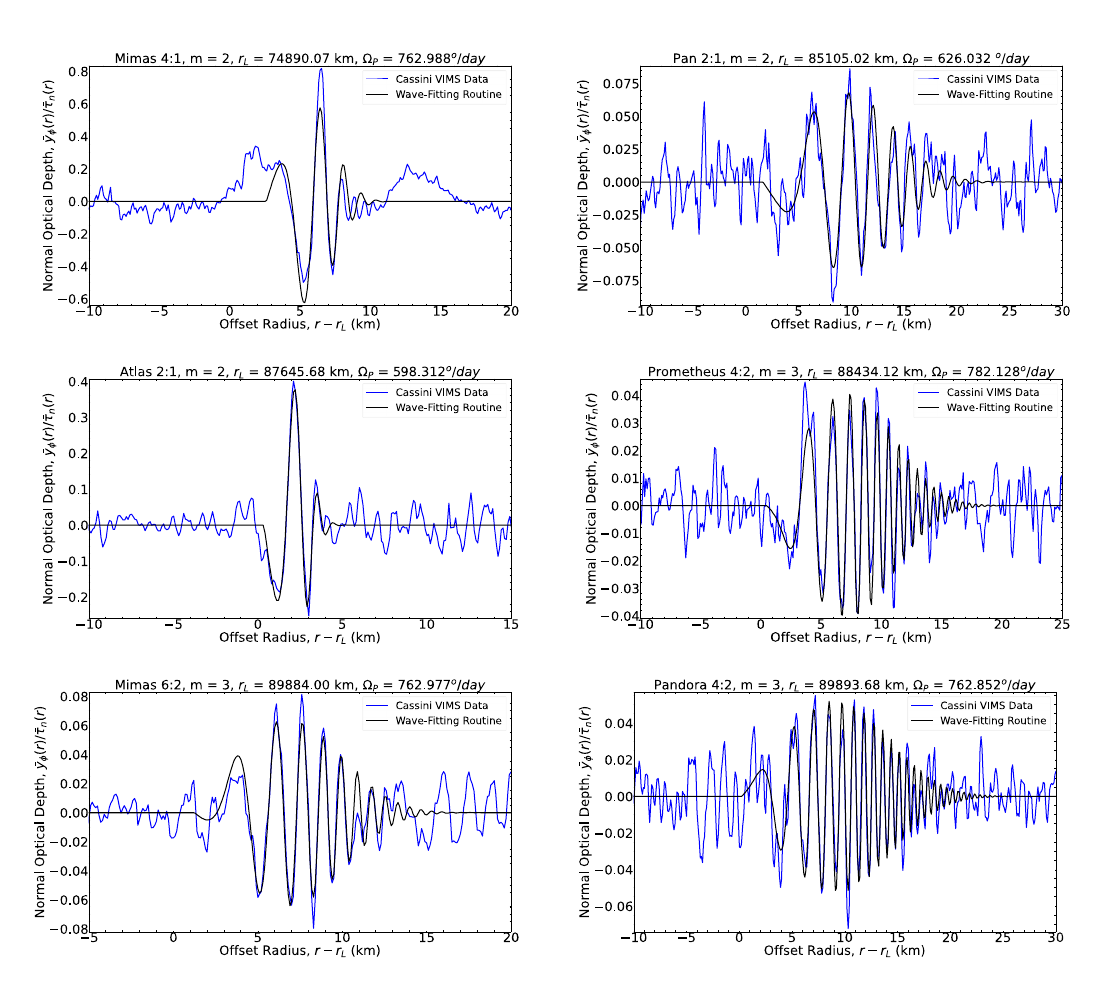}
\caption{Reconstructed fractional optical depth profiles of the C-ring density waves generated by satellite resonances (in blue), along with best-fit wave models (in black). Note the models fit the data well in all cases.} \label{fig:satres}
\end{figure}

\begin{figure}[tbp]
\hspace{0in}
\centering
\includegraphics[width=1\linewidth]{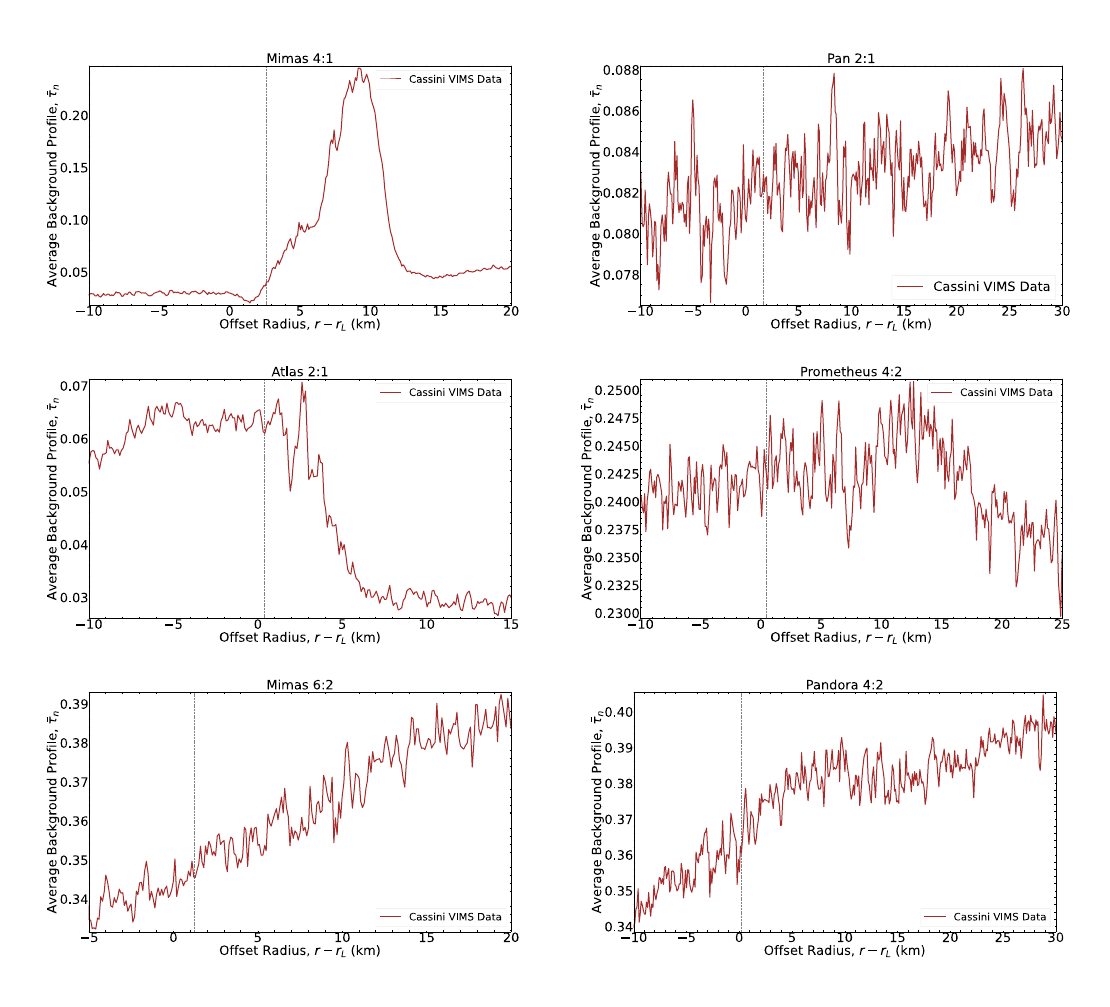}
\caption{The background optical depth profiles of the C-ring density waves shown in Figure~\ref{fig:satres}.} \label{fig:satback}
\end{figure}

Figure~\ref{fig:satres} shows profiles of the fractional optical variations associated with all six of the satellite-generated waves, along with the  corresponding best-fitting linear density wave model described in Section~\ref{fit}. The fit parameters for these models, along with assumed values of the pattern speed and the resonant radii, are provided in Table~\ref{Tab:fitpars}. Finally, Figure~\ref{fig:satback} shows the corresponding background optical depth profiles for these waves. 

In all cases, the linear models match the data quite well. The one exception is perhaps the Mimas 4:1 wave, where the model shows a longer wavelength than the data after the first two cycles. This is most likely because the ring's optical depth rises steeply across this wave (see Figure~\ref{fig:satback}), which may correspond to a surface mass density gradient across the wave that is not captured by our model. It is worth pointing out that while we used a radially-variable $\bar\tau_n(r)$ to normalize the observed $\tau$, the fitted model assumes a constant value of $\sigma_0$, which may not be appropriate in this case (see Equation~\ref{sigmasigma0}).

Finally, we can compare the physical parameters derived from these fits to earlier measurements and theoretical predictions. Table~\ref{tab:satparcomp} compares our estimates of the ring's surface mass density $\sigma_0$ and effective viscosity (parameterized in terms of the dimensionless damping length $\xi_D$) with the values obtained by \citeA{BailliColwellLissauerEsposito} and \citeA{Hedman_2022}. Meanwhile, Table~\ref{tab:satampcomp} compares our estimates of the gravitational perturbation $\Psi_L$ with the predicted values computed using the formulas for the satellite torques in \citeA{TiscarenoandHarris2018} and using current masses for the satellites \cite{Cooperetal2015, ThomasHelfenstein2020, Jacobson2022}. These torques were then converted to potentials using the following equation \cite{Hedman_2022},

\begin{equation}
    \Psi_{L} = \sqrt{\frac{3|m - 1|n_{L}^{2}}{|m|\pi^{2}} \left|\frac{\mathcal{T}}{\sigma_{0}}\right|},
\end{equation}
where $n_{L}^{2} = {G M_{P}}/{r_{L}^{3}}$.

Table~\ref{tab:satparcomp} shows that our estimates of the ring's surface mass density and dimensionless damping lengths are broadly consistent with previously published values. For the Mimas 6:2 and Pandora 4:2 waves, our values for both these parameters are within 20\% of the previously published values. Similarly, our estimate of the surface mass densities for the Atlas 2:1 and Prometheus 4:2 waves are consistent with the value derived in \citeA{BailliColwellLissauerEsposito}
and \citeA{Hedman_2022}, respectively. The biggest discrepancies visible in this table are that our surface mass density value from the Mimas 4:1 wave is about twice the \citeA{BailliColwellLissauerEsposito} value, while the damping lengths for the Mimas 4:1 and Atlas 2:1 waves are about half the \citeA{BailliColwellLissauerEsposito} values. These differences may in part be due to the steep optical depth gradients underlying these waves (see Figure~\ref{fig:satback}), which may be causing localized variations in the ring properties that are not accounted for by standard models that assume the background surface mass density is constant. This could cause the two fits to yield different parameter values if they have different sensitivities to different parts of the wave. Taking the discrepancies among these different estimates as a rough estimate of their overall accuracy, we can say that our estimates of the ring parameters are probably accurate to within a factor of around 2 in regions with strong optical depth gradients, and to within about 20\% in other regions. 

Regarding the perturbation amplitudes, Table~\ref{tab:satampcomp} shows a remarkable degree of consistency between our estimates of these parameters and the theoretical predictions. While  the term in the potential responsible for each wave has minimal uncertainty because the masses of the satellites are well measured, the  relationship between the gravitational potential amplitude and the wave amplitude assumes that the ring has a uniform background surface mass density, responds linearly to the perturbation, and that the Fresnel integral can be well approximated by a cosine. All these assumptions and approximations could potentially affect the estimates of the gravitational potential amplitudes, but the data from the satellite waves indicate that these factors do not dramatically alter these numbers. While the Mimas 4:1 and Atlas 2:1 are again the waves that show the largest discrepancies, they are only 23\% and 33\% off from predictions, respectively. The other waves yield amplitudes within 15\% of predictions. This demonstrates that our procedures do provide robust estimates of the gravitational perturbations and suggests that under a wide range of conditions we can reliably estimate these amplitudes to within 40\%. 

\subsection{Parameters for the waves generated by resonances with planetary normal modes}
\label{planetres}

The above analysis of the satellite-driven waves shows that our model and the real occultation data are consistent and likely produce sensible estimates of both the ring's properties and the amplitudes of the perturbing potentials. 
We can therefore confidently apply these techniques to the spiral density waves excited by Saturn.

\begin{figure}[tbp]
\hspace{0in}
\centering
\includegraphics[width=1\linewidth]{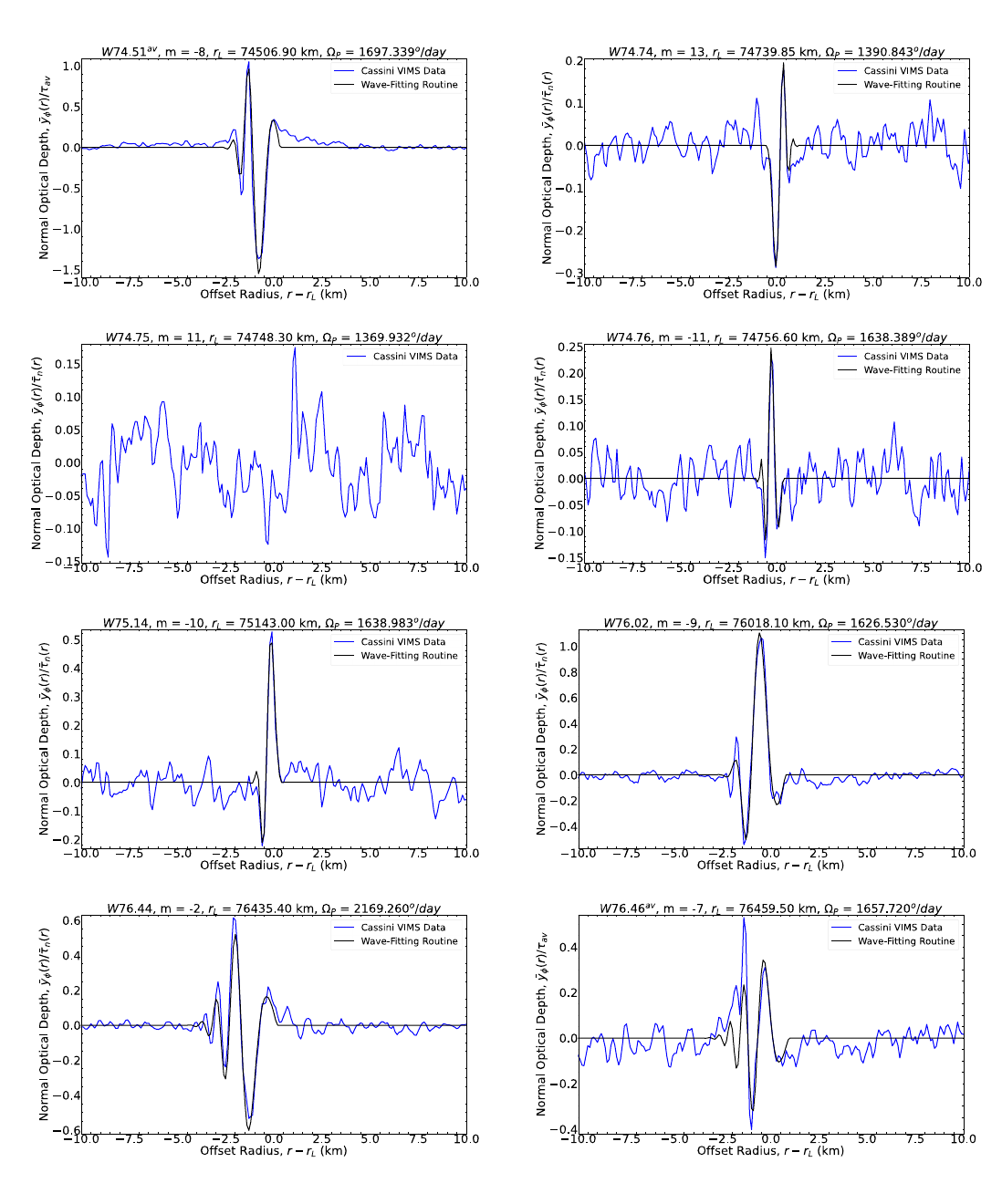}
\caption{Reconstructed fractional optical depth profiles of the C-ring density waves generated by planetary normal modes (in blue), along with best-fit wave models (in black). Note that no fit could be performed for W74.75.} \label{fig:planres1}
\end{figure}

\begin{figure}[tbp]
\hspace{0in}
\centering
\includegraphics[width=1\linewidth]{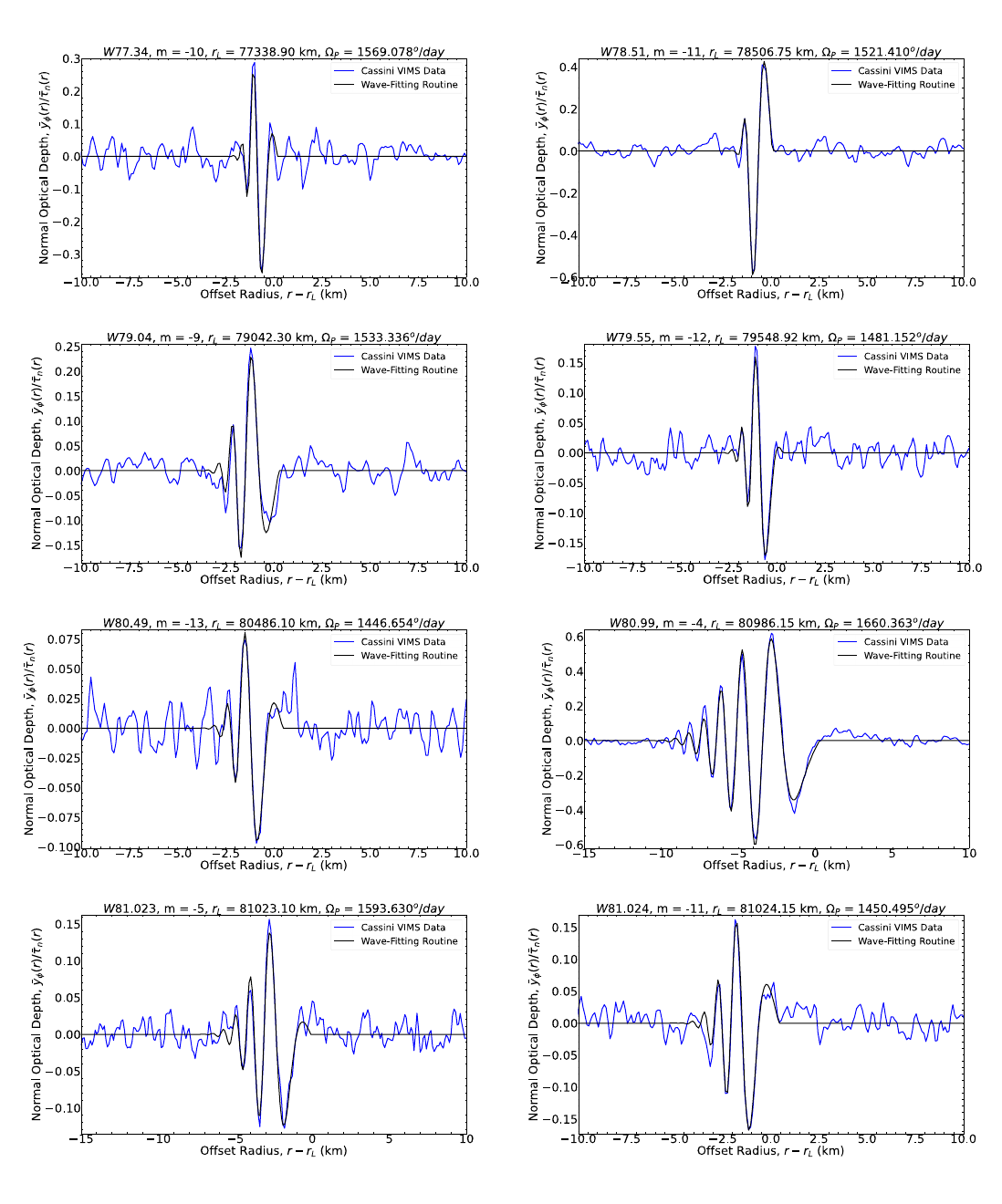}
\caption{Reconstructed fractional optical depth profiles of the C-ring density waves generated by planetary normal modes (in blue), along with best-fit wave models (in black).} \label{fig:planres2}
\end{figure}

\begin{figure}[tbp]
\hspace{0in}
\centering
\includegraphics[width=1\linewidth]{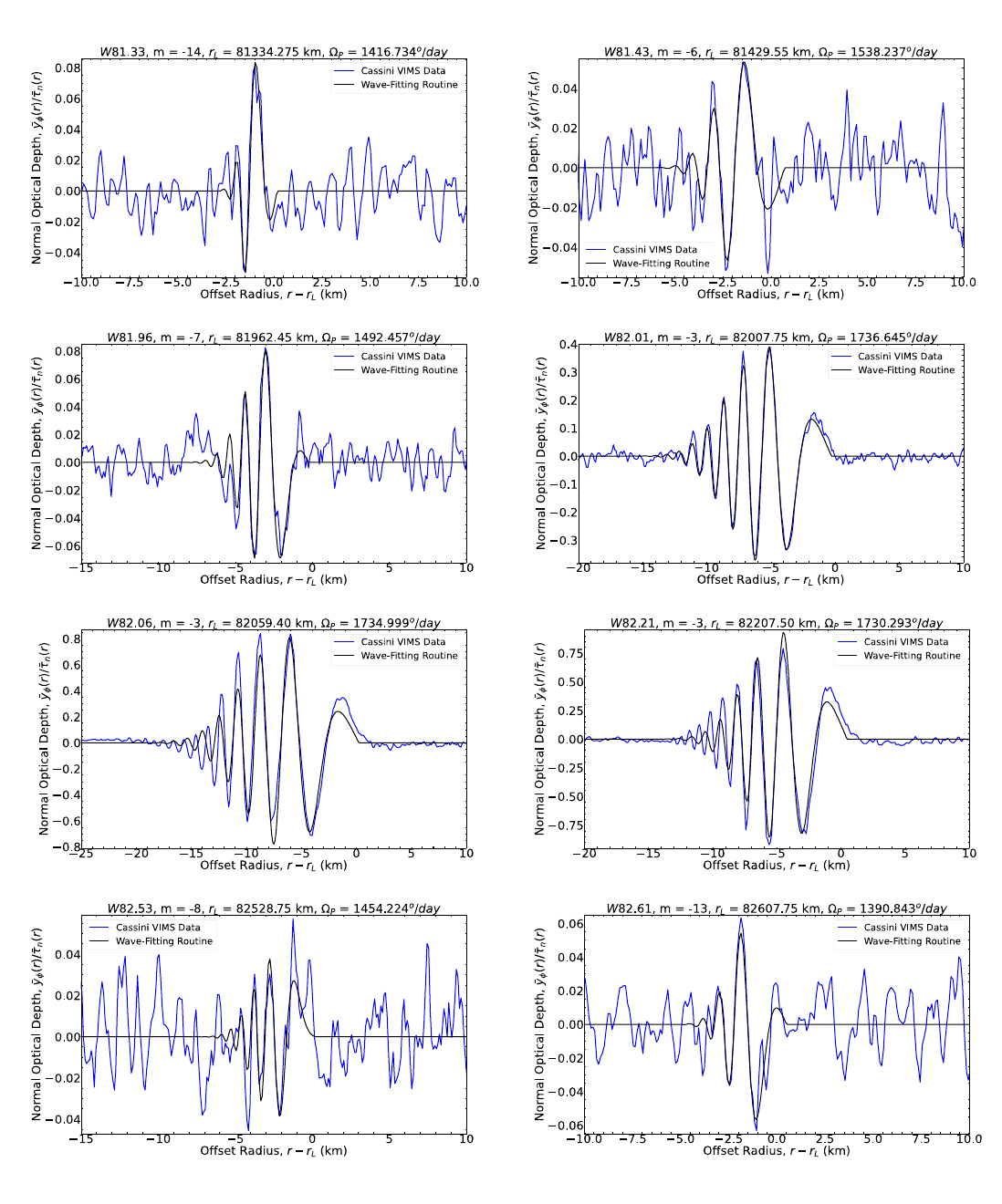}
\caption{Reconstructed fractional optical depth profiles of the C-ring density waves generated by planetary normal modes (in blue), along with best-fit wave models (in black).} \label{fig:planres3}
\end{figure}

\begin{figure}[tbp]
\hspace{0in}
\centering
\includegraphics[width=1\linewidth]{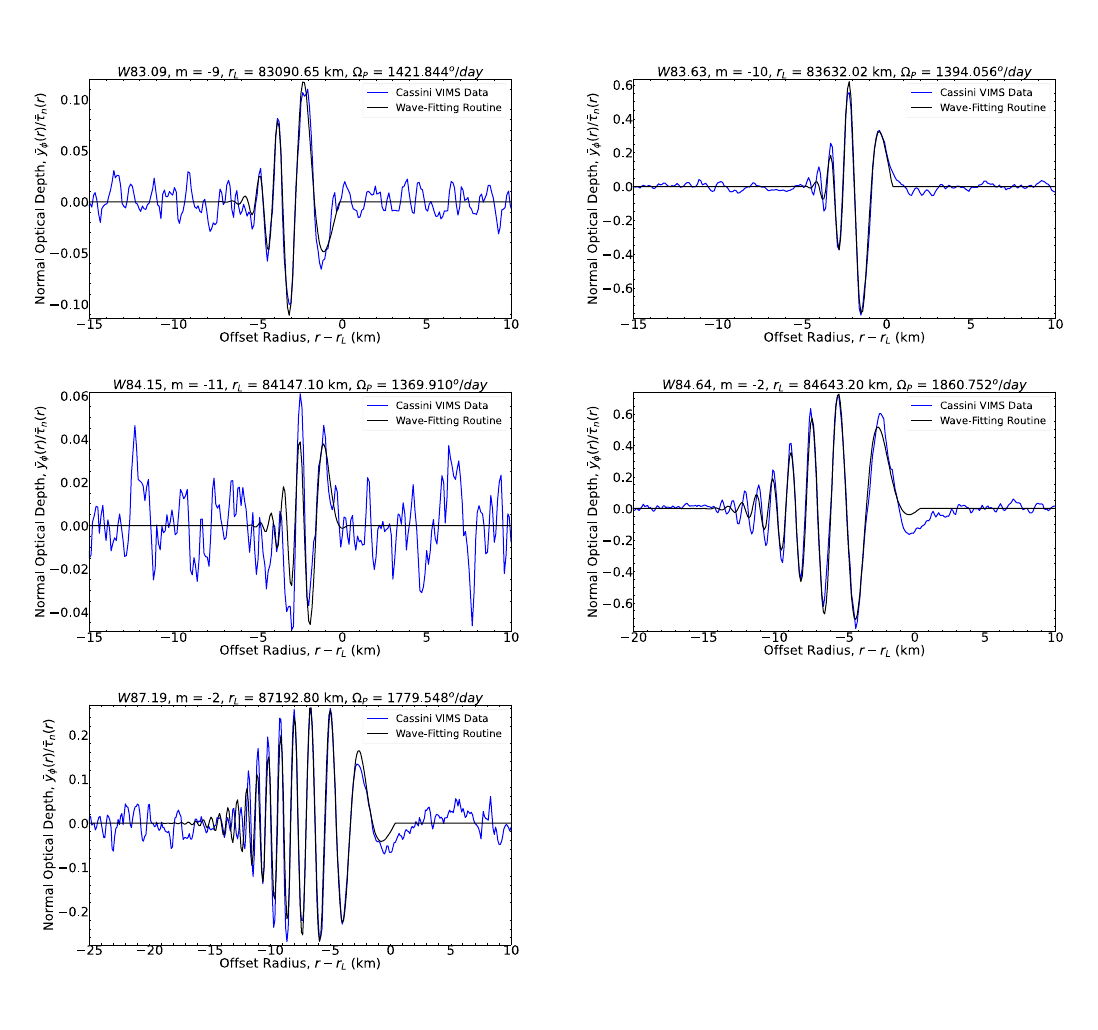}
\caption{Reconstructed fractional optical depth profiles of the C-ring density waves generated by planetary normal modes (in blue), along with best-fit wave models (in black) (continuation).} \label{fig:planres4}
\end{figure}

\begin{figure}[tbp]
\hspace{0in}
\centering
\includegraphics[width=1\linewidth]{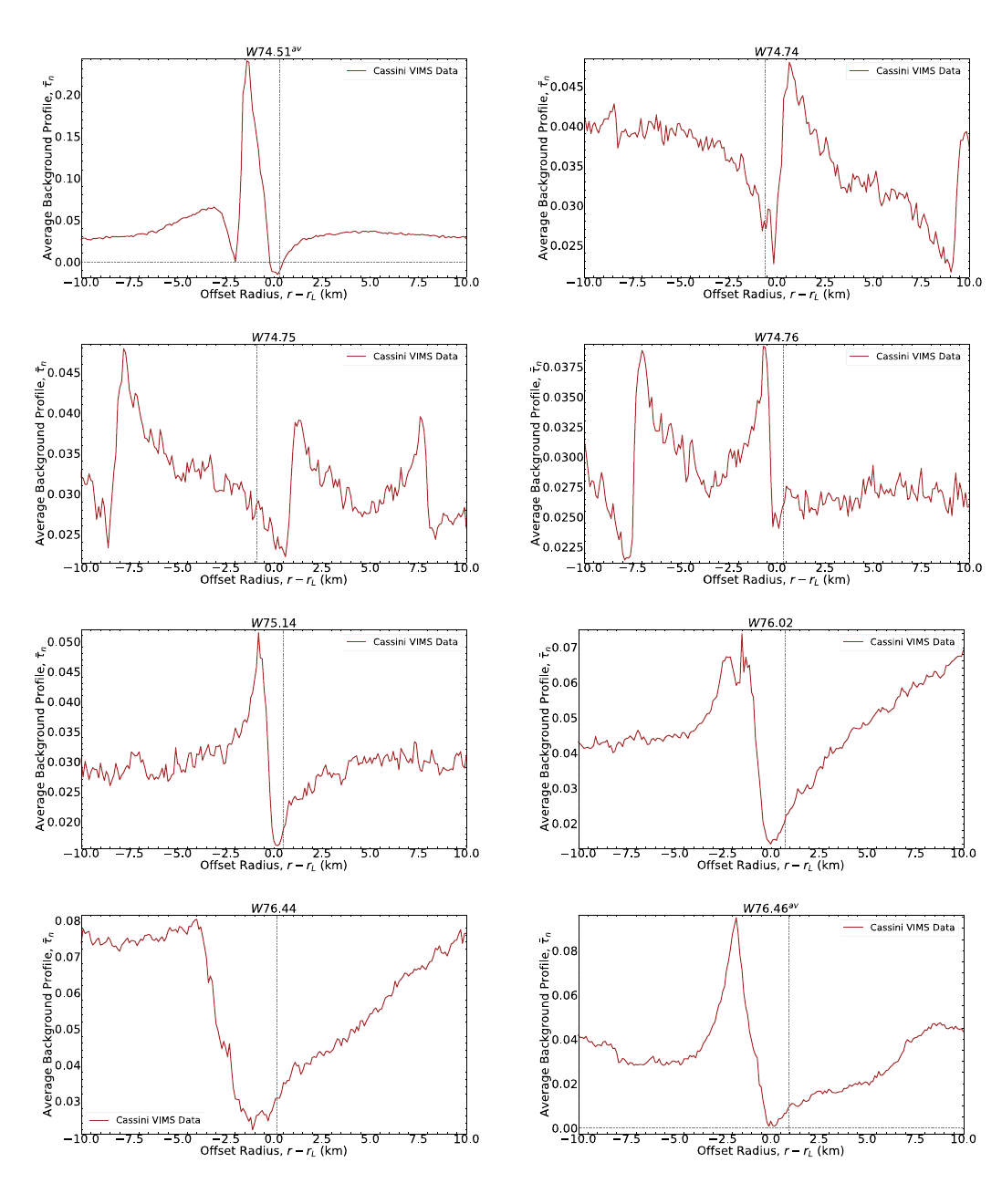}
\caption{Background optical depth profiles of the C-ring density waves shown in Figure~\ref{fig:planres1}.} \label{fig:planback1}
\end{figure}

\begin{figure}[tbp]
\hspace{0in}
\centering
\includegraphics[width=1\linewidth]{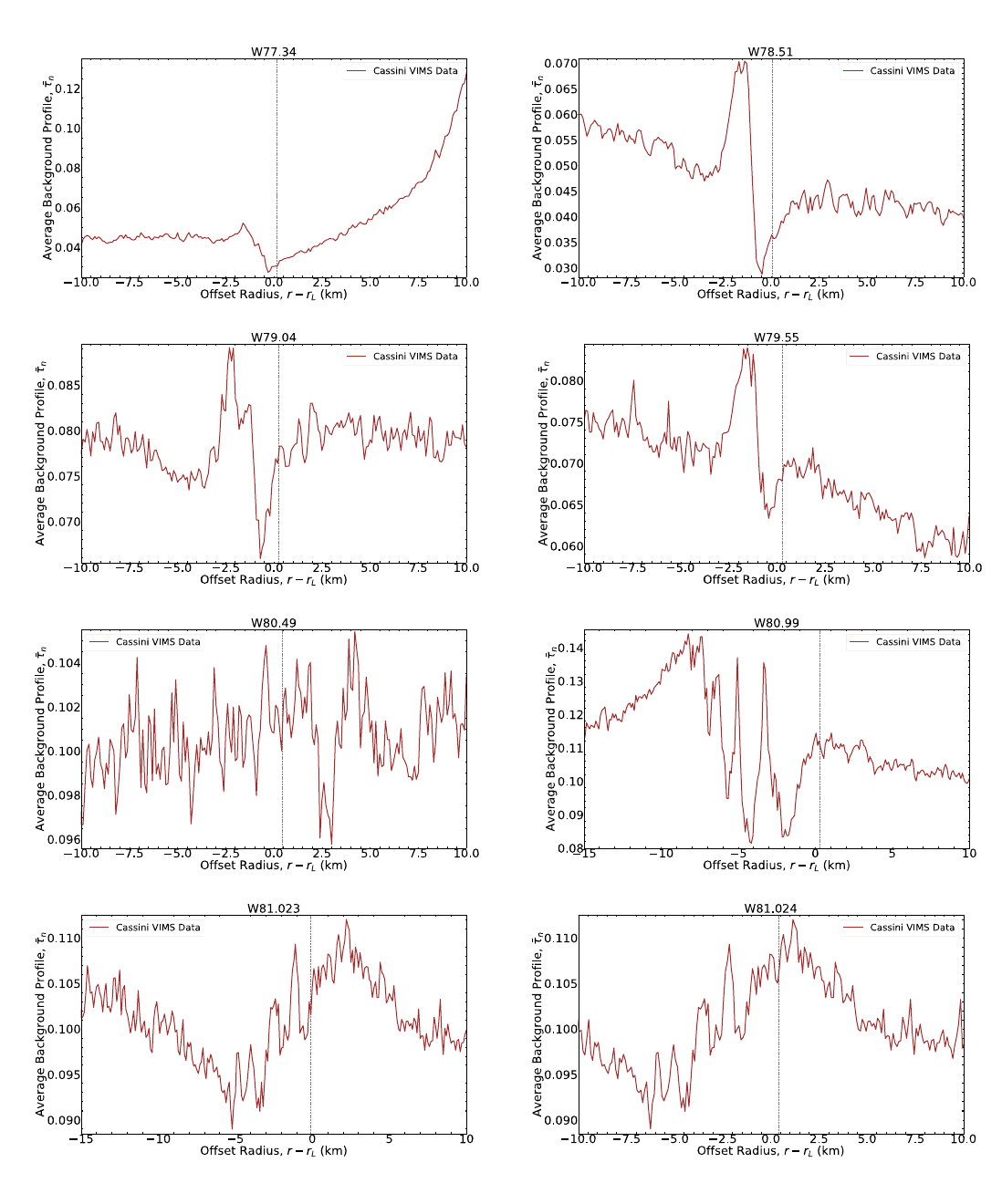}    
\caption{Background optical depth profiles of the C-ring density waves shown in Figure~\ref{fig:planres2}.} \label{fig:planback2}
\end{figure}

\begin{figure}[tbp]
\hspace{0in}
\centering
\includegraphics[width=1\linewidth]{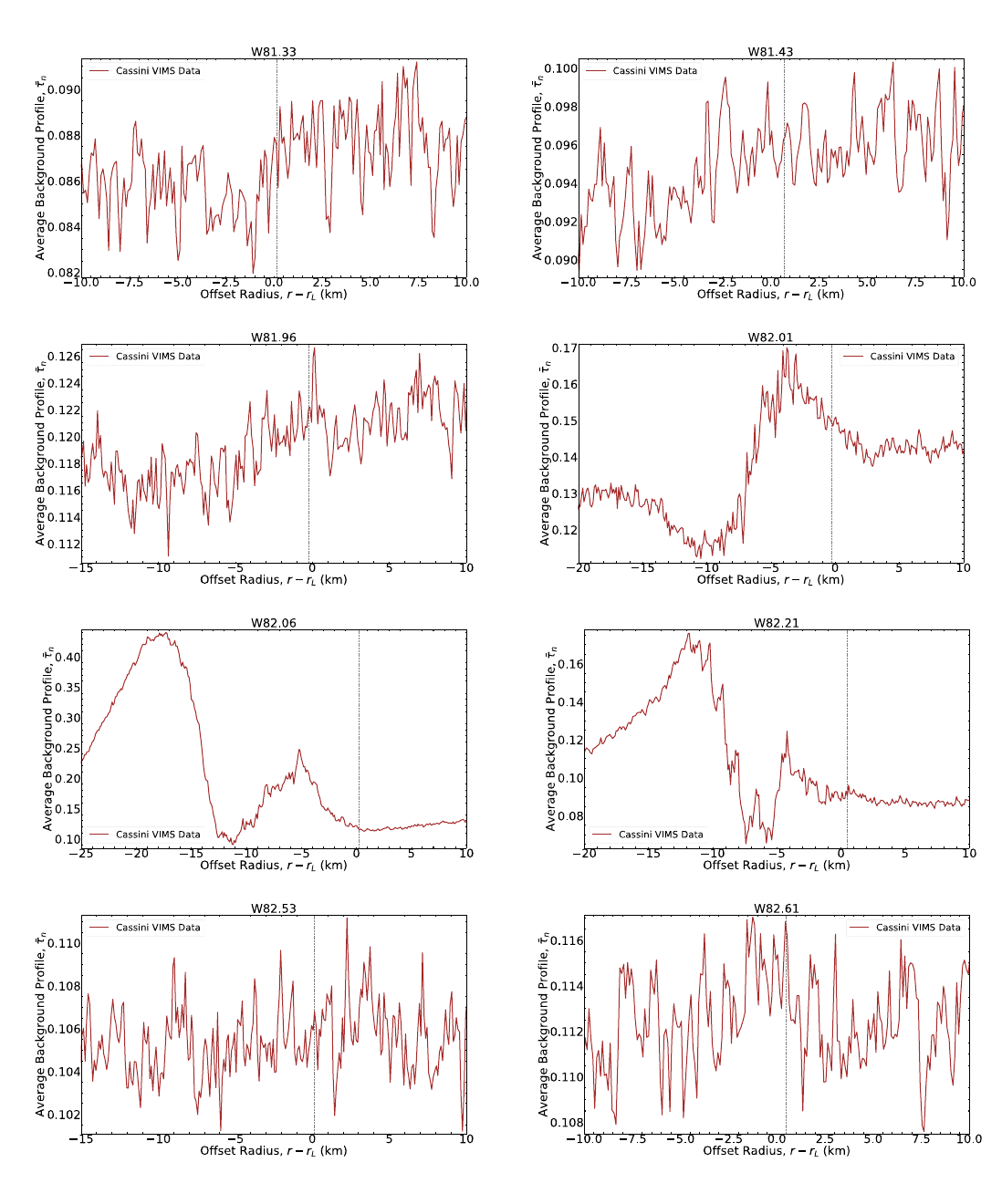}
\caption{Background optical depth profiles of the C-ring density waves shown in Figure~\ref{fig:planres3}.} \label{fig:planback3}
\end{figure}

\begin{figure}[tbp]
\hspace{0in}
\centering
\includegraphics[width=1\linewidth]{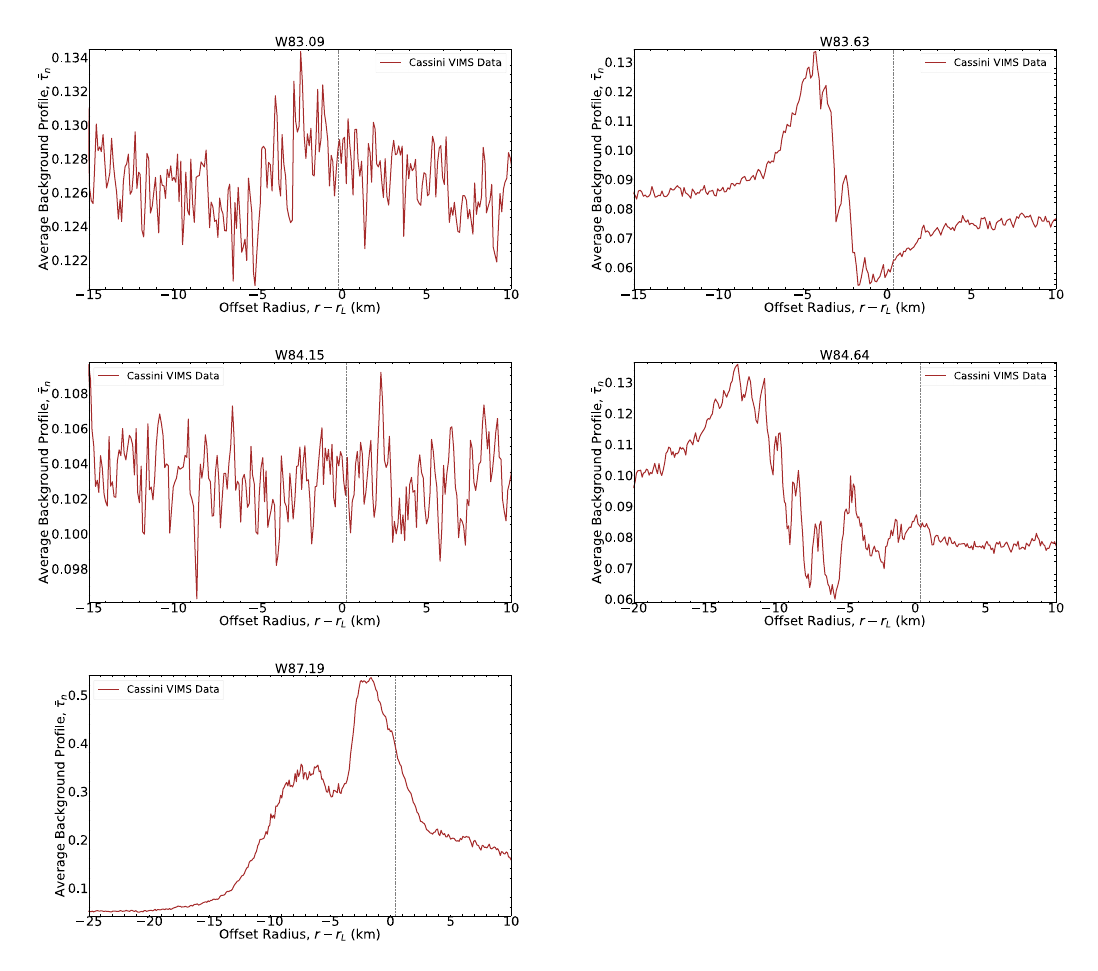}
\caption{Background optical depth profiles of the C-ring density waves shown in Figure~\ref{fig:planres4}.} \label{fig:planback4}
\end{figure}

Figures~\ref{fig:planres1} -~\ref{fig:planres4} compare the reconstructed profiles of the 29 currently known density waves generated by planetary normal modes, along with our best-fit linear density wave models. Meanwhile Figures~\ref{fig:planback1} -~\ref{fig:planback4} show the background optical depth variations underlying each of these waves. Interestingly, these often show a peak in optical depth just interior to a dip that is just interior to the resonance location (marked with a dotted line), a pattern that could reflect the torques applied by the wave on the rings \cite{Shu1985ApJ, Borderies1986,Borderies1989, Tajeddine2017}. The wave-fit parameters and assumed values of $m$, pattern speed, and resonant radius are all tabulated in Table~\ref{Tab:fitpars}, while the derived ring parameters are given in Table~\ref{Tab:ringpars} and the estimated amplitudes of the planetary normal modes are provided in Table~\ref{Tab:ampparsu}.

We were unable to fit the wave designated W74.75. As can be seen in Figure~\ref{fig:planres1}, the signal-to-noise on this wave was the lowest of all the detected features. While there is a peak in the profile that is likely associated with this wave, we found that when we tried to fit this feature, the fit parameters were extremely sensitive to slight changes in the assumed parameter ranges. Also, when we removed selected profiles to estimate uncertainties in the fit parameters, we found the fits were even more unstable. We therefore do not regard these fits as reliable and so do not consider this wave further.

For all of the other waves, we observe at least one wave cycle and are able to obtain reasonably stable fits. However, we should note that we only observe one wave cycle with W74.74, W74.76, and W75.14, while W74.51, W76.02, W76.46, W77.34,  W78.51, W79.55 and W81.33 only show about 1.5 cycles. Since many of these waves are also found on top of relatively sharp optical depth gradients, the fit parameters for these waves could be more uncertain than their formal error bars suggest.  On the other hand, W80.49, W81.43, and W84.15 show two or more wave cycles with relatively low signal-to-noise, so we expect that our estimated uncertainties for these waves will be more accurate.   

For the stronger waves, W76.44, W79.04, W81.023, W81.024, W81.96, W82.01, W83.09, W83.63 are all fit by the linear density wave model very well throughout their entire extent. By contrast, the fits to waves W80.99, W82.06, W82.21, W84.64 and W87.19 begin to depart from the observed profiles on the inner edges of these waves. This is probably partially due to the steep optical depth gradients underlying these waves. It is also possible that some of these waves are strong enough that they may be nonlinear. Both of these phenomena could potentially introduce some additional systematic uncertainties in the fit parameters. 

We can estimate the magnitudes of any systematic uncertainties in our parameter estimates by considering selected pairs of waves that should measure the same parameters. First of all, the waves W81.023 and W81.024 are two waves with very different $m$-values and pattern speeds that overlap each other, and so should give the same values for the ring's surface mass density and effective viscosity. Indeed, the surface mass densities derived from W81.023 and W81.024 are 2.52$\pm$0.29 and 3.07$\pm$0.04 g/cm$^2$, respectively, and the corresponding viscosity values are 14.9$\pm$3.9 and 17.1$\pm$1.8 cm$^2$/s. Both these numbers agree to within 20\% and therefore demonstrate that for these waves the systematic uncertainties in these parameters (such as those due to background surface mass density gradients and nonlinearities in the wave) are relatively small. While these are among the waves that best fit the linear density wave model, small discrepancies between the model and the data can still be seen in the leftmost cycle of each wave, so this demonstrates that such discrepancies are unlikely to have large effects on these fit parameters. 

The other pair of waves worth considering in this context is W74.74 and W82.61, both generated by a planetary f-mode with $\ell=15$ and $m=13$ \cite{French2021Icar}. These waves should therefore yield the same normal mode amplitude $\mathscr{A}'_{\ell m0}$. In actuality, the amplitude derived from W74.74 is (0.49$\pm$0.06)$\times10^{-10}$ while that derived from W82.61 is (1.79$\pm$0.24)$\times10^{-10}$. These numbers therefore differ from each other by roughly a factor of 3-4. In this context, we note that W74.74 is one of the waves that is most likely to exhibit systematic errors, and so we conclude that in the worst case, our uncertainties in the normal mode amplitudes may be as large as a factor of 3-4. However, given that W74.74 is one of the most difficult waves to fit, we expect that systematic uncertainties for most of the other waves are substantially smaller than this. Indeed, for the stronger waves, these uncertainties are almost certainly less than 40\% given the above results for the satellite-generated waves.

\begin{sidewaystable}
\caption{Data for Wave-fitting routine for satellite and non-satellite resonances.}
\label{Tab:fitpars}

\small
\begin{tabular}{lccccccccccccr}
\hline
Wave ID & Radial Range (km) & $r_{L}$ (km) &  $\ell$ &  $m$ & $\Omega_{P}$ $(^{o}/day)$ & $A_{L}$ & $\xi_{D}$ & $\phi_{L}(rad)$ & $x_{r}(km)$ & $r_{f}(km)$ \\
\hline

Mimas 4:1 & 74880.00 - 74895.00 & 74890.070 & - & 2 & 762.988 & 0.2296 $\pm$ 0.0304 & 2.7964 $\pm$ 0.0650 & -4.1833 $\pm$ 0.8610 & 2.6076 $\pm$ 1.1094 & 1.5673 $\pm$ 0.3184 \\
Pan 2:1 & 85090.00 - 85130.00 & 85105.020 & - & 2 & 626.032 & 0.0156 $\pm$ 0.0018 & 4.3698 $\pm$ 0.2086 & -0.5997 $\pm$ 0.1444 & 1.6271 $\pm$ 0.1274 &  2.623 $\pm$ 0.0234 \\
Atlas 2:1 & 87640.00 - 87650.00 & 87645.680 & - & 2 & 598.312 & 0.1737 $\pm$ 0.0093 & 2.2532 $\pm$ 0.0719 & -0.7984 $\pm$ 0.1046 & 0.3587 $\pm$ 0.0665 & 1.0435 $\pm$ 0.0267 \\
Prometheus 4:2 & 88420.00 - 88440.00 & 88434.120 & - & 3 & 782.128 & 0.0071 $\pm$ 0.0002 & 5.8033 $\pm$ 0.0784 &  0.2515 $\pm$ 0.2233 & 0.4825 $\pm$ 0.0919 & 1.6943 $\pm$ 0.0098 \\
Mimas 6:2 & 89870.00 - 89890.00 & 89884.000 & - & 3 & 762.977 & 0.0133 $\pm$ 0.0007 & 4.8613 $\pm$ 0.1481 & -1.7239 $\pm$ 0.6202 & 1.2506 $\pm$ 0.3037 & 1.6639 $\pm$ 0.0411 \\
Pandora 4:2 & 89887.00 - 89897.00 & 89893.680 & - & 3 & 762.852 & 0.0076 $\pm$ 0.0004 & 6.9149 $\pm$ 0.1152 &  3.0367 $\pm$ 0.0554 & 0.2389 $\pm$ 0.0157 &  1.8740 $\pm$ 0.0021 \\
\hline

$W74.51^{av}$ & 74501.00 - 74509.00 & 74506.900 & 12 & -8 & 1697.339 & 0.7321 $\pm$ 0.1108 & 2.1456 $\pm$ 0.1342 & 4.6659 $\pm$ 0.7724 & 0.3166 $\pm$ 0.2853 & 0.7195 $\pm$ 0.0721 \\
W74.74 & 74736.00 - 74743.00 & 74739.850 & 15 & 13 & 1390.843 & 0.1425 $\pm$ 0.0154 & 2.0611 $\pm$ 0.0807 & 0.7970 $\pm$ 0.2314 & -0.6040 $\pm$ 0.0518 & 0.4427 $\pm$ 0.0121 \\
W74.75 & 74746.00 - 74748.00 & 74748.300 & 11 & 11 & 1369.932 & - & - & - & - & - \\
W74.76 & 74752.00 - 74762.00 & 74756.600 & 19 & -11 & 1638.389 & 0.1256 $\pm$ 0.0019 & 2.0489 $\pm$ 0.0221 & -4.4533 $\pm$ 0.1133 & 0.3150 $\pm$ 0.0155 & 0.3780 $\pm$ 0.0046 \\
W75.14 & 75142.00 - 75144.00 & 75143.000 & 16 & -10 & 1638.983 & 0.2850 $\pm$ 0.0506 & 1.7833 $\pm$ 0.1584 & 0.8287 $\pm$ 2.1398 & 0.4988 $\pm$ 0.1288 & 0.5115 $\pm$ 0.0326 \\
W76.02A & 76016.00 - 76018.00 & 76018.100 & 13 & -9 & 1626.530 & 0.5925 $\pm$ 0.0948 & 1.9168 $\pm$ 0.1019 & 1.4467 $\pm$ 0.1950 & 0.7582 $\pm$ 0.0547 & 0.9133 $\pm$ 0.0270 \\
W76.44 & 76433.00 - 76436.00 & 76435.400 & 2 & -2 & 2169.260 & 0.2385 $\pm$ 0.0216 & 2.5803 $\pm$ 0.1171 & 4.9082 $\pm$ 0.1986 & 0.1663 $\pm$ 0.0828 & 0.9174 $\pm$ 0.0218 \\
$W76.46^{av}$ & 76457.00 - 76462.00 & 76459.500 & 9 & -7 & 1657.720 & 0.1277 $\pm$ 0.0254 & 2.8263 $\pm$ 0.2237 & 1.9068 $\pm$ 0.7003 & 0.9266 $\pm$ 0.1985 & 0.7791 $\pm$ 0.0359 \\
W77.34 & 77337.00 - 77339.00 & 77338.900 & 14 & -10 & 1569.078 & 0.1607 $\pm$ 0.0151 & 2.3072 $\pm$ 0.0942 & -1.6493 $\pm$ 0.0813 & 0.1831 $\pm$ 0.0181 & 0.5413 $\pm$ 0.0112 \\
W78.51 & 78503.00 - 78509.00 & 78506.750 & 15 & -11 & 1521.410 & 0.4304 $\pm$ 0.0240 & 1.7504 $\pm$ 0.0707 & -1.0106 $\pm$ 0.2709 & 0.0731 $\pm$ 0.0379 & 0.6044 $\pm$ 0.0119 \\
W79.04 & 79039.00 - 79045.00 & 79042.300 & 11 & -9 & 1533.336 & 0.0918 $\pm$ 0.0098 & 2.5458 $\pm$ 0.0967 & 2.4913 $\pm$ 0.1622 & 0.2571 $\pm$ 0.0559 & 0.7867 $\pm$ 0.0157 \\
W79.55 & 79546.00 - 79550.00 & 79548.920 & 16 & -12 & 1481.152 & 0.0741 $\pm$ 0.0023 & 2.4900 $\pm$ 0.0706 & -2.051 $\pm$ 0.2579 & 0.3061 $\pm$ 0.0996 & 0.6337 $\pm$ 0.0266 \\
W80.49 & 80484.00 - 80488.00 & 80486.100 & 17 & -13 & 1446.654 & 0.0391 $\pm$ 0.0089 & 2.4988 $\pm$ 0.2456 & 4.7554 $\pm$ 0.9148 & 0.4563 $\pm$ 0.3338 & 0.8404 $\pm$ 0.0765 \\
W80.99 & 80983.00 - 80989.00 & 80986.150 & 4 & -4 & 1660.363 & 0.1788 $\pm$ 0.0047 & 3.4663 $\pm$ 0.0285 & 3.1083 $\pm$ 0.0592 & 0.2776 $\pm$ 0.0214 & 1.5785 $\pm$ 0.0036 \\
W81.023A & 81018.00 - 81030.00 & 81023.100 & 5 & -5 & 1593.630 & 0.0466 $\pm$ 0.0071 & 3.0439 $\pm$ 0.2059 & -1.7411 $\pm$ 0.7521 & -0.1020 $\pm$ 0.3434 & 1.1566 $\pm$ 0.0657 \\
W81.024B & 81018.00 - 81030.00 & 81024.150 & 13 & -11 & 1450.495 & 0.0609 $\pm$ 0.0031 &  2.8584 $\pm$ 0.1010 & -1.0549 $\pm$ 0.1058 & 0.4127 $\pm$ 0.0371 & 0.9022 $\pm$ 0.0055 \\
W81.33 & 81333.00 - 81336.00 & 81334.275 & 18 & -14 & 1416.734 & 0.0392 $\pm$ 0.0043 & 2.1636 $\pm$ 0.1663 & 1.5506 $\pm$ 0.4063 & 0.1597 $\pm$ 0.2460 & 0.7242 $\pm$ 0.0896 \\
W81.43 & 81420.00 - 81430.00 & 81429.550 & 6 & -6 & 1538.237 & 0.0202 $\pm$ 0.0010 & 2.6967 $\pm$ 0.0339 & 2.1199 $\pm$ 0.2241 & 0.7123 $\pm$ 0.2484 & 1.2235 $\pm$ 0.0769 \\
W81.96 & 81959.00 - 81965.00 & 81962.450 & 7 & -7 & 1492.457 & 0.0257 $\pm$ 0.0003 & 3.1679 $\pm$ 0.0621 & -1.7799 $\pm$ 0.2438 & -0.2557 $\pm$ 0.0941 & 1.2142 $\pm$ 0.0188 \\
W82.01 & 82004.00 - 82009.00 & 82007.750 & 3 & -3 & 1736.645 & 0.1037 $\pm$ 0.0013 & 3.7869 $\pm$ 0.0097 & -0.7882 $\pm$ 0.0280 & -0.2753 $\pm$ 0.0219 & 1.9370 $\pm$ 0.0046 \\
W82.06 & 82055.00 - 82065.00 & 82059.400 & 3 & -3 & 1734.999 & 0.2172 $\pm$ 0.0036 & 3.7664 $\pm$ 0.0354 & -0.9522 $\pm$ 0.1419 & 0.2334 $\pm$ 0.1503 & 2.5493 $\pm$ 0.0279 \\
W82.21 & 82187.50 - 82207.51 & 82207.500 & 3 & -3 & 1730.293 & 0.2610 $\pm$ 0.0047 & 3.5927 $\pm$ 0.0536 & -0.8118 $\pm$ 0.0626 & 0.4771 $\pm$ 0.0728 & 1.9758 $\pm$ 0.0167 \\
W82.53 & 82510.00 - 82530.00 & 82528.750 & 8 & -8 & 1454.224 & 0.0116 $\pm$ 0.0034 & 3.4267 $\pm$ 0.4323 & 0.5827 $\pm$ 0.1901 & 0.1471 $\pm$ 0.0352 & 1.0587 $\pm$ 0.0167 \\
W82.61 & 82606.00 - 82608.00 & 82607.750 & 15 & -13 & 1390.843 & 0.0225 $\pm$ 0.0032 & 2.6792 $\pm$ 0.1388 & -1.6474 $\pm$ 0.2423 & 0.4970 $\pm$ 0.1359 & 1.0241 $\pm$ 0.0373 \\
W83.09 & 83086.00 - 83096.00 & 83090.650 & 9 & -9 & 1421.844 & 0.0422 $\pm$ 0.0008 & 2.8842 $\pm$ 0.0643 & -4.0288 $\pm$ 0.0836 & -0.2030 $\pm$ 0.0480 & 1.1956 $\pm$ 0.0120 \\
W83.63 & 83612.02 - 83637.02 & 83632.020 & 10 & -10 & 1394.056 & 0.2903 $\pm$ 0.0117 & 2.6202 $\pm$ 0.0493 & -0.9250 $\pm$ 0.0847 & 0.3441 $\pm$ 0.0381 & 1.0515 $\pm$ 0.0061 \\
W84.15 & 84140.00 - 84150.00 & 84147.100 & 11 & -11 & 1369.910 & 0.0151 $\pm$ 0.0038 & 3.0965 $\pm$ 0.1972 & 1.0096 $\pm$ 0.5604 & 0.2688 $\pm$ 0.2613 & 0.9923 $\pm$ 0.0696 \\
W84.64 & 84630.00 - 84650.00 & 84643.200 & 2 & -2 & 1860.752 & 0.1882 $\pm$ 0.0038 & 3.9241 $\pm$ 0.0400 & 1.2028 $\pm$ 0.0865 & 0.3489 $\pm$ 0.0872 & 2.0113 $\pm$ 0.0151 \\
W87.19 & 87170.00 - 87210.00 & 87192.800 & 2 & -2 & 1779.548 & 0.0513 $\pm$ 0.0017 & 5.2747 $\pm$ 0.0285 & -4.4093 $\pm$ 0.1868 & 0.3865 $\pm$ 0.0869 & 1.8036 $\pm$ 0.0114 \\
\hline
\end{tabular}
\vspace{-0.45cm}
\end{sidewaystable}

\section{Discussion}
\label{discussion}

The above analysis provides over 30 estimates of the C-ring's surface mass density and effective viscosity, as well as almost 30  normal mode amplitudes. While detailed studies of these measurements will likely require further investigation, we will discuss some implications of these measurements for the rings and the planet below. 

\subsection{Radial trends in ring properties across the C-ring}

\begin{figure}
  \hspace{0in}\includegraphics[width=1.0\textwidth]{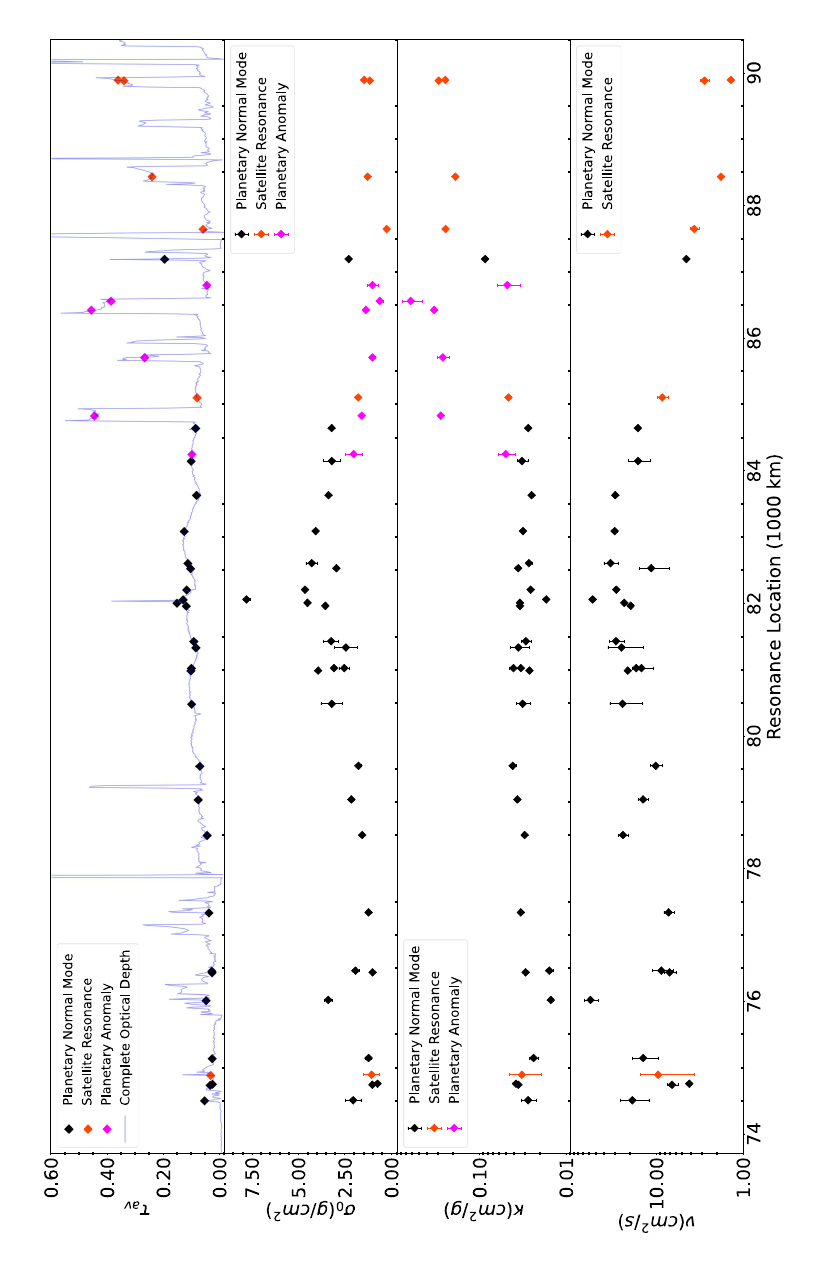}
  \caption{A panel of plots showing the average optical depth, surface mass density, mass extinction coefficient, and viscosity for satellite resonances and planetary normal modes, including planetary anomalies from \protect\citeA{Hedman_2022}.  The trace of the normal optical depth in the first panel is sourced from OPUS (Outer Planets Unified Search), a service of the Ring-Moon Systems Node of NASA's Planetary Data System (PDS). }
  \label{fig:ringplots}
\end{figure}

Figure~\ref{fig:ringplots} provides an overview of the ring parameters derived from both the satellite and planetary waves analyzed in this study. The top panel shows the normal optical depth, $\tau_n$, derived from a single VIMS occultation, while the second panel shows our estimates of the surface mass density, $\sigma_0$. The third panel shows the ratio of the above two quantities, $\kappa=\tau_n/\sigma_0$, which is the mass extinction coefficient. Finally, the bottom panel shows the estimated ring viscosity, $\nu$.

The surface mass density shows a generally increasing trend from around 1 g/cm$^2$ at  74,000 km to around 4 g/cm$^2$ at 82,000 km. On top of this overall trend, there are a couple of notable high points around 76,000 km and 82,000 km, which correspond to the waves W76.02 and W82.06, respectively. By contrast, exterior to 82,000 km, there is a somewhat steady decrease in surface mass density out to around 90,000 km, where the surface mass density falls back to around 1 g/cm$^2$.

We can better understand the implications of these large-scale trends in the ring's surface mass density by considering the corresponding trends in the ring's mass extinction coefficient $\kappa$, which quantifies the efficiency of light extinction per unit mass. Figure~\ref{fig:ringplots} shows that $\kappa$ is relatively constant around 0.03 cm$^2$/g between the radial range of 74,000 km to 85,000 km. However, exterior to 85,000 km, $\kappa$ is typically much larger, reaching values as high as 0.3 cm$^2$/g. Note that most of these high opacities are in regions of elevated optical depths called plateaux, but the Atlas 2:1 wave yields a comparably high $\kappa$ despite not being in a plateau. 

This change in the ring's mass extinction coefficient reflects a dramatic change in either the mass density or the typical size of the particles in the ring. If the ring particles have a mass density $\rho$ and the differential number density of those ring particles per unit ring area is given by the function $n(s)$, then the surface mass density of the rings is given by:

\begin{equation}
\sigma_0=\int \frac{4\pi}{3}\rho s^3 n(s) ds. 
\end{equation}
In addition, so long as the optical depth is sufficiently low, $\tau_n$ is just the total cross section per unit ring area of all the ring particles:

\begin{equation}
\tau_n=\int  \pi s^2 n(s) ds. 
\end{equation}
Hence the mass extinction coefficient is given by the ratio, assuming
density is independent of particle size,

\begin{equation}
\kappa=\frac{3\int s^2 n(s) ds}{4\rho \int s^3 n(s) ds}. 
\end{equation}
If we assume the particles follow a power-law size distribution with index $q$ (i.e. $n(s) \propto s^{-q}$) that is cut-off at a minimum size $s_{min}$ and a maximum size $s_{max}$, then this ratio can be represented as follows (for a case when $q \ne 3$) \cite{ZebkerMaroufTyler, BailliColwellLissauerEsposito, Zhang2017, Jerousek2020}:

\begin{equation}
    \kappa = \frac{3(4 - q)}{4(3 - q)}\left(\frac{s_{\rm max}^{3-q} - s_{\rm min}^{3-q}}{s_{\rm max}^{4-q} - s_{\rm min}^{4-q}}\right)\frac{1}{\rho}=\frac{3}{4s_{\rm av}}\frac{1}{\rho},
\label{eq:rhoeq}
\end{equation}
where $s_{\rm av}$ is the effective average ring particle size, and is given as:

\begin{equation}
 s_{\rm av} = \frac{(3-q)}{(4-q)} \frac{s_{\rm max}^{4-q}-s_{\rm min}^{4-q}}{s_{\rm max}^{3-q}-s_{\rm min}^{3-q}}.
 \label{eq:rhoeq2}
\end{equation}

Note that $s_{\rm av}$ does not exactly equal the effective average particle size $a_{\rm eff}$ used in previous analyses of occultation statistics \cite{showalter1990saturn, colwell2018particle,Jerousek2020} because the definitions of $a_{\rm eff}$ and $s_{\rm av}$ involve different moments of the particle size distribution. Specifically, $a_{\rm eff}$ depends on the fourth moment of the particle size distribution, rather than the third moment used in the computation of $\sigma_0$, and so tends to be somewhat larger than $s_{\rm av}$.

The roughly constant values of $\kappa$ interior to 85,000 km therefore imply the particles in this part of the ring have roughly the same mass density and average particle size. On the other hand, the much higher values of $\kappa$ seen in the outer C-ring imply a greatly reduced average particle size and/or typical particle mass density. 

\begin{figure}
    \centering
    \includegraphics[width=1\textwidth]{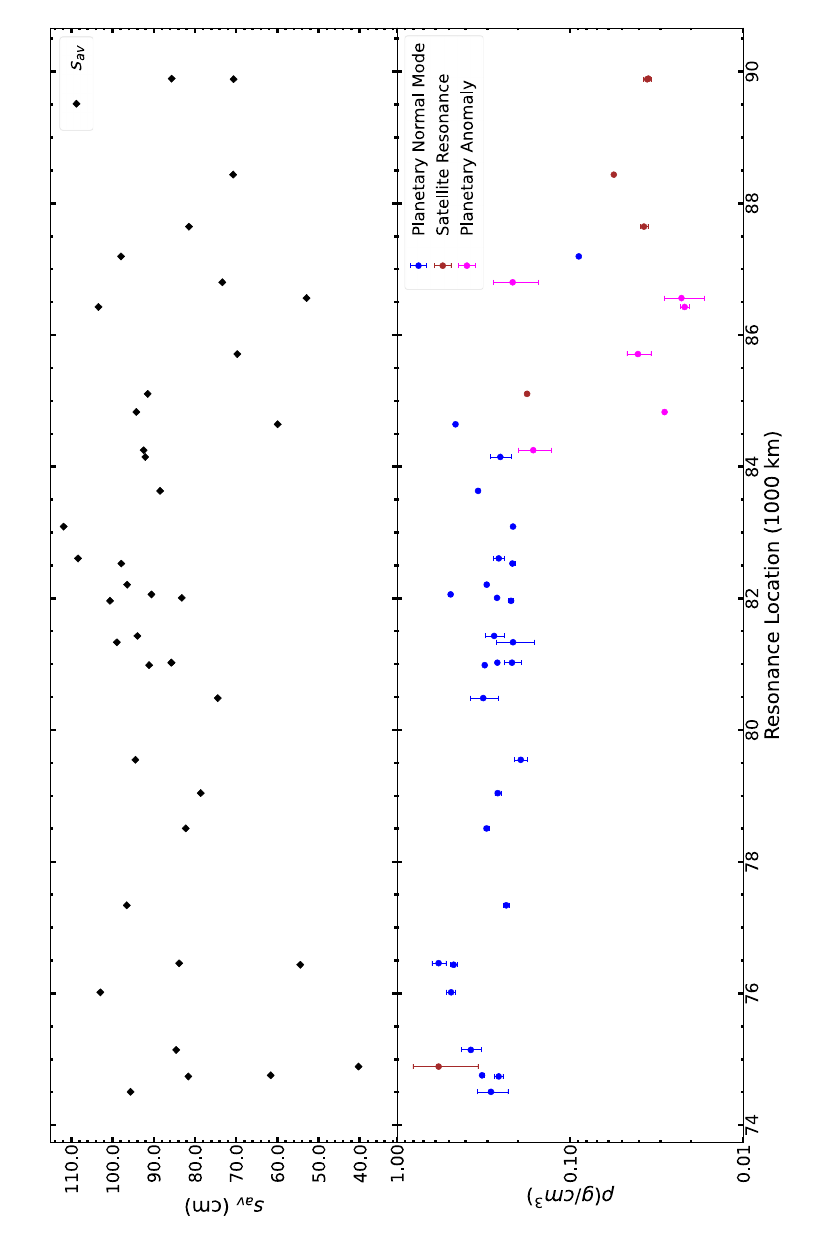}
    \caption{Plot showing trends in the C-ring's average effective particle size \protect\cite{Jerousek2020} and inferred average particle mass density as a function of radius across the rings.}
    \label{fig:densplot}
\end{figure}

Fortunately, prior work provides a means to independently estimate the typical particle sizes across the C-ring. In particular,
\citeA{Jerousek2020} fit Cassini occultation data obtained by multiple instruments at ultraviolet through radio wavelengths in order to estimate the power-law size distribution parameters $q$, $s_{\rm min}$, and $s_{\rm max}$. These parameters can then be used to compute the effective particle size in Equation~\ref{eq:rhoeq2}. In practice, \citeA{Jerousek2020}  evaluated these parameters in 10-km wide regions in the C-ring, so we simply evaluated $s_{\rm av}$ at locations closest to each of the waves. The resulting values can be found in Table~\ref{Tab:partdens}. Note that throughout the entire C-ring the effective average particle size ranges between 40 and 110 cm, and does not show strong trends with radius (see also Figure~\ref{fig:densplot}).

We then used these $s_{\rm av}$ values and the estimates for $\kappa$ in our work to compute the C-ring particle density $\rho$. Table ~\ref{Tab:partdens} shows the results from the implementation of the procedure described above in this sub-section, while Figure~\ref{fig:densplot} illustrates these variations across the C-ring.

Between 74,500 km and 84,000 km, we find that the typical particle densities range between 0.2 g/cm$^3$ and  0.6 g/cm$^3$, with most of the numbers being around 0.3 g/cm$^3$. However, between 84,000 and 89,000 km most of the particle density estimates are much lower, with the outermost four satellite waves and four of the anomaly-generated waves \cite{Hedman_2022} in this region giving typical particle densities between 0.02 and 0.06 g/cm$^3$. The innermost wave that shows such a low particle density is the wave designated PA2, which exists in close proximity to regions exhibiting much higher particle densities.  In this case the differences in particle densities could arise because PA2 occurs in a plateau while other nearby waves occur in the background C ring. By contrast, the waves outside 87,000 km all yield low particle densities regardless of whether they are found in plateaux or the background C ring.

Assuming the particles in the C-ring are composed primarily of water ice \cite{Zhang2017, Zhang2019, Miller2024}, our estimates of the particle densities interior to 84,000 km indicate that in this region the particles have porosities ranging between 40\% and 80\%. This is consistent with earlier analyses of the ring's radio-wave scattering properties that indicated the C-ring particles had porosities around 75\% \cite{Zhang2017, Zhang2019}. 
More remarkably, our data indicates that between 84,000 km and 90,000 km the porosity of the ring particles is typically of order 95\%, comparable to the porosity of freshly fallen, wind-blown snow \cite{clifton2008shear}. 

It is important to note that uncertainties in the C-ring's particle size distribution can influence these estimates of the particle densities and porosities. Recently, \citeA{Green2024Icar} analyzed Ultraviolet Imaging Spectrograph (UVIS) occultation data statistics that are more sensitive to larger particles than the wavelength-dependent optical depths used by\citeA{Jerousek2020}. Both studies show good agreement for the power-law exponent (q) and minimum particle radius ($a_{\rm min}$). However, in their studies of specific regions within the background C ring, \citeA{Green2024Icar} achieved a better fit by incorporating an ``ankle" into the size distribution. This ankle increases the number density of particles with radii between $a_{\rm bend}$ and $a_{\rm max}$. Assuming parameters consistent with the average background C ring parameters, this bent power law yields opacities approximately half as large as those found by \citeA{Jerousek2020}, implying the particle densities should be increased by roughly a factor of two. Doubling the derived average particle density in the inner C ring yields a particle density of roughly around 0.6 g/cm$^3$, corresponding to porosities of ~40\%. A factor of two increase in particle density in the outer C ring also remains consistent with extremely low-density particles with $\sim$90\% porosity.

Another notable trend that supports the idea that the particles in the outer C-ring are extremely porous involves our estimates of the ring's effective kinematic viscosity $\nu$. As shown in Figure~\ref{fig:ringplots}, there is a strong decrease in the viscosity in the same region where the opacity rises and the inferred particle density drops. For low optical depth rings like the C-ring,  $\nu$ depends on the root-mean square random velocity of the ring particles $c$ via the following equation \cite{araki1986dynamics, WisdomTremaine, 2009sfch.book..375C}

\begin{equation}
    \nu = k_{1}\frac{c^{2}}{n}\left(\frac{\tau}{1+\tau^{2}}\right) + k_{2}n D^{2}\tau.
\end{equation}
Here, $\nu$ signifies the kinematic viscosity, $\tau$ represents the local normal optical depth, $n$ stands for the orbital mean motion, and $D$ denotes the particle diameter. The constants $k_{1}$ and $k_{2}$ are characteristic coefficients typically on the order of unity. The sharp decrease in viscosity around 84,000 km therefore implies a similarly large decrease in the random velocities of the ring particles. Extremely porous particles would likely have extremely low coefficients of restitution \cite{Toyoda2024Icarus}, which would naturally reduce the velocity dispersion of the ring particles \cite{Schmidt2009, Hedman_2018}. The reasons behind and implications of the extremely low particle density and viscosity in the outer C-ring remain unclear and warrant further investigation in future studies.

In addition, Figure~\ref{fig:ringplots} shows significant radial gradients in viscosity throughout the entire C ring. Generally, viscosity gradients in planetary rings are expected to influence radial mass transport \cite{Stewart1984, Daisaka2001Icarus, Sega2024Icarus}. Further studies of these viscosity gradients using appropriate long-term evolution codes that include viscous processes \cite{Salmon2010Icarus, Estrada2015Icarus, Estrada2023Icarus} could therefore provide new insights into the long-term evolution of the structure and surface mass density distribution of the C ring.

\begin{table}
\caption{Data table for the effective average ring particle size and bulk particle density (from Planetary Normal Modes, Satellite Resonances and Planetary Anomalies).}
\label{Tab:partdens}
\centering
\begin{tabular}{lccccccr}
\hline
Wave ID & Radial Range (km) & $r_{L}$ (km) & $\ell$ & m & $s_{\rm av}$ (cm) & $\rho$ (g/cm$^3$) \\
\hline

Mimas 4:1 & 74880.00 - 74895.00 & 74890.070 & - & 2 & 40.294 & 0.5745 $\pm$ 0.2334 \\
Pan 2:1 & 85090.00 - 85130.00 & 85105.020 & - & 2 & 91.446 & 0.1772 $\pm$ 0.0032 \\
Atlas 2:1 & 87640.00 - 87650.00 & 87645.680 & - & 2 & 81.445 & 0.0374 $\pm$ 0.0019 \\
Prometheus 4:2 & 88420.00 - 88440.00 & 88434.120 & - & 3 & 70.739 & 0.0559 $\pm$ 0.0006 \\
Mimas 6:2 & 89870.00 - 89890.00 & 89884.000 & - & 3 & 70.627 & 0.0359 $\pm$ 0.0018 \\
Pandora 4:2 & 89887.00 - 89897.00 & 89893.680 & - & 3 & 85.632 & 0.0353 $\pm$ 0.0001 \\
\hline

$W74.51^{av}$ & 74501.00 - 74509.00 & 74506.900 & 12 & -8 & 95.615 & 0.2863 $\pm$ 0.0574 \\
W74.74 & 74736.00 - 74743.00 & 74739.850 & 15 & 13 & 81.588 & 0.2583 $\pm$ 0.0142 \\
W74.75 & 74746.00 - 74748.00 & 74748.300 & 11 & 11 & - & - \\
W74.76 & 74752.00 - 74762.00 & 74756.600 & 19 & -11 & 61.597 & 0.3220 $\pm$ 0.0078 \\
W75.14 & 75142.00 - 75144.00 & 75143.000 & 16 & -10 & 84.539 &  0.3743 $\pm$ 0.0476 \\
W76.02A & 76016.00 - 76018.00 & 76018.100 & 13 & -9 & 102.938 & 0.4865 $\pm$ 0.0288 \\
W76.44 & 76433.00 - 76436.00 & 76435.400 & 2 & -2 & 54.423 & 0.4702 $\pm$ 0.0223 \\
$W76.46^{av}$ & 76457.00 - 76462.00 & 76459.500 & 9 & -7 & 83.817 & 0.5745 $\pm$ 0.0530 \\
W77.34 & 77337.00 - 77339.00 & 77338.900 & 14 & -10 & 96.522 & 0.2335 $\pm$ 0.0097 \\
W78.51 & 78503.00 - 78509.00 & 78506.750 & 15 & -11 & 82.209 & 0.3033 $\pm$ 0.0120 \\
W79.04 & 79039.00 - 79045.00 & 79042.300 & 11 & -9 & 78.589 & 0.2615 $\pm$ 0.0104 \\
W79.55 & 79546.00 - 79550.00 & 79548.920 & 16 & -12 & 94.430 & 0.1926 $\pm$ 0.0162 \\
W80.49 & 80484.00 - 80488.00 & 80486.100 & 17 & -13 & 74.474 & 0.3171 $\pm$ 0.0577 \\
W80.99 & 80983.00 - 80989.00 & 80986.150 & 4 & -4 & 91.103 & 0.3112 $\pm$ 0.0014 \\
W81.023A & 81018.00 - 81030.00 & 81023.100 & 5 & -5 & 85.714 & 0.2162 $\pm$ 0.0246 \\
W81.024B & 81018.00 - 81030.00 & 81024.150 & 13 & -11 & 85.714 & 0.2631 $\pm$ 0.0032 \\
W81.33 & 81333.00 - 81336.00 & 81334.275 & 18 & -14 & 98.924 & 0.2136 $\pm$ 0.0529 \\
W81.43 & 81420.00 - 81430.00 & 81429.550 & 6 & -6 & 93.934 & 0.2745 $\pm$ 0.0345 \\
W81.96 & 81959.00 - 81965.00 & 81962.450 & 7 & -7 & 100.599 & 0.2194 $\pm$ 0.0068 \\
W82.01 & 82004.00 - 82009.00 & 82007.750 & 3 & -3 & 83.185 & 0.2641 $\pm$ 0.0012 \\
W82.06 & 82055.00 - 82065.00 & 82059.400 & 3 & -3 & 90.544 & 0.4893 $\pm$ 0.0107 \\
W82.21 & 82187.50 - 82207.51 & 82207.500 & 3 & -3 & 96.428 & 0.3030 $\pm$ 0.0051 \\
W82.53 & 82510.00 - 82530.00 & 82528.750 & 8 & -8 & 97.855 & 0.2151 $\pm$ 0.0068 \\
W82.61 & 82606.00 - 82608.00 & 82607.750 & 15 & -13 & 108.356 & 0.2580 $\pm$ 0.0188 \\
W83.09 & 83086.00 - 83096.00 & 83090.650 & 9 & -9 & 111.853 & 0.2133 $\pm$ 0.0043 \\
W83.63 & 83612.02 - 83637.02 & 83632.020 & 10 & -10 & 88.431 & 0.3397 $\pm$ 0.0039 \\
W84.15 & 84140.00 - 84150.00 & 84147.100 & 11 & -11 & 92.044 & 0.2528 $\pm$ 0.0355 \\
W84.64 & 84630.00 - 84650.00 & 84643.200 & 2 & -2 & 59.946 & 0.4586 $\pm$ 0.0069 \\
W87.19 & 87170.00 - 87210.00 & 87192.800 & 2 & -2 & 97.924 & 0.0890 $\pm$ 0.0011 \\
\hline

PA1 & 84200.00 - 84300.00 & 84250.000 & - & 3 & 92.433 & 0.1632 $\pm$ 0.0347 \\
PA2 & 84780.00 - 84880.00 & 84830.000 & - & 3 & 94.220 & 0.0284 $\pm$ 0.0004 \\
PA3 & 85660.00 - 85760.00 & 85710.000 & - & -1 & 69.724 & 0.0405 $\pm$ 0.0064 \\
PA4 & 86380.00 - 86470.00 & 86425.000 & - & 3 & 103.402 & 0.0218 $\pm$ 0.0013 \\
PA5 & 86520.00 - 86600.00 & 86560.000 & - & 3 & 52.903 & 0.0227 $\pm$ 0.0059 \\
PA6 & 86750.00 - 86850.00 & 86800.000 & - & 3 & 73.326 & 0.2146 $\pm$ 0.0616 \\
\hline
\end{tabular}

\textbf{Note:} PA1 to PA6 represent the planetary anomalies featured in \cite{Hedman_2022}, Table 2 therein.
\end{table}

\begin{sidewaystable}
\caption{Data table for average optical depth, surface mass density mass extinction coefficient and viscosity for satellite resonances and planetary normal modes.}
\label{Tab:ringpars}
\centering
\begin{tabular}{lccccccccccr}
\hline
Wave ID & Radial Range (km) & $r_{L}$ (km) & $\ell$ & $m$ & $\tau_{av}$ & $\sigma_{0} (g/cm^{2})$ & $\kappa = \frac{\tau_{av}}{\sigma_{0}} (cm^{2}/g)$ & $\nu (cm^{2}/s)$\\
\hline

Mimas 4:1 & 74880.00 - 74895.00 & 74890.070 & - &  2 & 0.0343 & 1.0598 $\pm$ 0.4306 & 0.0324 $\pm$ 0.0132 & 9.5348 $\pm$ 5.8497 \\
Pan 2:1 & 85090.00 - 85130.00 & 85105.020 & - &  2 & 0.0824 & 1.7798 $\pm$ 0.0317 & 0.0463 $\pm$ 0.0008 & 8.5279 $\pm$ 1.2422 \\
Atlas 2:1 & 87640.00 - 87650.00 & 87645.680 & - &  2 & 0.0617 & 0.2504 $\pm$ 0.0128 & 0.2462 $\pm$ 0.0126 & 3.6411 $\pm$ 0.4465 \\
Prometheus 4:2 & 88420.00 - 88440.00 & 88434.120 & - &  3 & 0.2417 &  1.2740 $\pm$ 0.0148 & 0.1898 $\pm$ 0.0022 & 1.7960 $\pm$ 0.0792 \\
Mimas 6:2 & 89870.00 - 89890.00 & 89884.000 & - &  3 & 0.3409 &  1.1512 $\pm$ 0.0569 & 0.2961 $\pm$ 0.0146 & 2.7785 $\pm$ 0.3269 \\
Pandora 4:2 & 89887.00 - 89897.00 & 89893.680 & - &  3 & 0.3619 & 1.4596 $\pm$ 0.0033 & 0.2479 $\pm$ 0.0006 & 1.3788 $\pm$ 0.0691 \\
\hline

$W74.51^{av}$ & 74501.00 - 74509.00 & 74506.900 & 12 & -8 & 0.0562 & 2.0519 $\pm$ 0.4112 & 0.0274 $\pm$ 0.0055 & 18.9263 $\pm$ 6.7063 \\
W74.74 & 74736.00 - 74743.00 & 74739.850 & 15 & 13 & 0.0364 & 1.0228 $\pm$ 0.0560 & 0.0356 $\pm$ 0.0020 & 6.5820 $\pm$ 0.9439 \\
W74.75 & 74746.00 - 74748.00 & 74748.300 & 11 & 11 & 0.0292 & - & - & - \\
W74.76 & 74752.00 - 74762.00 & 74756.600 & 19 & -11 & 0.0282 & 0.7451 $\pm$ 0.0181 & 0.0378 $\pm$ 0.0009 & 4.1699 $\pm$ 0.2034 \\
W75.14 & 75142.00 - 75144.00 & 75143.000 & 16 & -10 & 0.0290 & 1.2248 $\pm$ 0.1559 & 0.0237 $\pm$ 0.0030 & 14.1703 $\pm$ 4.6450 \\
W76.02A & 76016.00 - 76018.00 & 76018.100 & 13 & -9 & 0.0508 & 3.3898 $\pm$ 0.2005 & 0.0150 $\pm$ 0.0009 & 57.3702 $\pm$ 10.4715 \\
W76.44 & 76433.00 - 76436.00 & 76435.400 & 2 & -2 & 0.0294 & 1.0040 $\pm$ 0.0477 & 0.0293 $\pm$ 0.0014 & 7.0243 $\pm$ 1.0797 \\
$W76.46^{av}$ & 76457.00 - 76462.00 & 76459.500 & 9 & -7 & 0.0300 & 1.9283 $\pm$ 0.1778 & 0.0156 $\pm$ 0.0014 & 8.7552 $\pm$ 2.4059 \\
W77.34 & 77337.00 - 77339.00 & 77338.900 & 14 & -10 & 0.0407 & 1.2226 $\pm$ 0.0508 & 0.0333 $\pm$ 0.0014 & 7.2156 $\pm$ 0.9917 \\
W78.51 & 78503.00 - 78509.00 & 78506.750 & 15 & -11 & 0.0471 & 1.5661 $\pm$ 0.0619 & 0.0301 $\pm$ 0.0012 & 24.1692 $\pm$ 3.2607 \\
W79.04 & 79039.00 - 79045.00 & 79042.300 & 11 & -9 & 0.0785 & 2.1517 $\pm$ 0.0859 & 0.0365 $\pm$ 0.0015 & 14.1891 $\pm$ 1.8271 \\
W79.55 & 79546.00 - 79550.00 & 79548.920 & 16 & -12 & 0.0729 & 1.7691 $\pm$ 0.1486 & 0.0412 $\pm$ 0.0035 & 10.1429 $\pm$ 1.5421 \\
W80.49 & 80484.00 - 80488.00 & 80486.100 & 17 & -13 & 0.1015 & 3.1977 $\pm$ 0.5818 & 0.0318 $\pm$ 0.0058 & 24.4831 $\pm$ 9.8382 \\
W80.99 & 80983.00 - 80989.00 & 80986.150 & 4 & -4 & 0.1040 & 3.9304 $\pm$ 0.0179 &  0.0265 $\pm$ 0.0001 & 21.3265 $\pm$ 0.5456 \\
W81.023A & 81018.00 - 81030.00 & 81023.100 & 5 & -5 & 0.1023 & 2.5274 $\pm$ 0.2871 & 0.0405 $\pm$ 0.0046 & 14.8562 $\pm$ 3.9362 \\
W81.024B & 81018.00 - 81030.00 & 81024.150 & 13 & -11 & 0.1023 & 3.0758 $\pm$ 0.0373 & 0.0333 $\pm$ 0.0004 & 17.0547 $\pm$ 1.8346 \\
W81.33 & 81333.00 - 81336.00 & 81334.275 & 18 & -14 & 0.0866 & 2.4399 $\pm$ 0.6039 & 0.0355 $\pm$ 0.0088 & 25.1908 $\pm$ 11.0098 \\
W81.43 & 81420.00 - 81430.00 & 81429.550 & 6 & -6 & 0.0941 & 3.2342 $\pm$ 0.4063 & 0.0291 $\pm$ 0.0037 & 29.1486 $\pm$ 5.6024 \\
W81.96 & 81959.00 - 81965.00 & 81962.450 & 7 & -7 & 0.1205 &  3.5465$\pm$ 0.1096 &  0.0340 $\pm$ 0.0011 & 19.7645 $\pm$ 1.4795 \\
W82.01 & 82004.00 - 82009.00 & 82007.750 & 3 & -3 & 0.1537 & 4.5032 $\pm$ 0.0213 & 0.0341 $\pm$ 0.0002 & 23.4128 $\pm$ 0.2449 \\
W82.06 & 82055.00 - 82065.00 & 82059.400 & 3 & -3 & 0.1317 & 7.7803 $\pm$ 0.1703 &  0.0169 $\pm$ 0.0003 & 54.1606 $\pm$ 2.3422 \\
W82.21 & 82187.50 - 82207.51 & 82207.500 & 3 & -3 & 0.1191 & 4.6401 $\pm$ 0.0784 & 0.0257 $\pm$ 0.0004 & 28.9210 $\pm$ 1.4883 \\
W82.53 & 82510.00 - 82530.00 & 82528.750 & 8 & -8 & 0.1052 & 2.9511 $\pm$ 0.0930 & 0.0356 $\pm$ 0.0011 & 11.4504 $\pm$ 4.3673 \\
W82.61 & 82606.00 - 82608.00 & 82607.750 & 15 & -13 & 0.1148 & 4.2788 $\pm$ 0.3115 & 0.0268 $\pm$ 0.0020 & 33.6675 $\pm$ 6.3952 \\
W83.09 & 83086.00 - 83096.00 & 83090.650 & 9 & -9 & 0.1279 & 4.0696 $\pm$ 0.0820 & 0.0314 $\pm$ 0.0006 & 30.2127 $\pm$ 2.2168 \\
W83.63 & 83612.02 - 83637.02 & 83632.020 & 10 & -10 & 0.0842 & 3.3740 $\pm$ 0.0390 & 0.0250 $\pm$ 0.0003 & 29.6733 $\pm$ 1.7528 \\
W84.15 & 84140.00 - 84150.00 & 84147.100 & 11 & -11 & 0.1031 & 3.1983 $\pm$ 0.4486 & 0.0322 $\pm$ 0.0045 & 16.2324 $\pm$ 4.6136 \\
W84.64 & 84630.00 - 84650.00 & 84643.200 & 2 & -2 & 0.0875 & 3.2086 $\pm$ 0.0483 & 0.0273 $\pm$ 0.0004 & 16.3015 $\pm$ 0.6194 \\
W87.19 & 87170.00 - 87210.00 & 87192.800 & 2 & -2 & 0.1972 & 2.2912 $\pm$ 0.0290 & 0.0861 $\pm$ 0.0011 & 4.4932 $\pm$ 0.1122 \\
\hline
\end{tabular}
\vspace{-0.75cm}

\end{sidewaystable}

\subsection{Planetary Normal Mode Spectrum}\label{sec:modes}

\begin{figure}
\centering
\includegraphics[width=1.05\textwidth]{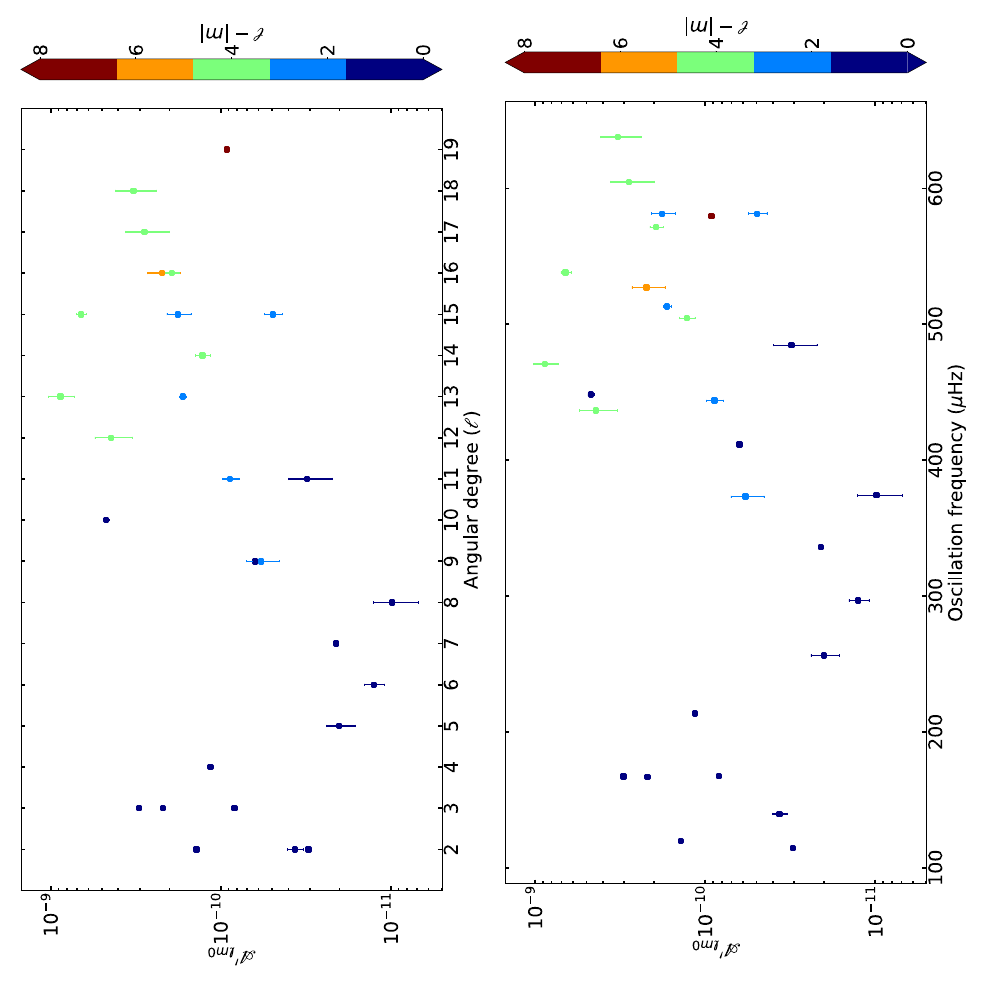}
\caption{
Plots showing Saturn's normal-mode amplitudes $\mathscr{A}'_{\ell mn}$ (from Equation~\ref{almneq}) as a function of either angular degree $\ell$ (top) or  oscillation frequency $\sigma_{\ell|m|0}/ 2\pi$ (from Table~\ref{Tab:ampparsu}) (bottom). Note that at $\ell = 2,3$ and frequencies less than 200 $\mu$Hz, the f-modes are apparently influenced by interactions with g-modes, and hence $n=0$ is something of an oversimplification (see text).}
\label{fig:saturn_normal_mode_amplitudeI}
\end{figure}

\begin{figure}
\centering
\includegraphics[width=1.05\textwidth]{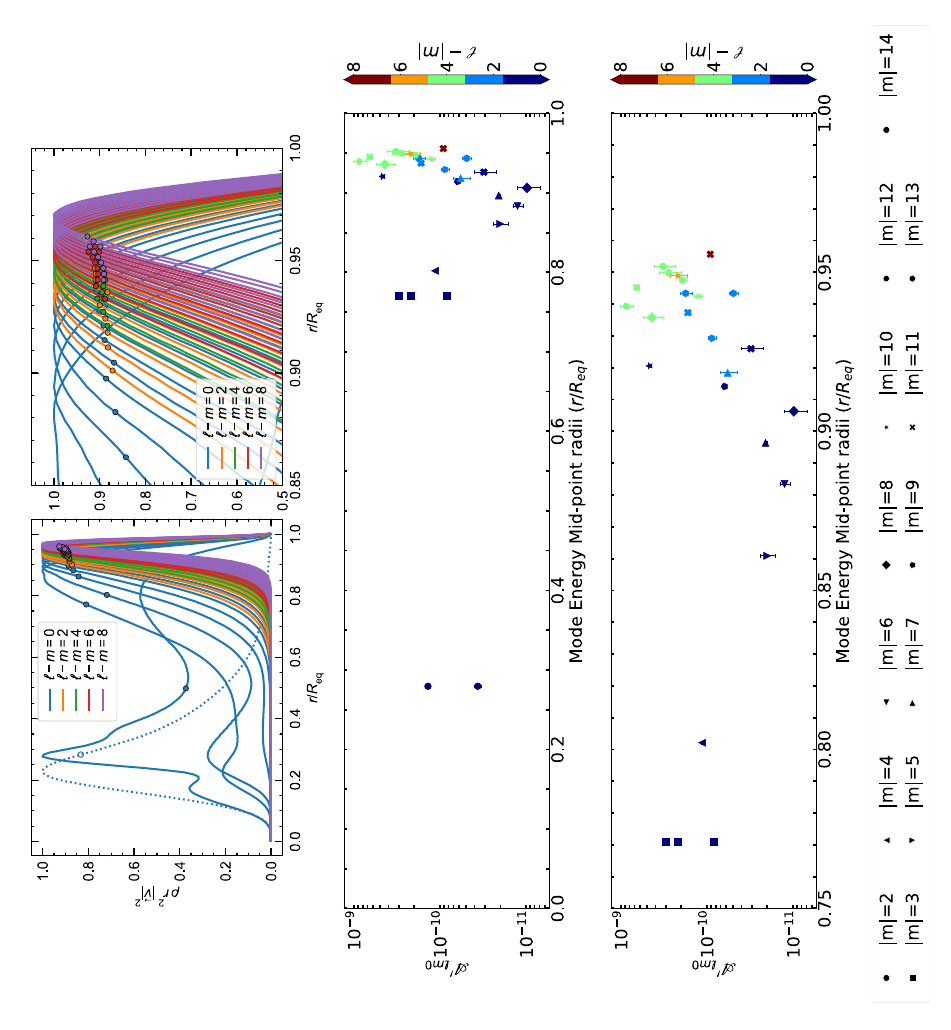}
\caption{(Top) Plots of the relative mode energy versus radius within Saturn for a variety of planetary normal modes within a representative model for Saturn's interior from \protect\cite{Mankovich2021}. Curves are color-coded by $\ell-|m|$ and modes with higher $\ell$ are more strongly concentrated near the planet's surface. Points on each curve mark the mode energy mid-point. For $m=2$ we show the $n=1$ g-mode (dashed line) in addition to the f-mode.
(Bottom) Observed normal mode amplitudes versus predicted mode energy mid-point radii (One panel covering the entire planet while the other shows a zoom-in on the region between $0.75 R_{\rm s}$ to $1.0 R_{\rm s}$). Note that most of the data fall along a clear trend where the mode amplitude increases with proximity to the planet's surface. The exceptions to this trend correspond to modes with $\ell=2-4$, which all extend deep inside the planet. Note that the three $m=3$ waves are expected to stem from three Saturn modes with distinct spatial structures; here these are all plotted with the energy midpoint radius $\sim0.77 R_{\rm s}$ corresponding to this model's $\ell=|m|=3$ f-mode.}
\label{fig:saturn_normal_mode_amplitudeII}
\end{figure}

Figure~\ref{fig:saturn_normal_mode_amplitudeI} shows plots of the normal mode amplitudes for Saturn $\mathscr{A}'_{\ell m0}$ derived from the observed density waves versus the angular degree $\ell$ and oscillation frequency, color coded by $\ell-|m|$, which is the number of latitudes where the mode has zero amplitude (see Section~\ref{Theory} above). This figure shows that the observed normal mode amplitudes range between 10$^{-9}$ and 10$^{-11}$, and  that the normal mode amplitude is a complex function of both $\ell$ and $\ell-|m|$. 

Working from left to right in these plots, we can first note that the modes with $\ell=2,3,4$ and oscillation frequencies below 220 $\mu$Hz include relatively high amplitudes in excess of 10$^{-10}$. By contrast, all the modes with $\ell=5,6,7,8$ (which have frequencies between 250 and 400 $\mu$Hz) all have rather low amplitudes around $1-2\times10^{-11}$. Note that the only modes we have data on, with these low $\ell$ modes, have $\ell-|m|=0$ because most of the modes with other values of $\ell-|m|$ would produce resonances interior to the C-ring's inner edge. For $\ell>9$, by contrast, we detected modes with a range of $\ell-|m|$ values. Interestingly, the strongest detectable modes between $\ell=10$ and $\ell=19$ have amplitudes that are either comparable to or even exceed those with $\ell<5$. Also, in this same $\ell$ range, the dominant signal seems to transition from the sectoral $\ell-|m|=0$ modes to the $\ell-|m|=4$ modes. In addition, none of the $\ell-|m|=2$ modes in this range appear to be as strong as the strongest $\ell-|m|=0$ or $\ell-|m|=4$ waves. (With only a single wave generated by each of the $\ell-|m|=6$ and 8 modes, the trends involving those sorts of modes are still obscure). 

\begin{sidewaystable}
\centering
\begin{tabular}{lccccccccccr}  
\hline
Wave ID & Radial Range (km) & $r_{L}$ (km) & $\ell$ & $m$ & $\Omega_{P}(^{o}/day)$ & $\sigma_{\ell|m|0}/2\pi = |m|\Omega_{P}/ 2\pi  $ ($\mu Hz$) & $\mathscr{A}'_{\ell m0}\times 10^{-10}$ \\
\hline

$W74.51^{av}$ & 74501.00 - 74509.00 & 74506.900 & 12 & -8 & 1697.339 & 436.556 & 4.402 $\pm$ 1.106 \\
W74.74 & 74736.00 - 74743.00 & 74739.850 & 15 & 13 & 1390.843 & 581.303 & 0.495 $\pm$ 0.060 \\
W74.75 & 74746.00 - 74748.00 & 74748.300 & 11 & 11 & 1369.932 & 484.476 & - \\
W74.76 & 74752.00 - 74762.00 & 74756.600 & 19 & -11 & 1638.389 & 579.416 & 0.922 $\pm$ 0.026 \\
W75.14 & 75142.00 - 75144.00 & 75143.000 & 16 & -10 & 1638.983 & 526.933 & 2.210 $\pm$ 0.483 \\
W76.02A & 76016.00 - 76018.00 & 76018.100 & 13 & -9 & 1626.530 & 470.636 & 8.769 $\pm$ 1.496 \\
W76.44 & 76433.00 - 76436.00 & 76435.400 & 2 & -2 & 2169.260 & 139.483 & 0.366 $\pm$ 0.037 \\
$W76.46^{av}$ & 76457.00 - 76462.00 & 76459.500 & 9 & -7 & 1657.720 & 373.070 & 0.579 $\pm$ 0.127 \\
W77.34 & 77337.00 - 77339.00 & 77338.900 & 14 & -10 & 1569.078 & 504.459 & 1.283 $\pm$ 0.132 \\
W78.51 & 78503.00 - 78509.00 & 78506.750 & 15 & -11 & 1521.410 & 538.047 & 6.614 $\pm$ 0.452 \\
W79.04 & 79039.00 - 79045.00 & 79042.300 & 11 & -9 & 1533.336 & 443.671 & 0.882 $\pm$ 0.101 \\
W79.55 & 79546.00 - 79550.00 & 79548.920 & 16 & -12 & 1481.152 & 571.428 & 1.943 $\pm$ 0.174 \\
W80.49 & 80484.00 - 80488.00 & 80486.100 & 17 & -13 & 1446.654 & 604.629 & 2.809 $\pm$ 0.816 \\
W80.99 & 80983.00 - 80989.00 & 80986.150 & 4 & -4 & 1660.363 & 213.523 & 1.147 $\pm$ 0.030 \\
W81.023A & 81018.00 - 81030.00 & 81023.100 & 5 & -5 & 1593.630 & 256.176 & 0.201 $\pm$ 0.038 \\
W81.024B & 81018.00 - 81030.00 & 81024.150 & 13 & -11 & 1450.495 & 512.968 & 1.675 $\pm$ 0.087 \\
W81.33 & 81333.00 - 81336.00 & 81334.275 & 18 & -14 & 1416.734 & 637.672 & 3.268 $\pm$ 0.886 \\
W81.43 & 81420.00 - 81430.00 & 81429.550 & 6 & -6 & 1538.237 & 296.726 & 0.126 $\pm$ 0.017 \\
W81.96 & 81959.00 - 81965.00 & 81962.450 & 7 & -7 & 1492.457 & 335.877 & 0.209 $\pm$ 0.007 \\
W82.01 & 82004.00 - 82009.00 & 82007.750 & 3 & -3 & 1736.645 & 167.499 & 0.834 $\pm$ 0.012 \\
W82.06 & 82055.00 - 82065.00 & 82059.400 & 3 & -3 & 1734.999 & 167.341 & 3.026 $\pm$ 0.083 \\
W82.21 & 82187.50 - 82207.51 & 82207.500 & 3 & -3 & 1730.293 & 166.887 & 2.188 $\pm$ 0.054 \\
W82.53 & 82510.00 - 82530.00 & 82528.750 & 8 & -8 & 1454.224 & 374.026 & 0.098 $\pm$ 0.029 \\
W82.61 & 82606.00 - 82608.00 & 82607.750 & 15 & -13 & 1390.843 & 581.303 & 1.788 $\pm$ 0.284 \\
W83.09 & 83086.00 - 83096.00 & 83090.650 & 9 & -9 & 1421.844 & 411.411 & 0.629 $\pm$ 0.017 \\
W83.63 & 83612.02 - 83637.02 & 83632.020 & 10 & -10 & 1394.056 & 448.189 & 4.706 $\pm$ 0.197 \\
W84.15 & 84140.00 - 84150.00 & 84147.100 & 11 & -11 & 1369.910 & 484.469 & 0.310 $\pm$ 0.089 \\
W84.64 & 84630.00 - 84650.00 & 84643.200 & 2 & -2 & 1860.752 & 119.646 & 1.389 $\pm$ 0.035 \\
W87.19 & 87170.00 - 87210.00 & 87192.800 & 2 & -2 & 1779.548 & 114.425 & 0.305 $\pm$ 0.011 \\
\hline
\end{tabular}
\vspace{-0.1cm}
\caption{Data table for Saturn's normal mode amplitude ($\mathscr{A}'_{\ell m0}$) values and corresponding oscillation frequencies for each Planetary resonance.
Note that $\mathscr{A}'_{\ell m0}$ was computed at $n=0$, which corresponds to f-modes.}
\label{Tab:ampparsu}
\end{sidewaystable}

While theoretical work will be needed to fully understand these trends in Saturn's normal mode amplitudes, preliminary insights can be gleaned from the relationship between our measured amplitudes and the effective radial depth probed by each mode. In particular, we consider the mode kinetic energy density $\rho|\vec v|^2$, where $\vec v$ is the velocity perturbation vector associated with a given mode and $\rho$ is the local background density in the planet. Integrating this quantity over the volume of the planet gives the kinetic energy of the mode. The top panels of Figure~\ref{fig:saturn_normal_mode_amplitudeII} show mode energy densities, weighted by $r^2$ from the spherical volume element, as a function of equatorial radius coordinate in a representative Saturn interior model from \citeA{Mankovich23}. These include equatorially symmetric (even $\ell-m$) prograde modes spanning $m=2-14$ and $\ell=2-22$, calculated using the code developed by \cite{Dewberry2021}. The interior model is stably stratified in the region $0<r<0.65 R_{\rm eq}$ and hence hosts g-modes, one of which ($\ell=m=2$, $n=1$) is included in Figure~\ref{fig:saturn_normal_mode_amplitudeII}; the remainder are f-modes. To assign a characteristic depth to each of these modes, we compute its ``energy midpoint radius," that is, the radius at which the mode energy integral reaches half its total value. Energy midpoint radii are marked by circular symbols in the top panels of Figure~\ref{fig:saturn_normal_mode_amplitudeII}.

If we plot our normal mode amplitudes versus these nominal mode energy midpoints, we see that most of the data points fall along a trend where the amplitude increases as the mode energy midpoint approaches the planet's surface. This suggests that these modes are probably being generated by processes operating at fairly shallow depth. This shallow source of excitation could include various atmospheric processes like impacts and storms \cite{Markham2018, Wu2019ApJ}. In that context, it is important to note that the highest amplitude modes have mode energy midpoints between 0.92 and 0.95, which corresponds to depths between 3000 km and 8000 km, and pressures of order 10$^4$ - 10$^5$ bars \cite{Kaspi2020}. This is within the region of the planet that likely exhibits differential rotation \cite{Galanti2019, Iess19, Mankovich23}, and is much greater than the expected penetration depth of cometary impactors \cite{Boslough1994}. However, the pressures at this depth could be comparable to the expected depth of the ``rock storms" that have been proposed to excite normal mode oscillations inside Jupiter and Saturn \cite{Markham2018}. 

At the same time, the sectoral modes with $\ell=m=2,3,4$ all fall well above the trend one would expect based on extrapolating from the $\ell>5$ modes. Importantly, the energy profiles for all these modes exhibit local maxima within $0.3$ planetary radii of Saturn's center. This correlation suggests that these modes may be excited deep inside the planet, perhaps in the stably stratified layer responsible for producing the multiple $\ell=m=2,3$ modes \cite{Mankovich2021}.

The physical process behind such a deep excitation mechanism is not clear. In the \citeA{Mankovich2021} model, these deep layers are stably stratified principally by a strong heavy element gradient, and the highly uncertain helium gradient following from H-He immiscibility may also play a role \cite{Fortney2003}. In either case, a deep source of mode excitation may involve fluid instabilities associated with the double-diffusive configuration established by the competing temperature and composition gradients in this region. This deep source of excitation may also be sensitive to the distribution of heavier elements deep inside Saturn since it is the molecular weight gradient in the diffuse core that makes it stable against convection and able to support g-modes, unlike the outer convective regions. Note that the relatively low amplitude of the $\ell=m=5$ mode also extends fairly deeply into the planet, so this mode may also help constrain where and how these deeper modes are excited. 

We note that the amplitudes we infer from observations of $m=2, 3$ waves may be sensitive to the fact that in rapidly rotating planets, gravitational perturbation produced by a single oscillation mode is not strictly limited to a single spherical harmonic. Instead, modes in rapid rotators (like Saturn) can involve simultaneous perturbations in multiple spherical harmonic degrees \cite{Dewberry2021}. This coupling across spherical harmonics is weak for fundamental oscillation modes with high values of $\ell$ and $m$, so we do not expect it to significantly affect our interpretation of the majority of the ring wave amplitudes. The same may not be true for the multiple $m=2$ and $m=3$ waves with closely spaced frequencies, the leading explanation for which involves higher degree gravito-inertial modes that commonly have more complicated structure \cite{Fuller_2014,Mankovich2021,Dewberry2021,Friedson2023}. However, we highlight that if some of the observed $m=2, 3$ modes with finely split frequencies were to involve higher degree gravitational perturbations at Saturn’s surface, the fall-off of gravitational potentials as $(R/r)^{\ell+1}$ in the external vacuum implies that our estimates of amplitudes for such modes would be lower bounds. The possible involvement of higher degree gravito-inertial modes therefore only strengthens our finding that Saturn’s deeper seated oscillations exhibit at least an order of magnitude larger amplitudes than might be expected from the trend observed for higher degrees.

\section{Conclusions}
\label{conclusion}

This study of density waves in Saturn's C-ring has provided new insights into both the planet's interior and its rings. We identify interesting trends in the ring's surface mass density and effective viscosity which suggest that the particles in the outer C-ring have extremely low mass densities and extremely high porosities. At the same time, trends in the amplitudes of the planetary normal modes indicate that two different mechanisms are responsible for exciting these oscillations, one that operates deep in Saturn's interior and another that occurs within a few thousand kilometers of its surface. Additionally, we have observed systematic differences in the amplitudes of modes with different latitudinal wave numbers (the combination $\ell - |m|$ of spherical harmonic degree $\ell$ and order $m$). These differences suggest that the oscillations' latitudinal structure may play a significant role in their excitation mechanisms.

\section*{Acknowledgements}
This research has made use of data obtained from the recent Cassini Mission (2004-2017) and was supported in part by NASA Cassini Data Analysis Grant NNX17AF85G. We also wish to thank M. Tiscareno, M. El Moutamid, B. Idini, J. Fuller, M. Marley, L. Quick and J. Barnes for helpful conversations. 

The authors confirm that there are no real or perceived financial conflicts of interest related to this work.

\section*{Data Availability Statement}  

The data that support the findings of this study are openly available in VERSO (University of Idaho) at \url{https://doi.org/10.60841/000000005} and specified by DOI 10.60841/000000005 \cite{AfigboVictor2024DfUW}. These data were derived from raw Cassini occultation data described in detail in \citeA{cassini_vims_profiles2} and are archived in the Planetary Data System at locations specified by DOIs 10.17189/1520275 and 10.17189/1522419 \cite{cassini_vims_cube_collection, cassini_vims_profiles2}. Calculations and figures were made using Python Programming Language version 3.11.5, available from the Python Software Foundation at \url{https://www.python.org/}.

\appendix

\section{Derivation of the Reconstruction constant, $\mathscr{C}$}
\label{reconstructionconstant}

The following procedures outline the steps involved in the determination of the constant used in the signal reconstruction, $\mathscr{C}$. To start with, using a normal convolution technique on a given signal, $\mathcal{Y}(R)$, we could evaluate the FFT as, 

\begin{equation}
    \mathcal{\tilde{Y}}(k) = \int_{-\infty}^{+\infty} \mathcal{Y}(R) e^{ikR} \, dR.
\label{constant1}
\end{equation}

Performing the inverse FFT on the signal $\mathcal{\tilde{Y}}(k)$, and using the products $\hat{y}(k)\hat\psi(sk)$, can yield the desired wavelet transform $\mathcal{\tilde{Y}}(s,r)$ \cite{1998BAMS...79...61T}. This can take the form of the actual convolution outlined below as,

\begin{equation}
   \mathcal{\tilde{Y}}(s,r) = \frac{1}{2\pi} \int_{-\infty}^{+\infty} \mathcal{\tilde{Y}}(k)\hat\psi(sk) e^{-ikr} \, dk.
\label{constant2}
\end{equation}

We can express the integral Equation~\ref{constant2} in terms of the integral components of Equation~\ref{constant1} as given below,

\begin{equation}
 \mathcal{\tilde{Y}}(s,r) = \frac{1}{2\pi} \int_{-\infty}^{+\infty} \mathcal{Y}(R) \int_{-\infty}^{+\infty} e^{-ik(r-R)} \pi^{-1/4}e^{-(sk-\omega_0)^2/2} \, dk\, dR.
\label{constant3}
\end{equation}
We can simplify this to have it in the form below,

\begin{equation}
 \mathcal{\tilde{Y}}(s,r) = \frac{1}{2\pi^{5/4}} \int_{-\infty}^{+\infty} \mathcal{Y}(R) \left\{\int_{-\infty}^{+\infty} e^{-ik(r-R)} e^{-(sk-\omega_0)^2/2} \, dk\right\}\, dR.
\label{constant4}
\end{equation}

Here, let $\mathcal{I}_{1} = \int_{-\infty}^{+\infty}e^{-ik(r-R)}e^{-(sk-\omega_0)^2/2}\,dk$ in Equation~\ref{constant4}. This expression can be simplified and rephrased as $\mathcal{I}_1 = \int_{-\infty}^{+\infty} e^{-\frac{1}{2} \left[ 2ik(r-R) + (sk-\omega_0)^2 \right]} \, dk$.

The terms in square brackets can be simplified further by  completing the squares, as described below,

\begin{equation}
    [2ik(r-R)+(sk-\omega_0)^2] = \left\{sk-\left[\omega_{0} - \frac{i(r-R)}{s}\right]\right\}^{2} + \frac{2i\omega_{0}(r-R)}{s} + \frac{(r-R)^2}{s^2}.
\label{constant5}
\end{equation}
By substituting the simplified expression for Equation~\ref{constant5} into the $\mathcal{I_{1}}$ integral expression, including rearranging for terms not dependent on $k$, we have it as,

\begin{equation}
    \mathcal{I}_{1} = e^{-\frac{i\omega_{0}(r-R)}{s}} e^{-\frac{1}{2}\frac{(r-R)^2}{s^2}} \int_{-\infty}^{+\infty} e^{-\frac{1}{2}\left\{sk-\left[\omega_{0} - \frac{i(r-R)}{s}\right]\right\}^{2}} \, dk.
\label{constant8}
\end{equation}
Taking $\alpha = sk-\left[\omega_{0} - \frac{i(r-R)}{s}\right]$, and working through the steps algebraically, this results in,

\begin{equation}
    \mathcal{I}_{1} = e^{-\frac{i\omega_{0}(r-R)}{s}} e^{-\frac{1}{2}\frac{(r-R)^2}{s^2}} s^{-1} \int_{-\infty}^{+\infty} e^{-\frac{\alpha^2}{2}} \, d \alpha = \sqrt{2\pi}e^{-\frac{i\omega_{0}(r-R)}{s}} e^{-\frac{1}{2}\frac{(r-R)^2}{s^2}} s^{-1}.
\label{constant9}
\end{equation}

Recall, from Equation~\ref{constant4}, we make substitutions for $\mathcal{I}_{1}$, which become,

\begin{equation}
 \mathcal{\tilde{Y}}(s,r) = \frac{1}{\sqrt{2}\pi^{3/4}} \int_{-\infty}^{+\infty} \mathcal{Y}(R) e^{-\frac{i\omega_{0}(r-R)}{s}} e^{-\frac{1}{2}\frac{(r-R)^2}{s^2}} s^{-1} \, dR.
\label{constant11}
\end{equation}

To properly reconstruct the equivalent of the original signal, we use a model that scales as,

\begin{equation}
 \mathcal{Y}'(r) = \mathscr{C} \int_{0}^{+\infty} \mathcal{\tilde{Y}}(s,r)\, \frac{ds}{s}.
\label{constant12}
\end{equation}

In terms of the ingredients for Equation~\ref{constant11}, we rewrite and rearrange Equation~\ref{constant12} as in the format, 

\begin{equation}
 \mathcal{Y}'(r) = \frac{\mathscr{C}}{\sqrt{2}\pi^{3/4}} \int_{-\infty}^{+\infty} \mathcal{Y}(R) \left\{ \int_{0}^{+\infty} e^{-\frac{i\omega_{0}(r-R)}{s}} e^{-\frac{1}{2}\frac{(r-R)^2}{s^2}}\, \frac{ds}{s^2}\right\}\, dR.
\label{constant13}
\end{equation}

Here, we take $\mathcal{I}_{2} = \int_{0}^{+\infty} e^{-\frac{i\omega_{0}(r-R)}{s}} e^{-\frac{1}{2}\frac{(r-R)^2}{s^2}}\, \frac{ds}{s^2}$. Assume $t = \frac{1}{s}$, we can substitute into $\mathcal{I}_{2}$, which becomes,

\begin{equation}
    \mathcal{I}_{2} = \int_{0}^{+\infty} e^{-i\omega_{0}t(r-R)} e^{-\frac{1}{2}[t(r-R)]^2}\, dt
\label{constant14}
\end{equation}

\begin{figure}[h]
\centering
\includegraphics[width=1.0\textwidth]{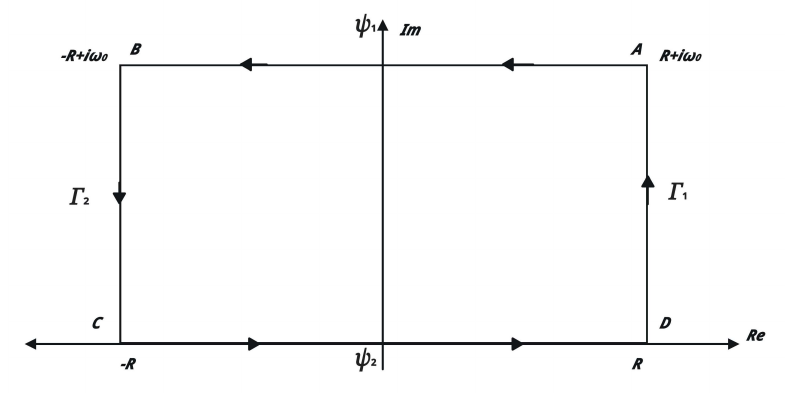}
\caption{Description of a rectangular contour, consisting of the real (horizontal) and imaginary (vertical) axis.}
\label{rectangularcontourdiagram}
\end{figure}

At this point, we would like to further express Equation~\ref{constant14} to best fit a variant of standard integrals. Let $\varrho = t(r-R)$. This implies that Equation~\ref{constant14} becomes, $\mathcal{I}_{2} = \frac{1}{(r-R)}\mathcal{I}_{3}$, where $\mathcal{I}_{3} = \int_{0}^{+\infty} e^{-i\omega_{0}\varrho} e^{-\frac{1}{2}\varrho^2}\, d\varrho$. To resolve $\mathcal{I}_{3}$, we use the concept of contour integration, specifically along a rectangular contour as described in Figure~\ref{rectangularcontourdiagram} \cite{1968hmfw.book.....A, Arfken2005}.

Given the complex function $f(z)=e^{-\frac{1}{2}z^2}$, it possesses several notable properties, but we are going to mention a few that are required for this procedure. For a start, \( f(z) \) is an entire function, meaning it is holomorphic everywhere in the complex plane.

Additionally, \( f(z) \) is not periodic; the exponential function \( e^{-\frac{1}{2}z^2} \) does not repeat values in a regular interval for any non-zero coefficient of $z^2$. The growth of \( f(z) \) depends on the coefficient of $z^2$ (or the real part of such coefficient, if it were a complex number). Since the coefficient of $z^2$ is greater than zero, \( f(z) \) rapidly decays to zero as \( |z| \to \infty \) in all directions. 

Importantly, \( f(z) = e^{-\frac{1}{2}z^2} \) has no zeros in the complex plane, as the exponential function is never zero for any finite \( z \). Using the Cauchy-Goursat theorem, we can say that,

\begin{equation}
  \oint_{C} f(z) \, dz = \int_{\Gamma_{1}} f(z) \, dz + \int_{\psi_{1}} f(z) \, dz + \int_{\Gamma_{2}} f(z) \, dz +  \int_{\psi_{2}} f(z) \, dz,
\label{constant15}
\end{equation}

where $\oint_{C} f(z) \, dz = 0$, according to the aforementioned theorem, Cauchy-Goursat \cite{1968hmfw.book.....A, Arfken2005}. Next, we consider the four length segments of the given rectangular contour with the given conditions, $C : \left\{ z = x \, \middle| \, -R \leq x \leq R \right\} \cup \left\{z = R+iy \, \middle| \, 0 \leq y \leq \omega_{0} \right\} \cup \left\{z = x + i\omega_{0} \, \middle| \,  -R \leq x \leq R\right\} \cup \left\{z = -R + iy \, \middle| \,  0 \leq y \leq \omega_{0}\right\}$. By substituting into Equation~\ref{constant15}, this becomes,

\begin{equation}
  0 = \int_{0}^{\omega_{0}} e^{-\frac{1}{2}(R+iy)^2} \, idy + \int_{R}^{-R} e^{-\frac{1}{2}( x + i\omega_{0})^2} \, dx + \int_{\omega_{0}}^{0} e^{-\frac{1}{2}(-R + iy)^2} \, idy +  \int_{-R}^{R} e^{-\frac{1}{2}x^2} \, dx.
\label{constant16}
\end{equation}

From Equation~\ref{constant16}, we can eliminate the line segments $\Gamma_1$ and $\Gamma_2$ by doing the following: using the integral for $\Gamma_2$, we use the reflection of the variable, say $y \rightarrow -y$. This implies, $-\int_{-\omega_{0}}^{0} e^{-\frac{1}{2}(-R + -iy)^2} \, idy = -\int_{0}^{\omega_{0}} e^{-\frac{1}{2}(R+iy)^2} \, idy$, which shows both line segments have the same magnitude but opposite direction of contour navigation. Now, we substitute into Equation~\ref{constant16}, with regards to the following condition,

\begin{equation}
  0 = \lim_{{R \to \infty}} \int_{R}^{-R} e^{-\frac{1}{2}( x + i\omega_{0})^2} \, dx + \lim_{{R \to \infty}} \int_{-R}^{R} e^{-\frac{1}{2}x^2} \, dx.
\label{constant17}
\end{equation}

This becomes,

\begin{equation}
   \lim_{{R \to \infty}} \int_{-R}^{R} e^{-\frac{1}{2}( x + i\omega_{0})^2} \, dx = \lim_{{R \to \infty}} \int_{-R}^{R} e^{-\frac{1}{2}x^2} \, dx.
\label{constant18}
\end{equation}

\begin{equation}
   \int_{-\infty}^{+\infty} e^{-\frac{1}{2}( x + i\omega_{0})^2} \, dx = \int_{-\infty}^{+\infty} e^{-\frac{1}{2}x^2} \, dx.
\label{constant19}
\end{equation}

Recall that, $\int_{-\infty}^{+\infty} e^{-\frac{1}{2}x^2} \, dx = \sqrt{2\pi}$. This implies that Equation~\ref{constant19} becomes,

\begin{equation}
   \int_{-\infty}^{+\infty} e^{-i\omega_{0}x}e^{-\frac{1}{2}x^{2}} dx = \sqrt{2\pi} e^{-\frac{1}{2}\omega_{0}^{2}}.
\label{constant20}
\end{equation}

Here, it is obvious that, $\mathcal{I}_{3} = \int_{0}^{+\infty} e^{-i\omega_{0}\varrho} e^{-\frac{1}{2}\varrho^2}\, d\varrho = \frac{1}{2}\int_{-\infty}^{+\infty} e^{-i\omega_{0}x}e^{-\frac{1}{2}x^{2}} dx = \frac{1}{2} \sqrt{2\pi} e^{-\frac{1}{2}\omega_{0}^{2}}$. This leaves us with $\mathcal{I}_{2} = \frac{1}{(r-R)}\frac{1}{2} \sqrt{2\pi} e^{-\frac{1}{2}\omega_{0}^{2}}$.

From Equation~\ref{constant13}, we go on to make substitutions for $\mathcal{I}_{2}$,

\begin{equation}
 \mathcal{Y}'(r) = \frac{\mathscr{C}}{\sqrt{2}\pi^{3/4}} \int_{-\infty}^{+\infty} \mathcal{Y}(R) \frac{1}{(r-R)}\frac{1}{2} \sqrt{2\pi} e^{-\frac{1}{2}\omega_{0}^{2}}\, dR.
\label{constant21}
\end{equation}

The main goal is simply to derive an expression for $\mathscr{C}$, the positive normalization term. The post-integral function given in Equation~\ref{constant21} is a variant of the normal Gaussian distribution, which we can standardize using the format below,

\begin{equation}
 \mathcal{Y}'(r) = \frac{\mathscr{C}\pi^{1/4}}{\sqrt{2}} \int_{-\infty}^{+\infty} \mathcal{Y}(R) \frac{1}{\sqrt{2\pi(r-R)^2}} e^{-\frac{[\omega_{0}(r-R)]^{2}}{2(r-R)^{2}}}\, dR.
\label{constant22}
\end{equation}

To best handle the integral in Equation~\ref{constant22}, there is a need for the ``nascent" delta function approach \cite{1968hmfw.book.....A, Brewster2018, Waters2018}. This technique is used in mathematical analysis, particularly in the context of integrals involving variants of real Dirac delta functions. The Dirac delta function, denoted \(\delta(r-R)\), is not a function in the traditional sense but rather a distribution. However, because \(\delta(r-R)\) is not a true function but an idealized concept, it can be approximated by a sequence of functions that are known as ``nascent delta functions." These are functions that approximate the Dirac delta function in the limiting case, as a parameter of such a function approaches zero \cite{MJLighthill1958}. Common examples include the Gaussian function, the sinc function, and other peaked functions.

For our case, we shall resolve the integrals involving the Dirac delta function, and then take the limit as the parameter, such as $\beta \triangleq (r-R)^2$ in our case, approaches zero \cite{Brewster2018, Waters2018}. From Equation~\ref{constant22}, we have, 

\begin{equation}
 \mathcal{Y}'(r) = \frac{\mathscr{C}\pi^{1/4}}{\sqrt{2}} \int_{-\infty}^{+\infty} \mathcal{Y}(R) \delta_{\beta}[\omega_{0}(r-R)]\, dR.
\label{constant23}
\end{equation}

Based on the condition, $\delta[\omega_{0}(r-R)] = \lim_{{\beta \to 0}}\delta_{\beta}[\omega_{0}(r-R)]\,, \forall \, 0 < |r - R| \ll 1$, we can convert our case study into a normal Dirac delta function. 

Using the scaling property of the delta function, $\delta[\omega_{0}(r-R)] = \frac{\delta(r-R)}{|\omega_{0}|}$ \cite{1968hmfw.book.....A, Arfken2005, Brewster2018, Waters2018}, we can also transform Equation~\ref{constant23} into the format, 

\begin{equation}
 \mathcal{Y}'(r) = \frac{\mathscr{C}\pi^{1/4}}{\sqrt{2}\omega_{0}} \int_{-\infty}^{+\infty} \mathcal{Y}(R)\delta(r-R)\, dR.
\label{constant24}
\end{equation}

Using the sifting property of the Dirac delta function for Equation~\ref{constant24} \cite{1968hmfw.book.....A, Arfken2005, Brewster2018, Waters2018} and assuming that $\mathcal{Y}'(r) = \mathcal{Y}(r)$, this becomes,

\begin{equation}
 \mathscr{C} = \frac{\sqrt{2}\omega_{0}}{\pi^{1/4}}.
\label{constant25}
\end{equation}
Equation~\ref{constant25} is the positive normalization constant that is applicable to our case, as seen in Equation~\ref{reconstruct2}.

\section{Detailed information about parameters used for density-wave analysis}

Tables~\ref{partab:C1} through~\ref{boundtabMAIN:C4} provide detailed information about the parameter settings used to analyze each of the density waves considered in this study. More specifically, Tables~\ref{obstab:C2} and~\ref{obstab2:C3} indicate which occultations were used for each wave, while Tables~\ref{partab:C1} and~\ref{boundtabMAIN:C4} provide the parameters used in the relevant wave-fitting routine in Python. Descriptions of the columns found in these tables are provided below.

\begin{itemize}
    \item \textbf{datfile}: This refers to the IDL save file containing the profiles relevant to our study. The data within this file form the basis of our analysis.
    
    \item \textbf{m}: The parameter `m' represents the signed azimuthal wavenumber of the wave. It provides crucial information about the spatial characteristics of the wave pattern.
    
    \item \textbf{omp}: Nominal Wave Pattern Speed (in degrees/day) - This parameter defines the expected speed at which the wave pattern propagates.
    
    \item \textbf{omr}: Range of Pattern Speeds (in degrees/day) - We consider a range of pattern speeds to comprehensively explore the behavior of the wave under different conditions.
    
    \item \textbf{xr}: Range of Radii for Analysis (in 1000 km) - This parameter defines the specific radial distances at which we analyze and plot wavelets, providing insights into the wave's distribution across distances.
    
    \item \textbf{rback}: Radius Range for Blank Region (in km) - Establishing a blank region around the wave aids in identifying and isolating profiles with optimal data quality.
    
    \item \textbf{rwave}: Radius Range Containing the Wave (in km) - This parameter specifies the radial range where the wave of interest is expected to be prominent.
    
    \item \textbf{frange}: Range of Spatial Frequencies - We compute wavelets across a defined range of spatial frequencies, offering a detailed view of the wave's spectral characteristics.
    
    \item \textbf{mthres}: Threshold Deviation for Optical Depth - This threshold guides the identification of acceptable profiles within the blank region defined by `rback'.
    
    \item \textbf{rthres}: Maximum Allowed Resolution - This parameter sets a limit on the resolution within the radial range of the wave (`rwave'), ensuring precision in our analysis.
    
    \item \textbf{rangedisp}: Range of Signals Shown in Power Wavelets - Defining the range of signals displayed in power wavelets helps in visualizing the strength and distribution of the wave's power.
    
    \item \textbf{badoccnames}: Names of Problematic Occultations - Identifying and removing problematic occultations, based on visual inspection, ensures the integrity of our dataset.
\end{itemize}

\begin{sidewaystable*}
\caption{Initial parameters required to execute each wave profile for Satellite and Planetary resonances.}
\label{partab:C1}
\hspace{-0.1in}{\resizebox{9.5in}{!}{\begin{tabular}{lccccccccccr}
\hline
Resonance Name & $datfile$ & $omr$ & $xr$ & $rback$ & $rwave$ & $wrange$ & $mthres$ & $rthres$ & $rangedisp$ \\
\hline
\hline
Mimas 4:1 & \text{bwaveinit2n\_shift\_generic\_74\_cxall\_e\_t\_e\_t\_041123.sav} & 0.1 & [74.855, 74.955] & [74896.00, 74898.00] & [74880.00, 74895.00] & [0.001, 15.000] & 0.1 & 1.0 & 0.010 \\
Pan 2:1 &\text{bwaveinit2n\_shift\_generic\_85\_cxall\_e\_t\_e\_t\_041123.sav} & 0.1 & [85.080, 85.200] & [85140.00, 85150.00] & [85090.00, 85130.00] & [0.001, 15.000] & 0.1 & 1.0 & 0.007 \\
Atlas 2:1 & \text{bwaveinit2n\_shift\_generic\_87\_cxall\_e\_t\_e\_t\_041123.sav} & 0.1 & [87.630, 87.680] & [87660.00, 87670.00] & [87640.00, 87650.00]  & [0.001, 15.000] & 0.1 & 1.0 & 0.010 \\
Prometheus 4:2 & \text{bwaveinit2n\_shift\_generic\_88\_cxall\_e\_t\_e\_t\_041123.sav} & 0.1 & [88.410, 88.470]   & [88450.00, 88460.00] & [88420.00, 88440.00] & [0.001, 15.000] & 0.1 & 1.0 & 0.010 \\
Mimas 6:2 & \text{bwaveinit2n\_shift\_generic\_89\_cxall\_e\_t\_e\_t\_041123.sav} & 0.1 & [89.820, 89.950]  & [89891.00, 89892.00] & [89870.00, 89890.00]  & [0.001, 15.000] & 0.1 & 1.0 & 0.010 \\
Pandora 4:2 & \text{bwaveinit2n\_shift\_generic\_89\_cxall\_e\_t\_e\_t\_041123.sav} & 0.1 & [89.850, 89.950] & [89898.00, 89899.00] & [89887.00, 89897.00] & [0.001, 15.000] & 0.1 & 1.0 & 0.010 \\
\hline
\hline
$W74.51^{av}$ & \text{bwaveinit2n\_shift\_generic\_74\_cxall\_e\_t\_e\_t\_041123.sav} & 0.1 & [74.450, 74.550] & [74530.00, 74540.00] & [74501.00, 74509.00] & [0.001, 15.000] & 0.1 & 1.0 & 0.010 \\
W74.74 & \text{bwaveinit2n\_shift\_generic\_74\_cxall\_e\_t\_e\_t\_041123.sav} & 0.1 & [74.710, 74.770] & [74744.00, 74746.00] & [74736.00, 74743.00] & [0.001, 15.000] & 0.1 & 1.0 & 0.010 \\
W74.75 & \text{bwaveinit2n\_shift\_generic\_74\_cxall\_e\_t\_e\_t\_041123.sav} & 0.1 & [74.710, 74.770] & [74749.00, 74750.00] & [74746.00, 74748.00] & [0.001, 15.000] & 0.1 & 1.0 & 0.010 \\
W74.76 & \text{bwaveinit2n\_shift\_generic\_74\_cxall\_e\_t\_e\_t\_041123.sav} & 0.1 & [74.720, 74.780] & [74760.00, 74761.00] & [74752.00, 74762.00] & [0.001, 15.000] & 0.1 & 1.0 & 0.010 \\
W75.14 & \text{bwaveinit2n\_shift\_generic\_75\_cxall\_e\_t\_e\_t\_041123.sav} & 0.1 & [75.100, 75.180] & [75145.00, 75146.00] & [75142.00, 75144.00] & [0.001, 15.000] & 0.1 & 1.0 & 0.010 \\
W76.02A & \text{bwaveinit2n\_shift\_generic\_76\_cxall\_e\_t\_e\_t\_041123.sav} & 0.1 & [76.000, 76.040] & [76025.00, 76026.00] & [76016.00, 76018.00] & [0.001, 15.000] & 0.1 & 1.0 & 0.010 \\
W76.44 & \text{bwaveinit2n\_shift\_generic\_76\_cxall\_e\_t\_e\_t\_041123.sav} & 0.1 & [76.400, 76.465] & [76437.00, 76438.00] & [76433.00, 76436.00] & [0.001, 15.000] & 0.1 & 1.0 & 0.010 \\
$W76.46^{av}$ & \text{bwaveinit2n\_shift\_generic\_76\_cxall\_e\_t\_e\_t\_041123.sav} & 0.1 & [76.440, 76.480] & [76463.00, 76464.00] & [76457.00, 76462.00] & [0.001, 15.000] & 0.1 & 1.0 & 0.010 \\
W77.34 & \text{bwaveinit2n\_shift\_generic\_77\_cxall\_e\_t\_e\_t\_041123.sav} & 0.1 & [77.300, 77.360] & [77340.00, 77341.00] & [77337.00, 77339.00] & [0.001, 15.000] & 0.1 & 1.0 & 0.010 \\
W78.51 & \text{bwaveinit2n\_shift\_generic\_78\_cxall\_e\_t\_e\_t\_041123.sav} & 0.1 & [78.450, 78.550] & [78510.00, 78511.00] & [78503.00, 78509.00] & [0.001, 15.000] & 0.1 & 1.0 & 0.010 \\
W79.04 & \text{bwaveinit2n\_shift\_generic\_79\_cxall\_e\_t\_e\_t\_041123.sav} & 0.1 & [79.010, 79.070] & [79046.00, 79047.00] & [79039.00, 79045.00] & [0.001, 15.000] & 0.1 & 1.0 & 0.010 \\
W79.55 & \text{bwaveinit2n\_shift\_generic\_79\_cxall\_e\_t\_e\_t\_041123.sav} & 0.1 & [79.520, 79.580] & [79551.00, 79552.00] & [79546.00, 79550.00] & [0.001, 15.000] & 0.1 & 1.0 & 0.010 \\
W80.49 & \text{bwaveinit2n\_shift\_generic\_80\_cxall\_e\_t\_e\_t\_041123.sav} & 0.1 & [80.450, 80.510] & [80489.00, 80490.00] & [80484.00, 80488.00] & [0.001, 15.000] & 0.1 & 1.0 & 0.010 \\
W80.99 & \text{bwaveinit2n\_shift\_generic\_80\_cxall\_e\_t\_e\_t\_041123.sav} & 0.1 & [80.950, 81.000] & [80998.00, 80999.00] & [80983.00, 80989.00] & [0.001, 15.000] & 0.1 & 1.0 & 0.010 \\
W81.023A & \text{bwaveinit2n\_shift\_generic\_81\_cxall\_e\_t\_e\_t\_041123.sav} & 0.1 & [81.000, 81.060] & [81090.00, 81095.00] & [81018.00, 81030.00] & [0.001, 15.000] & 0.1 & 1.0 & 0.010 \\
W81.024B & \text{bwaveinit2n\_shift\_generic\_81\_cxall\_e\_t\_e\_t\_041123.sav} & 0.1 & [81.000, 81.060] &  [81090.00, 81095.00] & [81018.00, 81030.00] & [0.001, 15.000] & 0.1 & 1.0 & 0.010 \\
W81.33 & \text{bwaveinit2n\_shift\_generic\_81\_cxall\_e\_t\_e\_t\_041123.sav} & 0.1 & [81.300, 81.370] & [81337.00, 81338.00] & [81333.00, 81336.00] & [0.001, 15.000] & 0.1 & 1.0 & 0.010 \\
W81.43 & \text{bwaveinit2n\_shift\_generic\_81\_cxall\_e\_t\_e\_t\_041123.sav} & 0.1
& [81.400, 81.500] & [81800.00, 81900.00] & [81420.00, 81430.00] & [0.001, 15.000] & 0.1 & 1.0 & 0.010 \\
W81.96 & \text{bwaveinit2n\_shift\_generic\_81\_cxall\_e\_t\_e\_t\_041123.sav} & 0.1 & [81.930, 81.990] & [81966.00, 81968.00] & [81959.00, 81965.00] & [0.001, 15.000] & 0.1 & 1.0 & 0.010 \\
W82.01 & \text{bwaveinit2n\_shift\_generic\_80-85\_cxall\_e\_080122.sav} & 0.1 & [81.950, 82.050] & [82020.00, 82030.00] & [82004.00, 82009.00] & [0.001, 15.000] & 0.1 & 1.0 & 0.010 \\
W82.06 & \text{bwaveinit2n\_shift\_generic\_82\_cxall\_e\_t\_e\_t\_041123.sav} & 0.1 & [82.005, 82.105] & [82070.00, 82080.00] & [82055.00, 82065.00] & [0.001, 20.000] & 0.1 & 1.0 & 0.010 \\
W82.21 & \text{bwaveinit2n\_shift\_generic\_82\_cxall\_e\_t\_e\_t\_041123.sav} & 0.1 & [82.150, 82.250] & [82400.00, 82450.00] & [82187.50, 82207.51] & [0.001, 15.000] & 0.1 & 1.0 & 0.010 \\
W82.53 & \text{bwaveinit2n\_shift\_generic\_82\_cxall\_e\_t\_e\_t\_041123.sav} & 0.1 & [82.490, 82.550] & [82531.00, 82532.00] & [82510.00, 82530.00] & [0.001, 15.000] & 0.1 & 1.0 & 0.010 \\
W82.61 & \text{bwaveinit2n\_shift\_generic\_82\_cxall\_e\_t\_e\_t\_041123.sav} & 0.1 & [82.570, 82.640] & [82610.00, 82611.00] & [82606.00, 82608.00] & [0.001, 15.000] & 0.1 & 1.0 & 0.010 \\
W83.09 & \text{bwaveinit2n\_shift\_generic\_83\_cxall\_e\_t\_e\_t\_041123.sav} & 0.1 & [83.060, 83.120] & [83097.00, 83098.00] & [83086.00, 83096.00] & [0.001, 15.000] & 0.1 & 1.0 & 0.010 \\
W83.63 & \text{bwaveinit2n\_shift\_generic\_83\_cxall\_e\_t\_e\_t\_041123.sav} & 0.1 & [83.580, 83.680] & [83640.00, 83650.00] & [83612.02, 83637.02] & [0.001, 15.000] & 0.1 & 1.0 & 0.010 \\
W84.15 & \text{bwaveinit2n\_shift\_generic\_84\_cxall\_e\_t\_e\_t\_041123.sav} & 0.1 & [84.120, 84.170] & [84180.00, 84190.00] & [84140.00, 84150.00] & [0.001, 15.000] & 0.1 & 1.0 & 0.010 \\
W84.64 & \text{bwaveinit2n\_shift\_generic\_84\_cxall\_e\_t\_e\_t\_041123.sav} & 0.1 & [84.590, 84.690] & [84660.00, 84670.00] & [84630.00, 84650.00] & [0.001, 15.000] & 0.1 & 1.0 & 0.010 \\
W87.19 & \text{bwaveinit2n\_shift\_generic\_87\_cxall\_e\_t\_e\_t\_041123.sav} & 0.1 & [87.160, 87.225] & [87220.00, 87223.00] & [87170.00, 87210.00] & [0.001, 15.000] & 0.1 & 1.0 & 0.010 \\
\hline
\end{tabular}}}
$datfile$ = IDL file of the profiles (in $.sav$ extension). $omr$ = Range of pattern speeds to consider in degrees/dat. $xr$ = Range of radii to plot and compute wavelets (in 1000 km). $rback$ = Radius range of blank region for identifying good profiles (in km). $rwave$ = Radius range  containing the wave (in km). $wrange$ = Range of spatial wavelengths to compute in the wavelet. $mthres$ = Threshold deviation for normal optical depth in range rback. $rthres$ = Maximum allowed resolution in range $rwave$. $rangedisp$ = range of signals shown in power wavelets.
\end{sidewaystable*}

\begin{sidewaystable*}
\caption{List of 77 occultations analyzed in this study.}
\label{obstab:C2}
\hspace{-0.1in}{\resizebox{9.1in}{!}{\begin{tabular}{lcccccccccccccccccccccccccr} 
\hline
Star & Rev & $^a$ & Date & B($^{o}$)$^b$ & Mimas 4:1 & Pan 2:1 & Atlas 2:1 & Prometheus 4:2 & Mimas 6:2 & Pandora 4:2 & $W74.51^{av}$ & W74.74 & W74.75 & W74.76 & W75.14 & W76.02A & W76.44 & $W76.46^{av}$ & W77.34 & W78.51 & W79.04 & W79.55 & W80.49 & W80.99 \\ 
\hline
RHya & 036 & i & 2007-001 & -29.40 & \checkmark & \checkmark & \checkmark & \checkmark & \checkmark & \checkmark & \checkmark & \checkmark & \checkmark & \checkmark & \checkmark & \checkmark & \checkmark & \checkmark & \checkmark & \checkmark & \checkmark & \checkmark & \checkmark & \checkmark \\ 

$\alpha$Aur & 041 & i & 2007-082 & +50.88 & \checkmark & \checkmark & \checkmark & \checkmark & \checkmark & \checkmark & \checkmark & X & X & X & X & \checkmark & \checkmark & \checkmark & \checkmark & X & \checkmark & X & \checkmark & \checkmark \\ 

RCas & 065 & i & 2008-112 & +56.04 & N/A & N/A & N/A & N/A & N/A & N/A & N/A & \checkmark & \checkmark & N/A & \checkmark & N/A & \checkmark & \checkmark & \checkmark & \checkmark & \checkmark & N/A & \checkmark & \checkmark \\ 

$\gamma$Cru & 073 & i & 2008-174 & -62.35 & \checkmark & \checkmark & \checkmark & \checkmark & X & X & \checkmark & \checkmark & \checkmark & \checkmark & \checkmark & \checkmark & \checkmark & \checkmark & \checkmark & \checkmark & \checkmark & \checkmark & \checkmark & \checkmark \\ 

$\gamma$Cru & 078 & i & 2008-209 & -62.35 & \checkmark & \checkmark & \checkmark & \checkmark & \checkmark & \checkmark & \checkmark & \checkmark & \checkmark & \checkmark & \checkmark & \checkmark & \checkmark & \checkmark & \checkmark & \checkmark & \checkmark & \checkmark & \checkmark & \checkmark \\ 

$\beta$Gru & 078 & i & 2008-210 & -43.38 & N/A & N/A & \checkmark & \checkmark & \checkmark & \checkmark & N/A & N/A & N/A & N/A & N/A & N/A & N/A & N/A & N/A & N/A & N/A & N/A & N/A & N/A \\ 

$\gamma$Cru & 079 & i & 2008-216 & -62.35 & \checkmark & \checkmark & \checkmark & \checkmark & X & X & \checkmark & \checkmark & \checkmark & \checkmark & \checkmark & \checkmark & \checkmark & \checkmark & \checkmark & \checkmark & \checkmark & \checkmark & \checkmark & \checkmark \\ 

RSCnc & 080 & i & 2008-226 & +29.96 & N/A & \checkmark & X & \checkmark & \checkmark & \checkmark & N/A & N/A & N/A & N/A & N/A & N/A & N/A & N/A & N/A & \checkmark & \checkmark & \checkmark & \checkmark & \checkmark \\ 

RSCnc & 080 & e& 2008-226 & +29.96 & N/A & \checkmark & \checkmark & \checkmark & \checkmark & \checkmark & N/A & N/A & N/A & N/A & N/A & N/A & N/A & N/A & N/A & \checkmark & \checkmark & \checkmark & \checkmark & \checkmark \\ 

$\gamma$Cru & 081 & i & 2008-231 & -62.35 & \checkmark & \checkmark & \checkmark & \checkmark & \checkmark & N/A & \checkmark & \checkmark & \checkmark & \checkmark & \checkmark & \checkmark & \checkmark & \checkmark & N/A & \checkmark & \checkmark & \checkmark & \checkmark & \checkmark \\ 

$\gamma$Cru & 082 & i & 2008-238 & -62.35 & \checkmark & \checkmark & \checkmark & \checkmark & \checkmark & \checkmark & \checkmark & \checkmark & \checkmark & \checkmark & \checkmark & \checkmark & \checkmark & \checkmark & \checkmark & \checkmark & \checkmark & \checkmark & \checkmark & \checkmark \\ 

RSCnc  & 085 & i & 2008-263 & +29.96 & N/A & \checkmark & \checkmark & \checkmark & \checkmark & \checkmark & N/A & N/A & N/A & N/A & N/A & N/A & N/A & N/A & N/A & N/A & N/A & N/A & N/A & N/A \\ 

RSCnc  & 085 & e& 2008-263 & +29.96 & N/A & X & \checkmark & \checkmark & \checkmark & \checkmark & N/A & N/A & N/A & N/A & N/A & N/A & N/A & N/A & N/A & N/A & N/A & N/A & N/A & N/A \\ 

$\gamma$Cru & 086 & i & 2008-268 & -62.35 & N/A & \checkmark & \checkmark & \checkmark & \checkmark & \checkmark & N/A & N/A & N/A & N/A & \checkmark & \checkmark & \checkmark & \checkmark & \checkmark & \checkmark & \checkmark & \checkmark & \checkmark & \checkmark \\ 

RSCnc  & 087 & i & 2008-277 & +29.96 & N/A & \checkmark & \checkmark & \checkmark & \checkmark & \checkmark & N/A & N/A & N/A & N/A & N/A & N/A & N/A & N/A & N/A & N/A & N/A & N/A & N/A & N/A \\ 

RSCnc  & 087 & e& 2008-277 & +29.96 & N/A & \checkmark & \checkmark & \checkmark & \checkmark & \checkmark & N/A & N/A & N/A & N/A & N/A & N/A & N/A & N/A & N/A & N/A & N/A & N/A & N/A & N/A \\ 

$\gamma$Cru & 089 & i & 2008-290 & -62.35 & \checkmark & \checkmark & \checkmark & \checkmark & \checkmark & \checkmark & \checkmark & X & \checkmark & \checkmark & \checkmark & \checkmark & \checkmark & \checkmark & \checkmark & \checkmark & \checkmark & \checkmark & \checkmark & \checkmark \\ 

$\gamma$Cru & 093 & i & 2009-320 & -62.35 & \checkmark & \checkmark & \checkmark & \checkmark & \checkmark & \checkmark & \checkmark & \checkmark & \checkmark & \checkmark & \checkmark & \checkmark & \checkmark & \checkmark & \checkmark & \checkmark & \checkmark & \checkmark & \checkmark & \checkmark \\ 

$\gamma$Cru & 094 & i & 2008-328 & -62.35 & \checkmark & \checkmark & X & \checkmark & \checkmark & \checkmark & X & \checkmark & \checkmark & \checkmark & \checkmark & \checkmark & \checkmark & \checkmark & \checkmark & N/A & \checkmark & \checkmark & \checkmark & \checkmark \\ 

$\gamma$Cru & 096 & i & 2008-343 & -62.35 & \checkmark & N/A & N/A & N/A & \checkmark & \checkmark & N/A & \checkmark & \checkmark & \checkmark & \checkmark & \checkmark & N/A & N/A & N/A & N/A & N/A & N/A & N/A & N/A \\

$\gamma$Cru & 100 & i & 2009-012 & -62.35 & \checkmark & \checkmark & \checkmark & \checkmark & \checkmark & \checkmark & \checkmark & \checkmark & \checkmark & \checkmark & \checkmark & \checkmark & X & X & \checkmark & \checkmark & \checkmark & \checkmark & \checkmark & \checkmark \\ 

$\gamma$Cru & 102 & i & 2009-031 & -62.35 & \checkmark & \checkmark & \checkmark & \checkmark & \checkmark & \checkmark & \checkmark & \checkmark & \checkmark & \checkmark & \checkmark & \checkmark & \checkmark & \checkmark & \checkmark & \checkmark & \checkmark & \checkmark & \checkmark & \checkmark \\ 

$\beta$Peg & 104 & i & 2009-057 & +31.68 & N/A & N/A & \checkmark & N/A & \checkmark & \checkmark & N/A & \checkmark & \checkmark & \checkmark & N/A & \checkmark & N/A & N/A & N/A & N/A & \checkmark & \checkmark & N/A & \checkmark \\ 

RCas & 106 & i & 2009-082 & +56.04 & N/A & X & \checkmark & X & \checkmark & \checkmark & N/A & N/A & N/A & N/A & N/A & N/A & N/A & N/A & N/A & N/A & N/A & N/A & N/A & N/A \\ 

$\beta$Peg & 108 & i & 2009-095 & +31.68 & N/A & X & \checkmark & \checkmark & \checkmark & \checkmark & N/A & N/A & N/A & N/A & N/A & N/A & N/A & N/A & N/A & N/A & N/A & N/A & N/A & N/A \\ 

$\alpha$Sco & 115 & i & 2009-209 & -32.16 & N/A & \checkmark & \checkmark & \checkmark & \checkmark & \checkmark & N/A & N/A & N/A & N/A & N/A & N/A & N/A & N/A & N/A & N/A & X & \checkmark & \checkmark & \checkmark \\ 

$\beta$Peg & 170 & e & 2012-224 & +31.68 & N/A & \checkmark & \checkmark & \checkmark & X & \checkmark & N/A & N/A & N/A & N/A & N/A & N/A & N/A & N/A & N/A & \checkmark & \checkmark & \checkmark & \checkmark & \checkmark \\ 

$\beta$Peg & 172 & i & 2012-266 & +31.68 & \checkmark & X & \checkmark & \checkmark & \checkmark & \checkmark & \checkmark & \checkmark & \checkmark & \checkmark & \checkmark & \checkmark & \checkmark & \checkmark & \checkmark & \checkmark & \checkmark & \checkmark & \checkmark & \checkmark \\ 

$\lambda$Vel & 173 & i & 2012-292 & -43.81 & N/A & N/A & N/A & N/A & \checkmark & N/A & N/A & N/A & N/A & N/A & N/A & N/A & N/A & N/A & X & \checkmark & \checkmark & \checkmark & \checkmark & \checkmark \\ 

WHya & 179 & i & 2013-019 & -34.64 & \checkmark & \checkmark & \checkmark & \checkmark & \checkmark & \checkmark & \checkmark & \checkmark & \checkmark & \checkmark & \checkmark & \checkmark & \checkmark & \checkmark & \checkmark & \checkmark & \checkmark & \checkmark & \checkmark & \checkmark \\ 

WHya & 180 & i & 2013-033 & -34.64 & \checkmark & \checkmark & \checkmark & \checkmark & \checkmark & \checkmark & \checkmark & \checkmark & \checkmark & \checkmark & \checkmark & \checkmark & \checkmark & \checkmark & \checkmark & \checkmark & \checkmark & \checkmark & \checkmark & \checkmark \\ 

WHya & 181 & i & 2013-049 & -34.64 & \checkmark & \checkmark & \checkmark & \checkmark & \checkmark & \checkmark & \checkmark & \checkmark & \checkmark & \checkmark & \checkmark & \checkmark & \checkmark & \checkmark & \checkmark & \checkmark & \checkmark & \checkmark & \checkmark & \checkmark \\ 

RCas & 185 & i & 2013-091 & +56.04 & \checkmark & \checkmark & \checkmark & \checkmark & \checkmark & \checkmark & \checkmark & \checkmark & \checkmark & \checkmark & \checkmark & \checkmark & \checkmark & \checkmark & \checkmark & \checkmark & X & \checkmark & X & \checkmark \\ 

$\mu$Cep & 185 & e & 2013-090 & +59.90 & \checkmark & \checkmark & \checkmark & \checkmark & \checkmark & \checkmark & \checkmark & \checkmark & \checkmark & \checkmark & X & \checkmark & \checkmark & \checkmark & \checkmark & \checkmark & \checkmark & \checkmark & X & \checkmark \\ 

WHya & 186 & e& 2013-103 & -34.64 & N/A & \checkmark & \checkmark & \checkmark & \checkmark & \checkmark & N/A & N/A & N/A & N/A & N/A & N/A & N/A & N/A & N/A & N/A & N/A & N/A & \checkmark & \checkmark \\ 

$\gamma$Cru & 187 & i & 2013-112 & -62.35 & \checkmark & \checkmark & \checkmark & \checkmark & \checkmark & \checkmark & \checkmark & \checkmark & \checkmark & \checkmark & \checkmark & \checkmark & \checkmark & \checkmark & \checkmark & \checkmark & \checkmark & \checkmark & \checkmark & \checkmark \\ 

$\gamma$Cru & 187 & e& 2013-112 & -62.35 & \checkmark & \checkmark & \checkmark & \checkmark & \checkmark & \checkmark & \checkmark & \checkmark & \checkmark & \checkmark & \checkmark & \checkmark & \checkmark & \checkmark & \checkmark & X & \checkmark & \checkmark & \checkmark & \checkmark \\ 

WHya & 189 & e& 2013-132 & -34.64 & N/A & \checkmark & \checkmark & \checkmark & \checkmark & \checkmark & N/A & N/A & N/A & N/A & N/A & N/A & N/A & N/A & N/A & N/A & N/A & N/A & \checkmark & \checkmark \\ 

RCar & 191 & i &  2013-152 & -63.48 & \checkmark & \checkmark & \checkmark & \checkmark & X & X & \checkmark & \checkmark & \checkmark & \checkmark & \checkmark & \checkmark & \checkmark & \checkmark & \checkmark & X & \checkmark & \checkmark & \checkmark & \checkmark \\ 

RCas & 191 & i &  2013-149 & +56.04 & \checkmark & X & \checkmark & \checkmark & \checkmark & \checkmark & X & X & \checkmark & \checkmark & \checkmark & \checkmark & \checkmark & \checkmark & \checkmark & \checkmark & X & \checkmark & \checkmark & \checkmark \\ 

$\mu$Cep & 191 & i & 2013-148 & +59.90 & \checkmark & X & X & \checkmark & \checkmark & \checkmark & \checkmark & \checkmark & \checkmark & \checkmark & X & \checkmark & \checkmark & X & X & \checkmark & \checkmark & \checkmark & \checkmark & \checkmark \\ 

$\mu$Cep & 193 & i & 2013-172 & +59.90 & \checkmark & \checkmark & \checkmark & \checkmark & \checkmark & \checkmark & \checkmark & \checkmark & \checkmark & \checkmark & \checkmark & \checkmark & \checkmark & X & \checkmark & \checkmark & \checkmark & \checkmark & \checkmark & \checkmark \\ 

RCas & 194 & e& 2013-186 & +56.04 & X & \checkmark & \checkmark & \checkmark & \checkmark & \checkmark & X & \checkmark & \checkmark & \checkmark & \checkmark & \checkmark & \checkmark & \checkmark & \checkmark & \checkmark & \checkmark & \checkmark & \checkmark & \checkmark \\ 

2Cen & 194 & i & 2013-189 & -40.73 & \checkmark & N/A & \checkmark & \checkmark & N/A & \checkmark & \checkmark & \checkmark & \checkmark & \checkmark & \checkmark & \checkmark & \checkmark & \checkmark & \checkmark & \checkmark & \checkmark & \checkmark & \checkmark & \checkmark \\ 

2Cen & 194 & e& 2013-189 & -40.73 & N/A & N/A & \checkmark & N/A & N/A & N/A & N/A & N/A & N/A & N/A & N/A & N/A & N/A & N/A & X & \checkmark & \checkmark & \checkmark & \checkmark & \checkmark \\ 

RLyr & 198 & i & 2013-289 & +40.77 & \checkmark & \checkmark & N/A & \checkmark & \checkmark & \checkmark & \checkmark & \checkmark & \checkmark & \checkmark & \checkmark & \checkmark & X & X & \checkmark & \checkmark & \checkmark & \checkmark & \checkmark & \checkmark \\ 

L2Pup & 199 & e & 2013-327 & -41.91 & N/A & N/A & \checkmark & \checkmark & X & X & N/A & N/A & N/A & N/A & N/A & N/A & N/A & N/A & N/A & N/A & N/A & N/A & N/A & N/A \\ 

RLyr & 199 & i & 2013-337 & +40.77 & \checkmark & \checkmark & \checkmark & \checkmark & \checkmark & \checkmark & X & X & X & X & \checkmark & \checkmark & \checkmark & \checkmark & \checkmark & \checkmark & \checkmark & \checkmark & \checkmark & \checkmark \\ 

RLyr & 200 & i & 2014-003 & +40.77 & \checkmark & \checkmark & X & \checkmark & \checkmark & \checkmark & \checkmark & \checkmark & \checkmark & \checkmark & \checkmark & \checkmark & \checkmark & \checkmark & \checkmark & \checkmark & \checkmark & \checkmark & \checkmark & \checkmark \\ 

L2Pup & 201 & i & 2014-051 & -41.91 & X & X & X & X & X & X & X & X & X & X & X & X & X & X & X & X & X & X & X & X \\ 

RLyr & 202 & e & 2014-067 & +40.77 & X & \checkmark & \checkmark & \checkmark & \checkmark & \checkmark & \checkmark & \checkmark & \checkmark & \checkmark & X & \checkmark & \checkmark & \checkmark & \checkmark & \checkmark & \checkmark & \checkmark & \checkmark & \checkmark \\ 

RLyr & 202 & i & 2014-067 & +40.77 & N/A & N/A & \checkmark & \checkmark & \checkmark & \checkmark & N/A & N/A & N/A & N/A & N/A & N/A & N/A & N/A& N/A & N/A & N/A & N/A & N/A & N/A \\ 

L2Pup & 205 & e & 2014-175 & -41.91 & N/A & \checkmark & \checkmark & \checkmark & \checkmark & \checkmark & N/A & N/A & N/A & N/A & \checkmark & \checkmark & \checkmark & \checkmark & \checkmark & \checkmark & X & \checkmark & \checkmark & X \\ 

L2Pup & 205 & i & 2014-174 & -41.91 & N/A & N/A & \checkmark & \checkmark & \checkmark & \checkmark & N/A & N/A & N/A & N/A & N/A & N/A & N/A & N/A & N/A & N/A & N/A & N/A & N/A & N/A \\ 

L2Pup & 206 & e& 2014-206 & -41.91 & N/A & X & \checkmark & \checkmark & \checkmark & \checkmark & N/A & N/A & N/A & N/A & X & \checkmark & \checkmark & \checkmark & \checkmark & \checkmark & \checkmark & \checkmark & \checkmark & \checkmark \\ 

RLyr & 208 & e& 2014-262 & +40.77 & \checkmark & \checkmark & \checkmark & \checkmark & \checkmark & \checkmark & \checkmark & \checkmark & \checkmark & \checkmark & X & \checkmark & \checkmark & \checkmark & \checkmark & \checkmark & \checkmark & \checkmark & \checkmark & \checkmark \\ 

WHya & 236 & i & 2016-148 & -34.64 & N/A & N/A & N/A & N/A & \checkmark & \checkmark & N/A & N/A & N/A & N/A & N/A & N/A & N/A & N/A & N/A & N/A & N/A & N/A & N/A & N/A \\ 

$\rho$Per & 239 & e & 2016-221 & +45.27 & N/A & \checkmark & \checkmark & \checkmark & \checkmark & \checkmark & N/A & N/A & N/A & N/A & N/A & N/A & N/A & N/A & N/A & N/A & N/A & N/A & N/A & N/A \\ 

$\alpha$Sco & 241 & e& 2016-243 & -32.16 & \checkmark & \checkmark & \checkmark & \checkmark & \checkmark & \checkmark & \checkmark & \checkmark & \checkmark & \checkmark & \checkmark & \checkmark & \checkmark & \checkmark & \checkmark & \checkmark & \checkmark & \checkmark & \checkmark & \checkmark \\ 

$\alpha$Sco & 243 & e& 2016-267 & -32.16 & \checkmark & \checkmark & \checkmark & \checkmark & \checkmark & \checkmark & \checkmark & \checkmark & \checkmark & \checkmark & \checkmark & \checkmark & \checkmark & \checkmark & \checkmark & \checkmark & \checkmark & \checkmark & \checkmark & \checkmark \\ 

RCas & 243 & i & 2016-268 & +56.04 & N/A & \checkmark & \checkmark & \checkmark & X & X & X & X & X & X & \checkmark & X & \checkmark & \checkmark & \checkmark & \checkmark & \checkmark & \checkmark & \checkmark & \checkmark \\ 

$\alpha$Sco & 245 & e& 2016-287 & -32.16 & \checkmark & \checkmark & \checkmark & \checkmark & \checkmark & \checkmark & \checkmark & \checkmark & \checkmark & \checkmark & \checkmark & \checkmark & \checkmark & \checkmark & \checkmark & \checkmark & \checkmark & \checkmark & \checkmark & \checkmark \\ 

$\alpha$Sco & 245 & i & 2016-287 & -32.16 & \checkmark & N/A & N/A & X & \checkmark & \checkmark & \checkmark & \checkmark & \checkmark & \checkmark & \checkmark & N/A & N/A & N/A & N/A & N/A & N/A & N/A & N/A & N/A \\

$\gamma$Cru & 245 & e& 2016-286 & -62.35 & N/A & \checkmark & \checkmark & \checkmark & \checkmark & \checkmark & N/A & N/A & N/A & N/A & N/A & N/A & N/A & N/A & N/A & N/A & \checkmark & \checkmark & \checkmark & \checkmark \\ 

$\gamma$Cru & 255 & i & 2017-001 & -62.35 & \checkmark & \checkmark & \checkmark & \checkmark & \checkmark & \checkmark & \checkmark & \checkmark & \checkmark & \checkmark & \checkmark & \checkmark & X & X & \checkmark & \checkmark & \checkmark & X & \checkmark & \checkmark \\ 

$\gamma$Cru & 264 & i & 2017-086 & -62.35 & N/A  & \checkmark & \checkmark & \checkmark & \checkmark & \checkmark & \checkmark & \checkmark & \checkmark & \checkmark & \checkmark & \checkmark & \checkmark & \checkmark & \checkmark & \checkmark & \checkmark & X & \checkmark & \checkmark \\

$\lambda$Vel & 268 & i & 2017-094 & -43.81 & \checkmark & N/A & \checkmark & N/A & N/A & \checkmark & \checkmark & \checkmark & N/A & \checkmark & \checkmark & X & \checkmark & \checkmark & \checkmark & \checkmark & \checkmark & \checkmark & N/A & \checkmark \\ 

$\gamma$Cru & 268 & i & 2017-095 & -62.35 & \checkmark & \checkmark & \checkmark & \checkmark & \checkmark & \checkmark & \checkmark & \checkmark & \checkmark & \checkmark & \checkmark & \checkmark & X & \checkmark & \checkmark & \checkmark & \checkmark & \checkmark & \checkmark & X \\ 

$\alpha$Ori & 268 & e& 2017-096 & +11.68 & \checkmark & \checkmark & \checkmark & \checkmark & \checkmark & \checkmark & \checkmark & \checkmark & \checkmark & \checkmark & \checkmark & \checkmark & \checkmark & \checkmark & \checkmark & \checkmark & \checkmark & \checkmark & \checkmark & \checkmark \\ 

VYCMa & 269 & i & 2017-100 & -23.43 & \checkmark & \checkmark & \checkmark & X & \checkmark & \checkmark & \checkmark & \checkmark & \checkmark & \checkmark & \checkmark & \checkmark & \checkmark & \checkmark & \checkmark & \checkmark & \checkmark & \checkmark & \checkmark & \checkmark \\ 

$\gamma$Cru & 269 & i & 2017-102 & -62.35 & \checkmark & \checkmark & \checkmark & \checkmark & \checkmark & \checkmark & \checkmark & \checkmark & \checkmark & \checkmark & \checkmark & \checkmark & \checkmark & \checkmark & \checkmark & \checkmark & \checkmark & \checkmark & \checkmark & \checkmark \\ 

$\alpha$Ori & 269 & e& 2017-104 & +11.68 & \checkmark & \checkmark & \checkmark & \checkmark & \checkmark & \checkmark & \checkmark & \checkmark & \checkmark & \checkmark & \checkmark & \checkmark & \checkmark & \checkmark & \checkmark & \checkmark & \checkmark & \checkmark & \checkmark & \checkmark \\ 

$\gamma$Cru & 276 & i & 2017-148 & -62.35 & N/A & N/A & N/A & N/A & N/A & \checkmark & \checkmark & N/A & \checkmark & N/A & \checkmark & N/A & \checkmark & \checkmark & \checkmark & X & \checkmark & N/A & N/A & \checkmark \\ 

$\alpha$Ori & 277 & i & 2017-155 & +11.68 & \checkmark & \checkmark & \checkmark & \checkmark & \checkmark & \checkmark & \checkmark & \checkmark & \checkmark & \checkmark & \checkmark & \checkmark & \checkmark & \checkmark & \checkmark & \checkmark & \checkmark & \checkmark & \checkmark & \checkmark \\

$\gamma$Cru & 282 & i & 2017-187 & -62.35 & N/A & N/A & N/A & N/A & N/A & N/A & \checkmark & \checkmark & \checkmark & \checkmark & \checkmark & \checkmark & \checkmark & N/A & \checkmark & N/A & \checkmark & N/A & \checkmark & N/A \\ 

$\gamma$Cru & 291 & i & 2017-245 & -62.35 & N/A & N/A & \checkmark & N/A & N/A & X & N/A & X & X & N/A & N/A & \checkmark & \checkmark & \checkmark & N/A & \checkmark & \checkmark & N/A & \checkmark & \checkmark \\ 

$\gamma$Cru & 292 & i & 2017-251 & -62.35 & \checkmark & N/A & \checkmark & N/A & N/A & \checkmark & N/A & N/A & \checkmark & N/A & \checkmark & \checkmark & \checkmark & N/A & N/A & N/A & \checkmark & \checkmark & \checkmark & \checkmark  \\ 
\hline
\end{tabular}}}

$^a$ i=ingress occultation, e= egress occultation. $^b$ Ring opening angle to star (positive indicates star is north of Saturn's equatorial plane). $^c$ Span of inertial longitudes, measured relative to the ascending node of the ring particles on the J2000 coordinate system. (\textbf{Note:} \checkmark = Occultation used during wave analysis. X = Occultation excluded during wave analysis. N/A = Occultation not applicable to wave analysis.) 
\end{sidewaystable*}

\begin{sidewaystable*}
\caption{List of 77 occultations analyzed in this study (continued).}
\label{obstab2:C3}
\hspace{-0.1in}{\resizebox{7.0in}{!}{\begin{tabular}{lccccccccccccccccccccr} 
\hline 
Star & Rev & $^a$ & Date & B($^{o}$)$^b$ & W81.023A & W81.024B & W81.33 & W81.43 & W81.96 & W82.01 & W82.06 & W82.21 & W82.53 & W82.61 & W83.09 & W83.63 & W84.15 & W84.64 & W87.19 \\ 
\hline
RHya & 036 & i & 2007-001 & -29.40 & \checkmark & \checkmark & \checkmark & X & \checkmark & \checkmark & \checkmark & \checkmark & \checkmark & \checkmark & \checkmark & \checkmark & \checkmark & \checkmark & \checkmark \\ 

$\alpha$Aur & 041 & i & 2007-082 & +50.88 & \checkmark & \checkmark & \checkmark & \checkmark & \checkmark & \checkmark & \checkmark & X & \checkmark & \checkmark & \checkmark & \checkmark & X & \checkmark & \checkmark \\ 

RCas & 065 & i & 2008-112 & +56.04 & N/A & N/A & \checkmark & N/A & \checkmark & \checkmark & \checkmark & N/A & N/A & \checkmark & N/A & N/A & N/A & N/A & N/A \\  

$\gamma$Cru & 073 & i & 2008-174 & -62.35 & \checkmark & \checkmark & \checkmark & \checkmark & \checkmark & \checkmark & \checkmark & \checkmark & \checkmark & \checkmark & \checkmark & \checkmark & \checkmark & \checkmark & \checkmark \\ 

$\gamma$Cru & 078 & i & 2008-209 & -62.35 & N/A & \checkmark & \checkmark & N/A & \checkmark & \checkmark & \checkmark & \checkmark & \checkmark & \checkmark & \checkmark & \checkmark & \checkmark & \checkmark & \checkmark \\ 

$\beta$Gru & 078 & i & 2008-210 & -43.38 & \checkmark & N/A & N/A & \checkmark & N/A & N/A & N/A & N/A & N/A & N/A & N/A & N/A & N/A & N/A & \checkmark \\ 

$\gamma$Cru & 079 & i & 2008-216 & -62.35 & \checkmark & \checkmark & \checkmark & \checkmark & \checkmark & \checkmark & \checkmark & \checkmark & \checkmark & \checkmark & \checkmark & \checkmark & \checkmark & \checkmark & \checkmark \\ 

RSCnc & 080 & i & 2008-226 & +29.96 & \checkmark & \checkmark & \checkmark & \checkmark & \checkmark & \checkmark & \checkmark & \checkmark & \checkmark & \checkmark & X & X & \checkmark & \checkmark & \checkmark \\ 

RSCnc & 080 & e& 2008-226 & +29.96 & \checkmark & \checkmark & \checkmark & X & \checkmark & \checkmark & \checkmark & \checkmark & \checkmark & X & \checkmark & \checkmark & \checkmark & \checkmark & \checkmark \\ 

$\gamma$Cru & 081 & i & 2008-231 & -62.35 & \checkmark & \checkmark & \checkmark & \checkmark & \checkmark & \checkmark & \checkmark & \checkmark & X & \checkmark & \checkmark & \checkmark & \checkmark & \checkmark & \checkmark \\ 

$\gamma$Cru & 082 & i & 2008-238 & -62.35 & \checkmark & \checkmark & \checkmark & \checkmark & \checkmark & \checkmark & \checkmark & \checkmark & \checkmark & \checkmark & \checkmark & \checkmark & \checkmark & \checkmark & \checkmark \\ 

RSCnc  & 085 & i & 2008-263 & +29.96 & N/A & N/A & N/A & N/A & N/A & N/A & \checkmark & \checkmark & \checkmark & \checkmark & X & \checkmark & \checkmark & \checkmark & \checkmark \\ 

RSCnc  & 085 & e& 2008-263 & +29.96 & N/A & N/A & N/A & N/A & N/A & N/A & \checkmark & \checkmark & \checkmark & \checkmark & \checkmark & \checkmark & \checkmark & \checkmark & \checkmark \\ 

$\gamma$Cru & 086 & i & 2008-268 & -62.35 & \checkmark & \checkmark & \checkmark & \checkmark & \checkmark & \checkmark & \checkmark & \checkmark & \checkmark & \checkmark & \checkmark & \checkmark & \checkmark & \checkmark & \checkmark \\ 

RSCnc  & 087 & i & 2008-277 & +29.96 & N/A & N/A & N/A & N/A & N/A & N/A & N/A & N/A & N/A & N/A & \checkmark & \checkmark & \checkmark & \checkmark & \checkmark \\ 

RSCnc  & 087 & e& 2008-277 & +29.96 & N/A & N/A & N/A & N/A & N/A & N/A & N/A & N/A & N/A & N/A & \checkmark & \checkmark & X & \checkmark & \checkmark \\ 

$\gamma$Cru & 089 & i & 2008-290 & -62.35 & \checkmark & \checkmark & \checkmark & \checkmark & \checkmark & \checkmark & \checkmark & \checkmark & X & \checkmark & \checkmark & \checkmark & \checkmark & \checkmark & \checkmark \\ 

$\gamma$Cru & 093 & i & 2009-320 & -62.35 & \checkmark & \checkmark & \checkmark & \checkmark & \checkmark & \checkmark & \checkmark & \checkmark & \checkmark & \checkmark & \checkmark & \checkmark & \checkmark & X & \checkmark \\ 

$\gamma$Cru & 094 & i & 2008-328 & -62.35 & \checkmark & \checkmark & \checkmark & \checkmark & \checkmark & \checkmark & \checkmark & \checkmark & X & \checkmark & \checkmark & \checkmark & \checkmark & \checkmark & N/A \\ 

$\gamma$Cru & 096 & i & 2008-343 & -62.35 & N/A & N/A & N/A & N/A & N/A & N/A & N/A & N/A & N/A & N/A & N/A & N/A & N/A & N/A & \checkmark \\

$\gamma$Cru & 100 & i & 2009-012 & -62.35 & \checkmark & \checkmark & \checkmark & \checkmark & \checkmark & \checkmark & \checkmark & \checkmark & \checkmark & \checkmark & \checkmark & \checkmark & \checkmark & \checkmark & \checkmark \\ 

$\gamma$Cru & 102 & i & 2009-031 & -62.35 & \checkmark & \checkmark & \checkmark & \checkmark & \checkmark & \checkmark & \checkmark & \checkmark & \checkmark & \checkmark & \checkmark & \checkmark & \checkmark & \checkmark & \checkmark \\ 

$\beta$Peg & 104 & i & 2009-057 & +31.68 & \checkmark & \checkmark & \checkmark & \checkmark & N/A & \checkmark & \checkmark & N/A & N/A & \checkmark & N/A & N/A & N/A & N/A & N/A \\ 

RCas & 106 & i & 2009-082 & +56.04 & \checkmark & \checkmark & \checkmark & \checkmark & \checkmark & N/A & \checkmark & X & \checkmark & \checkmark & \checkmark & \checkmark & \checkmark & \checkmark & \checkmark \\ 

$\beta$Peg & 108 & i &  2009-095 & +31.68 & N/A & N/A & N/A & N/A & N/A & N/A & N/A & N/A & N/A & N/A & N/A & N/A & N/A & N/A & \checkmark \\ 

$\alpha$Sco & 115 & i & 2009-209 & -32.16 & \checkmark & \checkmark & \checkmark & \checkmark & \checkmark & \checkmark & \checkmark & \checkmark & \checkmark & \checkmark & \checkmark & \checkmark & \checkmark & \checkmark & \checkmark \\ 

$\beta$Peg & 170 & e& 2012-224 & +31.68 & \checkmark & \checkmark & \checkmark & \checkmark & \checkmark & \checkmark & \checkmark & \checkmark & \checkmark & \checkmark & \checkmark & \checkmark & \checkmark & \checkmark & \checkmark \\ 

$\beta$Peg & 172 & i & 2012-266 & +31.68 & \checkmark & \checkmark & \checkmark & \checkmark & \checkmark & \checkmark & \checkmark & \checkmark & \checkmark & \checkmark & \checkmark & \checkmark & \checkmark & \checkmark & \checkmark \\ 

$\lambda$Vel & 173 & i & 2012-292 & -43.81 & \checkmark & \checkmark & \checkmark & \checkmark & \checkmark & \checkmark & N/A & N/A & N/A & \checkmark & N/A & N/A & N/A & \checkmark & N/A \\ 

WHya & 179 & i & 2013-019 & -34.64 & \checkmark & \checkmark & \checkmark & \checkmark & \checkmark & \checkmark & \checkmark & \checkmark & \checkmark & \checkmark & \checkmark & \checkmark & \checkmark & \checkmark & \checkmark \\ 

WHya & 180 & i & 2013-033 & -34.64 & \checkmark & \checkmark & \checkmark & X & \checkmark & \checkmark & \checkmark & \checkmark & \checkmark & \checkmark & \checkmark & \checkmark & \checkmark & \checkmark & \checkmark \\ 

WHya & 181 & i & 2013-049 & -34.64 & \checkmark & \checkmark & \checkmark & \checkmark & \checkmark & \checkmark & \checkmark & \checkmark & \checkmark & \checkmark & \checkmark & \checkmark & \checkmark & \checkmark & \checkmark \\ 

RCas & 185 & i& 2013-091 & +56.04 & X & X & \checkmark & X & X & \checkmark & \checkmark & X & X & X & \checkmark & X & X & X & X \\ 

$\mu$Cep & 185 & e& 2013-090 & +59.90 & \checkmark & \checkmark & \checkmark & \checkmark & \checkmark & \checkmark & \checkmark & X & \checkmark & \checkmark & \checkmark & \checkmark & X & \checkmark & \checkmark \\ 

WHya & 186 & e& 2013-103 & -34.64 & \checkmark & \checkmark & \checkmark & \checkmark & \checkmark & X & \checkmark & \checkmark & \checkmark & \checkmark & \checkmark & \checkmark & \checkmark & \checkmark & N/A \\ 

$\gamma$Cru & 187 & i & 2013-112 & -62.35 & \checkmark & \checkmark & \checkmark & \checkmark & \checkmark & \checkmark & \checkmark & \checkmark & \checkmark & \checkmark & \checkmark & \checkmark & \checkmark & \checkmark & \checkmark \\ 

$\gamma$Cru & 187 & e & 2013-112 & -62.35 & \checkmark & \checkmark & \checkmark & \checkmark & \checkmark & \checkmark & \checkmark & \checkmark & \checkmark & \checkmark & \checkmark & \checkmark & \checkmark & \checkmark & \checkmark \\ 

WHya & 189 & e& 2013-132 & -34.64 & \checkmark & \checkmark & \checkmark & \checkmark & \checkmark & \checkmark & \checkmark & \checkmark & \checkmark & \checkmark & \checkmark & \checkmark & \checkmark & \checkmark & \checkmark \\ 

RCar & 191 & i &  2013-152 & -63.48 & \checkmark & \checkmark & \checkmark & \checkmark & X & X & X & \checkmark & \checkmark & \checkmark & X & X & \checkmark & X & \checkmark \\ 

RCas & 191 & i &  2013-149 & +56.04 & \checkmark & \checkmark & \checkmark & \checkmark & \checkmark & \checkmark & \checkmark & \checkmark & \checkmark & \checkmark & \checkmark & \checkmark & \checkmark & \checkmark & \checkmark \\ 

$\mu$Cep & 191 & i & 2013-148 & +59.90 & \checkmark & \checkmark & \checkmark & \checkmark & \checkmark & \checkmark & \checkmark & \checkmark & \checkmark & X & X & \checkmark & \checkmark & \checkmark & \checkmark \\ 

$\mu$Cep & 193 & i & 2013-172 & +59.90 & \checkmark & \checkmark & \checkmark & \checkmark & \checkmark & \checkmark & \checkmark & X & \checkmark & X & \checkmark & \checkmark & \checkmark & \checkmark & \checkmark \\ 

RCas & 194 & e& 2013-186 & +56.04 & \checkmark & \checkmark & \checkmark & \checkmark & \checkmark & X & X & \checkmark & \checkmark & \checkmark & \checkmark & \checkmark & X & \checkmark & X \\ 

2Cen & 194 & i & 2013-189 & -40.73 & N/A & N/A & \checkmark & \checkmark & \checkmark & \checkmark & N/A & N/A & \checkmark & \checkmark & \checkmark & N/A & \checkmark & X & N/A \\ 

2Cen & 194 & e& 2013-189 & -40.73 & \checkmark & \checkmark & \checkmark & \checkmark & \checkmark & N/A & N/A & \checkmark & N/A & \checkmark & N/A & \checkmark & \checkmark & N/A & N/A \\ 

RLyr & 198 & i & 2013-289 & +40.77 & \checkmark & \checkmark & \checkmark & \checkmark & \checkmark & N/A & \checkmark & \checkmark & \checkmark & \checkmark & \checkmark & \checkmark & \checkmark & \checkmark & \checkmark \\ 

L2Pup & 199 & e & 2013-327 & -41.91 & N/A & N/A & N/A & N/A & N/A & N/A & N/A & N/A & N/A & N/A & N/A & N/A & N/A & N/A & \checkmark \\ 

RLyr & 199 & i & 2013-337 & +40.77 & \checkmark & \checkmark & \checkmark & \checkmark & \checkmark & \checkmark & \checkmark & \checkmark & \checkmark & \checkmark & X & \checkmark & \checkmark & \checkmark & \checkmark \\ 

RLyr & 200 & i & 2014-003 & +40.77 & \checkmark & \checkmark & \checkmark & \checkmark & \checkmark & \checkmark & \checkmark & \checkmark & \checkmark & X & \checkmark & \checkmark & \checkmark & \checkmark & \checkmark \\ 

L2Pup & 201 & i & 2014-051 & -41.91 & X & X & X & X & X & X & X & X & X & X & X & X & X & X & X \\ 

RLyr & 202 & e& 2014-067 & +40.77 & X & X & \checkmark & \checkmark & \checkmark & \checkmark & \checkmark & \checkmark & \checkmark & \checkmark & \checkmark & \checkmark & \checkmark & \checkmark & \checkmark \\ 

RLyr & 202 & i& 2014-067 & +40.77 & N/A & N/A & N/A & N/A & N/A & N/A & N/A & N/A & N/A & N/A & N/A & N/A & N/A & N/A & \checkmark \\ 

L2Pup & 205 & e& 2014-175 & -41.91 & \checkmark & \checkmark & \checkmark & \checkmark & \checkmark & \checkmark & \checkmark & \checkmark & \checkmark & \checkmark & \checkmark & \checkmark & \checkmark & \checkmark & \checkmark \\ 

L2Pup & 205 & i& 2014-174 & -41.91 & N/A & N/A & N/A & N/A & N/A & N/A & N/A & N/A & N/A & N/A & N/A & N/A & N/A & N/A & \checkmark \\ 

L2Pup & 206 & e& 2014-206 & -41.91 & \checkmark & \checkmark & \checkmark & \checkmark & \checkmark & \checkmark & \checkmark & X & \checkmark & \checkmark & \checkmark & \checkmark & \checkmark & \checkmark & X \\ 

RLyr & 208 & e & 2014-262 & +40.77 & \checkmark & \checkmark & \checkmark & \checkmark & \checkmark & \checkmark & \checkmark & \checkmark & \checkmark & \checkmark & \checkmark & \checkmark & \checkmark & \checkmark & \checkmark \\ 

WHya & 236 & i & 2016-148 & -34.64 & N/A & N/A & N/A & N/A & N/A & N/A & N/A & N/A & N/A & N/A & N/A & N/A & N/A & N/A & N/A \\ 

$\rho$Per & 239 & e & 2016-221 & +45.27 & N/A & N/A & N/A & N/A & N/A & N/A & N/A & N/A & N/A & N/A & \checkmark & \checkmark & \checkmark & \checkmark & \checkmark \\ 

$\alpha$Sco & 241 & e& 2016-243 & -32.16 & \checkmark & \checkmark & \checkmark & \checkmark & \checkmark & \checkmark & \checkmark & \checkmark & \checkmark & \checkmark & \checkmark & \checkmark & \checkmark & \checkmark & \checkmark \\ 

$\alpha$Sco & 243 & e& 2016-267 & -32.16 & \checkmark & \checkmark & \checkmark & \checkmark & \checkmark & \checkmark & \checkmark & \checkmark & \checkmark & \checkmark & \checkmark & \checkmark & \checkmark & \checkmark & \checkmark \\ 

RCas & 243 & i & 2016-268 & +56.04 & \checkmark & \checkmark & \checkmark & \checkmark & \checkmark & \checkmark & X & X & \checkmark & \checkmark & \checkmark & \checkmark & \checkmark & \checkmark & \checkmark \\ 

$\alpha$Sco & 245 & e& 2016-287 & -32.16 & \checkmark & \checkmark & \checkmark & \checkmark & \checkmark & \checkmark & \checkmark & \checkmark & \checkmark & \checkmark & \checkmark & \checkmark & \checkmark & \checkmark & \checkmark \\ 

$\alpha$Sco & 245 & i&  2016-287 & -32.16 & N/A & N/A & N/A & N/A & N/A & N/A & N/A & N/A & N/A & N/A & N/A & N/A & N/A & N/A & N/A \\

$\gamma$Cru & 245 & e& 2016-286 & -62.35 & \checkmark & \checkmark & \checkmark & \checkmark & \checkmark & \checkmark & \checkmark & \checkmark & \checkmark & \checkmark & \checkmark & \checkmark & \checkmark & \checkmark & \checkmark \\ 

$\gamma$Cru & 255 & i & 2017-001 & -62.35 & \checkmark & \checkmark & \checkmark & \checkmark & \checkmark & \checkmark & \checkmark & \checkmark & \checkmark & \checkmark & \checkmark & \checkmark & \checkmark & \checkmark & \checkmark \\ 

$\gamma$Cru & 264 & i & 2017-086 & -62.35 & \checkmark & \checkmark & \checkmark & \checkmark & \checkmark & \checkmark & \checkmark & \checkmark & \checkmark & \checkmark & \checkmark & \checkmark & \checkmark & \checkmark & \checkmark \\

$\lambda$Vel & 268 & i & 2017-094 & -43.81 & \checkmark & \checkmark & N/A & \checkmark & \checkmark & N/A & \checkmark & N/A & N/A & \checkmark & \checkmark & N/A & \checkmark & N/A & N/A \\ 

$\gamma$Cru & 268 & i & 2017-095 & -62.35 & \checkmark & \checkmark & \checkmark & \checkmark & \checkmark & \checkmark & \checkmark & \checkmark & \checkmark & \checkmark & \checkmark & \checkmark & \checkmark & \checkmark & \checkmark \\ 

$\alpha$Ori & 268 & e& 2017-096 & +11.68 & \checkmark & \checkmark & \checkmark & \checkmark & \checkmark & \checkmark & \checkmark & \checkmark & \checkmark & \checkmark & \checkmark & \checkmark & \checkmark & \checkmark & \checkmark \\ 

VYCMa & 269 & i & 2017-100 & -23.43 & \checkmark & \checkmark & \checkmark & \checkmark & \checkmark & X & X & \checkmark & \checkmark & \checkmark & \checkmark & \checkmark & \checkmark & \checkmark & \checkmark \\ 

$\gamma$Cru & 269 & i & 2017-102 & -62.35 & \checkmark & \checkmark & X & \checkmark &  \checkmark & \checkmark & \checkmark & \checkmark & \checkmark & \checkmark & \checkmark & N/A & X & \checkmark & \checkmark \\ 

$\alpha$Ori & 269 & e& 2017-104 & +11.68 & \checkmark & \checkmark & \checkmark & \checkmark & \checkmark & \checkmark & \checkmark & \checkmark &  \checkmark & \checkmark & \checkmark & \checkmark & \checkmark & \checkmark & \checkmark \\ 

$\gamma$Cru & 276 & i & 2017-148 & -62.35 & N/A & N/A & \checkmark & N/A & \checkmark & \checkmark & N/A & N/A & N/A & N/A & N/A & N/A & \checkmark & N/A & N/A \\ 

$\alpha$Ori & 277 & i& 2017-155 & +11.68 & \checkmark & \checkmark & \checkmark & \checkmark & \checkmark & \checkmark & \checkmark & \checkmark & \checkmark & \checkmark & \checkmark & \checkmark & \checkmark & \checkmark & \checkmark \\

$\gamma$Cru & 282 & i & 2017-187 & -62.35 & N/A & N/A & \checkmark & \checkmark & N/A & N/A & \checkmark & N/A & N/A & \checkmark & \checkmark & N/A & \checkmark & N/A & N/A \\ 

$\gamma$Cru & 291 & i & 2017-245 & -62.35 & N/A & N/A & \checkmark & \checkmark & \checkmark & \checkmark & N/A & N/A & N/A & \checkmark & N/A & N/A & N/A & N/A & N/A \\ 

$\gamma$Cru & 292 & i & 2017-251 & -62.35 & N/A & N/A & \checkmark & \checkmark & \checkmark & N/A & \checkmark & N/A & N/A & \checkmark & N/A & N/A & N/A & N/A & N/A \\ 
\hline
\end{tabular}}}
\end{sidewaystable*}

\begin{sidewaystable*}
\caption{Wave-fit bounds for Satellite and Planetary resonances.}
\label{boundtabMAIN:C4}
{\begin{tabular}{lccccccccccr}
\hline
Resonance Name & Radial Range (km) & $r_{L}$ (km) & $\ell$ & $m$ & $A_{L,b}$ & $\xi_{D,b}$ & $\phi_{L,b}$ (rad) & $x_{r,b}$ (km) & $r_{f,b}$ (km) \\
\hline
Mimas 4:1 & 74880.00 - 74895.00 &  74890.07 & - & 2 & (0, 1.0) & (1, 5) & (-2$\pi$, 2$\pi$) & (-10, 10) & (0, 4) \\
Pan 2:1 & 85090.00 - 85130.00 & 85105.02 & - & 2 & (0, 1.0) & (1, 15) & (-2$\pi$, 2$\pi$) & (0, 5) & (0, 4) \\
Atlas 2:1 & 87640.00 - 87650.00 & 87645.68 & - & 2 & (0, 1.0) & (1, 15) & (-2$\pi$, 2$\pi$) & (-10, 10) & (0, 2) \\
Prometheus 4:2 & 88420.00 - 88440.00 & 88434.12 & - & 3 & (0, 1.0) & (1, 15) & (-2$\pi$, 2$\pi$) & (-10, 10) & (0, 3) \\
Mimas 6:2 & 89870.00 - 89890.00 & 89884.00 & - & 3 & (0, 1.0) & (1, 15) & (-2$\pi$, 2$\pi$) & (-10, 10) & (0, 4) \\
Pandora 4:2 & 89887.00 - 89897.00 & 89893.68 & - & 3 & (0, 1.0) & (1, 15) & (0, 2$\pi$) & (-10, 10) & (0, 3) \\
\hline
$W74.51^{av}$ & 74501.00 - 74509.00 & 74506.900 & 12 & -8 & (0, 0.8) & (1, 30) & (0, 2$\pi$) &(-10, 10) &(0, 1) \\
W74.74 & 74736.00 - 74743.00 & 74739.850 & 15 & 13 & (0, 0.2) & (1, 10) & (-2$\pi$, 2$\pi$) & (-2, 1) & (0, 1) \\
W74.75 & 74746.00 - 74748.00 & 74748.300 & 11 & 11 & - & - & - & - & - \\
W74.76 & 74752.00 - 74762.00 & 74756.600 & 19 & -11 & (0, 1.0) & (0.5, 10) & (-2$\pi$, 2$\pi$) & (-1, 3) & (0, 1) \\
W75.14 & 75142.00 - 75144.00 & 75143.000 & 16 & -10 & (0, 0.8) & (1, 40) & (-2$\pi$, 2$\pi$) & (-10, 10) & (0, 1) \\
W76.02A & 76016.00 - 76018.00 & 76018.100 & 13 & -9 & (0, 0.6) & (1, 10) & (-2$\pi$, 2$\pi$) & (-10, 10) & (0, 2) \\
W76.44 & 76433.00 - 76436.00 & 76435.400 & 2 & -2 & (0, 0.6) & (1, 25) & (0, 2$\pi$) & (-10, 10) & (0, 2) \\
$W76.46^{av}$ & 76457.00 - 76462.00 & 76459.500 & 9 & -7 & (0, 0.8) & (1, 40) & (-$\pi$, $\pi$) & (-5, 10) & (0, 3) \\
W77.34 & 77337.00 - 77339.00 & 77338.900 & 14 & -10 & (0, 0.3) & (1, 35) & (-2$\pi$, 2$\pi$) & (-10, 10) & (0, 1) \\
W78.51 & 78503.00 - 78509.00 & 78506.750 & 15 & -11 & (0, 0.8) & (1, 20) & (-2$\pi$, 2$\pi$) & (-10, 10) & (0, 1) \\
W79.04 & 79039.00 - 79045.00 & 79042.300 & 11 & -9 & (0, 0.4) & (1, 25) & (0, 2$\pi$) & (-10, 10) & (0, 1) \\
W79.55 & 79546.00 - 79550.00 & 79548.920 & 16 & -12 & (0, 0.2) & (1, 25) & (-2$\pi$, 2$\pi$) & (-10, 10) & (0, 1) \\
W80.49 & 80484.00 - 80488.00 & 80486.100 & 17 & -13 & (0, 0.3) & (1, 25) & (0, 2$\pi$) & (-10, 10) & (0, 1) \\
W80.99 & 80983.00 - 80989.00 & 80986.150 & 4 & -4 & (0, 0.2) & (1, 25) & (0, 2$\pi$) & (-10, 10) & (0, 2) \\
W81.023A & 81018.00 - 81030.00 & 81023.100 & 5 & -5 & (0, 0.2) & (1, 25) & (-2$\pi$, 2$\pi$) & (-10, 10) & (0, 2) \\
W81.024B & 81018.00 - 81030.00 & 81024.150 & 13 & -11 & (0, 0.3) & (1, 25) & (-2$\pi$, 2$\pi$) & (-10, 10) & (0, 1) \\
W81.33 & 81333.00 - 81336.00 & 81334.275 & 18 & -14 & (0, 0.1) & (1, 15) & (-2$\pi$, 2$\pi$) & (-10, 10) & (0, 2) \\
W81.43 & 81420.00 - 81430.00 & 81429.550 & 6 & -6 & (0, 0.2) & (1, 30) & (-2$\pi$, 2$\pi$) & (-10, 10) & (0, 2) \\
W81.96 & 81959.00 - 81965.00 & 81962.450 & 7 & -7 & (0, 0.1) & (1, 30) & (-2$\pi$, 2$\pi$) & (-10, 10) & (0, 3) \\
W82.01 & 82004.00 - 82009.00 & 82007.750 & 3 & -3 & (0, 0.2) & (1, 5) & (-2$\pi$, 2$\pi$) & (-10, 10) & (0, 4) \\
W82.06 & 82055.00 - 82065.00 & 82059.400 & 3 & -3 & (0, 0.5) & (1, 40) & (-2$\pi$, 2$\pi$) & (-10, 10) & (0, 3) \\
W82.21 & 82187.50 - 82207.51 & 82207.500 & 3 & -3 & (0, 0.3) & (1, 6) & (-2$\pi$, 2$\pi$) & (-15, 15) & (0, 4) \\
W82.53 & 82510.00 - 82530.00 & 82528.750 & 8 & -8 & (0, 0.1) & (1, 15) & (-2$\pi$, 2$\pi$) & (-10, 10) & (0, 2) \\
W82.61 & 82606.00 - 82608.00 & 82607.750 & 15 & -13 & (0, 0.2) & (1, 25) & (-2$\pi$, 2$\pi$) & (-10, 10) & (0, 2) \\
W83.09 & 83086.00 - 83096.00 & 83090.650 & 9 & -9 & (0, 0.2) & (1, 25) & (-2$\pi$, 2$\pi$) & (-10, 10) & (0, 2) \\
W83.63 & 83612.02 - 83637.02 & 83632.020 & 10 & -10 & (0, 0.3) & (1, 20) & (-2$\pi$, 2$\pi$) & (-10, 10) & (0, 2) \\
W84.15 & 84140.00 - 84150.00 & 84147.100 & 11 & -11 & (0, 0.2) & (1, 10) & (-2$\pi$, 2$\pi$) & (-10, 10) & (0, 2) \\
W84.64 & 84630.00 - 84650.00 & 84643.200 & 2 & -2 & (0, 0.3) & (1, 25) & (-2$\pi$, 2$\pi$) & (-10, 10) & (0, 3) \\
W87.19 & 87170.00 - 87210.00 & 87192.800 & 2 & -2 & (0, 0.2) & (1, 40) & (-2$\pi$, 0) & (-10, 10) & (0, 3) \\
\hline
\end{tabular}}
\end{sidewaystable*}

\clearpage

\bibliography{reference}

\end{document}